# FORECASTING SEP EVENT PROPERTIES BY FITTING LIGHT CURVES WITH FUNCTIONAL FORMS

Dr. Stephen M. White
AFRL/RVBXD

**Technical Report**

**Geospace Environment Applications and Impacts Program**

**POC: Arron Layns, PM**

This page intentionally left blank.

# Forecasting SEP Event Properties by Fitting Light Curves with Functional Forms

Stephen White (AFRL/RVBXD)

2021 Jan 20

**Executive Summary**. The use of functional forms for predicting the profiles of solar energetic particle (SEP) events by fitting the early rise phase is explored. Three functional forms (lognormal, Weibull and inverse Weibull) are fitted to 410 $>$ 10 MeV proton datasets from GOES satellites, representing 203 distinct SEP events. Fits are made both to the flux intensity light curves as well as the smoother cumulative fluence time profiles. The fitted functions have amplitude and "zero-time" (onset) parameters in addition to the two "shape" parameters of the corresponding distribution functions. Fits to the full event durations are used to determine the longitude dependence of the shape parameters. These results show that lognormal and Weibull functions fit virtually all events, but inverse Weibull is less successful with the particular version of the Levenberg-Marquardt least-squares minimization technique being used. Different options for specifying the properties of an event by fitting the rise phase as a function of time are explored, with the goal of identifying the process that gives the earliest plausible forecasts. In order to facilitate successful fitting early in an event, the shape parameters need to be fixed, and we use values corresponding to the event longitude. For at least the first hour, the zero-time parameter needs to be fixed at the onset time or else fitting fails. The shape parameters cannot reliably be fitted until the event peak has been passed. Determination of an optimal forecasting process focusses on fitting some 78 relatively well-behaved events selected from the sample, on the grounds that events with more erratic time profiles are not likely to be fit successfully by functional forms with just two shape parameters. A prototype fitting process is identified that can forecast the peak of "good" events to within about a factor of 5 within 1 hour of event onset. This is the first step in identifying a light-curve fitting process that provides value for SEP forecasting.

## Introduction

The purpose of this report is to assess the feasibility of forecasting the profiles of solar energetic particle (SEP) events by fitting the rise phase of such events in real time with functional forms shown previously to provide reasonable representations of the light curves of (arguably idealized) SEP events. In this report we focus on data for protons with energies exceeding 10 MeV as measured by detectors on NOAA's GOES weather satellites.

As will be discussed below, the functional forms investigated each have 4 free parameters. Since SEP events can reach their peak within a few hours of onset, it is critical that a forecast be available as quickly as possible after event onset: since the standard GOES data have 5-minute resolution, this means that we wish to fit the rise with relatively few points, and therefore it is advantageous to be fitting as few parameters as possible. Each of the functional forms has 2 shape parameters, 1 amplitude parameter and 1 reference-time parameter: we explore whether it is possible to fix the shape parameters during the rise phase by fitting the parameters for a large number of SEP events and determining whether the shape parameters have a distribution that yields suitable options. Since east-limb SEP events generally are more gradual than well-connected west-limb events, we assume that it may be useful to alter the shape parameters depending on the longitude of the associated flare event, and first assess this question.

The intent is to identify the best method for using such fits to forecast the properties of future SEP events early in their rise phase, and to provide an assessment of the uncertainties involved. We describe the data used in the fitting, introduce the properties of the functional forms being considered, fit the full profiles of a large number of events and determine the dependence of the shape parameters on event longitude. We then use those results to investigate fitting the rise phase of events. We describe a prototype forecasting procedure and assess its uncertainties.

## Solar Energetic Particle Data

For the analysis presented here, we use the proton data collected by NOAA's GOES geosynchronous weather satellites. The event list covers the period 1986–2017, and mostly comes from NOAA's list of events that exceeded 10 pfu for >10 MeV protons, supplemented with some slightly smaller events compiled by Steve Kahler (AFRL). The standard NOAA 5-minute cadence text files were used to extract light curves for the periods of known SEP events, starting at least several hours before the nominal event onset and continuing to the nominal event end. NOAA generally has two satellites providing data, with one designated as the "primary" and the other serving as a "secondary" backup in case of issues with the primary (including Earth-blockage during eclipse season).

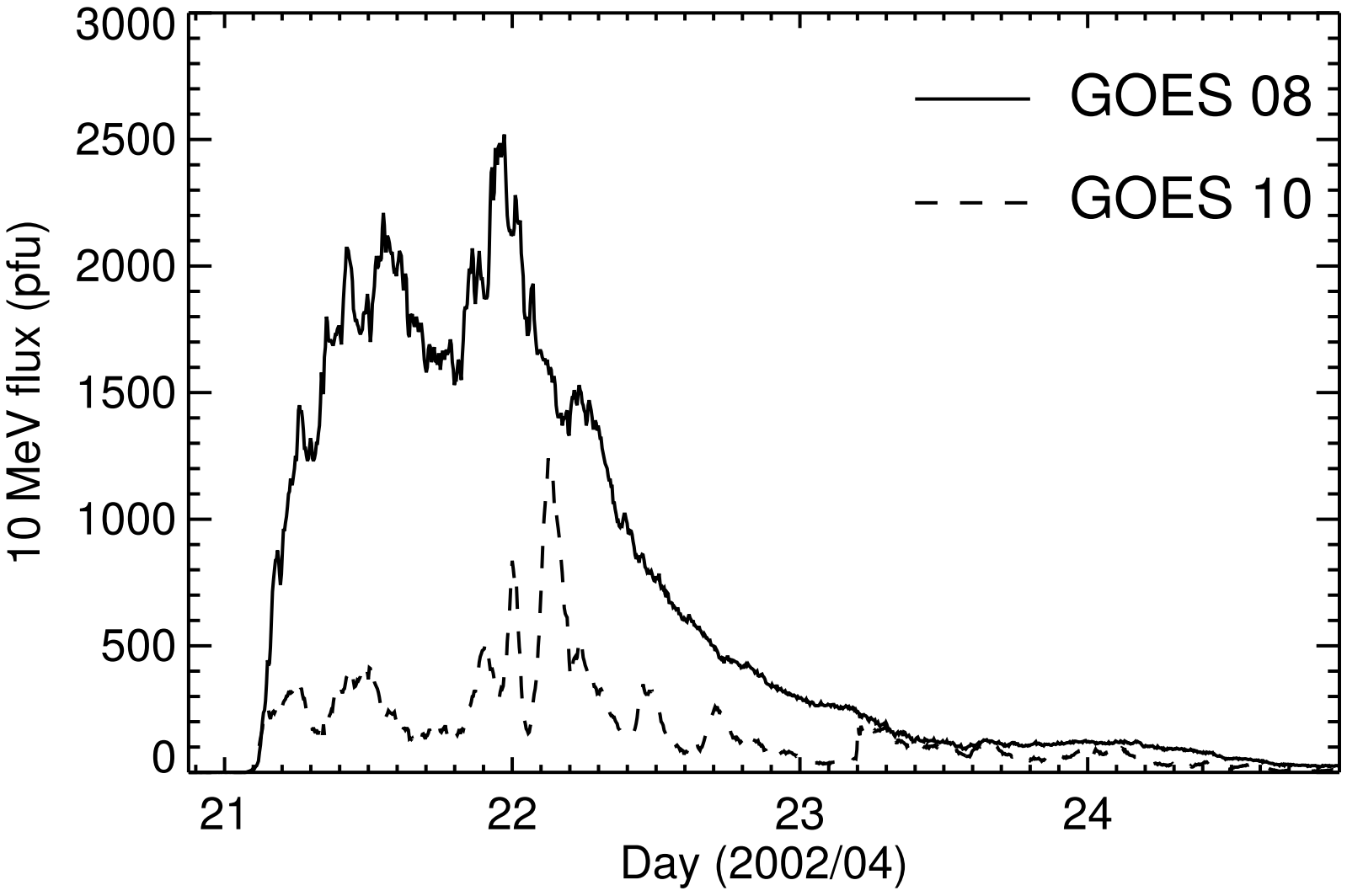

Fig. 1.— The SEP event of 2002 April 21 as observed by the detectors on GOES 08 (solid line) and GOES 10 (dashed line).

The datasets span the use of GOES satellites from GOES-06 to GOES-15, and the SEP detectors in use on the satellites have evolved over the years. In particular, GOES-11 and earlier satellites have a single detector value recorded, while later satellites (GOES-13 and 15, for our purposes) report values for both "east-facing" and "west-facing" detectors. Thanks to the structure of the magnetosphere, the actual west-facing detector is more sensitive to solar protons, while the east-facing detector is more sensitive to inner–radiation–belt particles. However, due to operational usage the spacecraft can change orientation such that the detector labelled "west-facing" is actually facing east. This is described in a document on the NGDC website:[1] "The EPEAD telemetry channels labelled 'E' look westward when the spacecraft is upright and eastward when the spacecraft is inverted", and inversely for the telemetry channels labelled 'W'. GOES-13 was upright, GOES-14 was inverted, and GOES-15 did a yaw flip twice a year at the equinoxes. The datasets from these satellites were selected for solar protons accordingly, and the results were checked against reported event fluxes for confirmation.

Typically there is data from both the primary and the secondary satellite (for a brief period data from 3 satellites are available). The two satellites are located at very different longitudes in the geosynchronous belt, and their energetic particle detectors can have very different orientations relative to the incoming SEP flux. Thus the two satellites can see quite different flux profiles, particularly when only one detector is available per satellite: an example is shown in Fig. 1, from the west-limb event on 2002 April 21.

[1] `https://ngdc.noaa.gov/stp/satellite/goes/doc/Note%20on%20GOES%2013-15%20Yaw%20Flip.pdf`

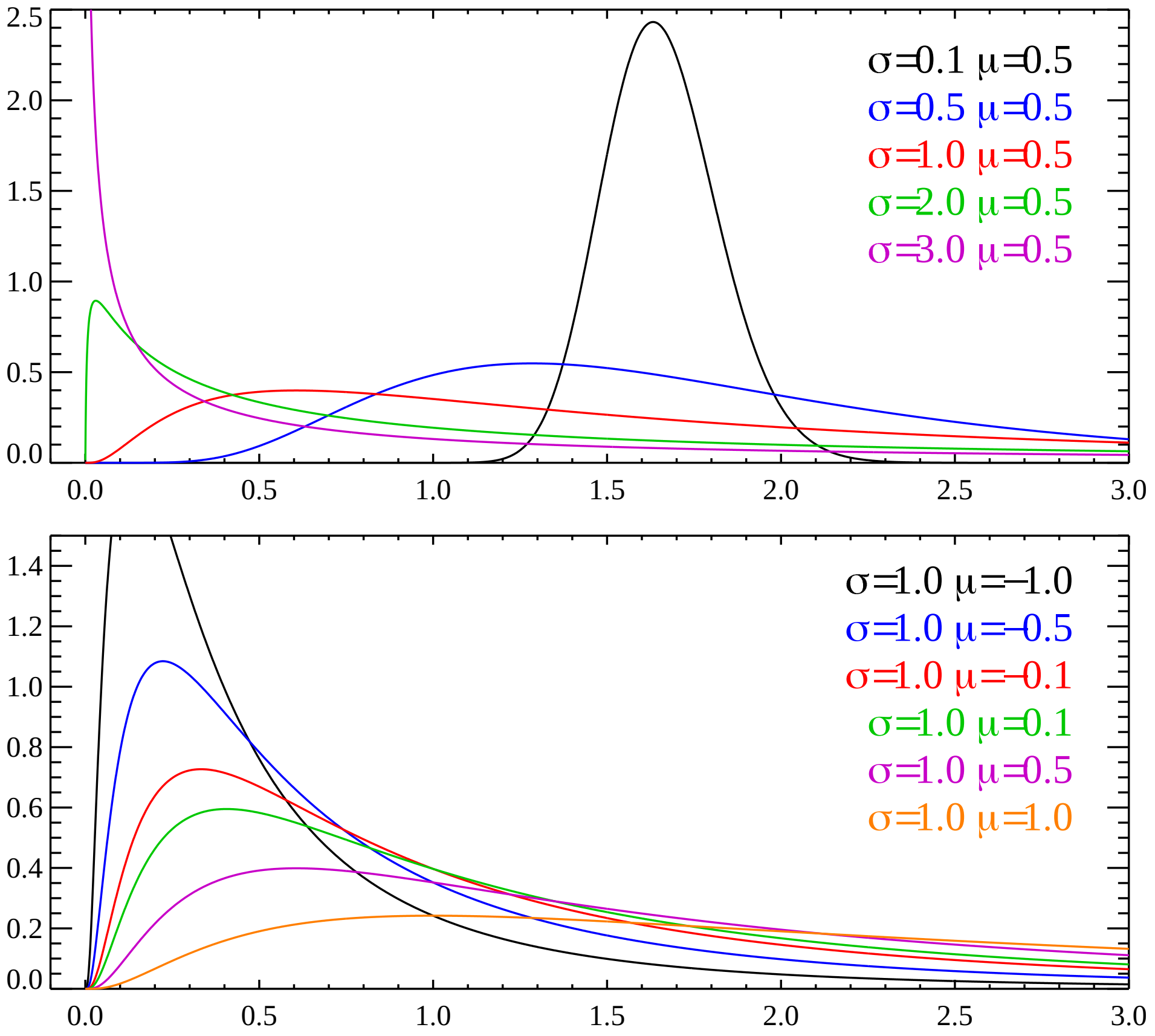

Fig. 2.— Changes in the shape of the lognormal function (top) as a function of $\sigma$ for fixed $\mu$, and (bottom) as a function of $\mu$ for fixed $\sigma$. Note that $\sigma$ can be required to be positive, whereas $\mu$ can be positive or negative.

The total number of unique SEP events in the dataset used here is 203, resulting in 410 light curves at >10 MeV. Some of the events do not reach NOAA's threshold of 10 pfu, but weaker events with good signal-to-noise are still useful for surveying event shapes. We note that the time periods for events during very active periods often had to be manipulated in order to separate events during the fitting procedure, and there will certainly be instances where such separation was not readily feasible and overlapping events may be contained within the one file. In other cases apparently distinct peaks were separated as distinct events. For our initial purpose of determining the longitudinal dependence of the parameters in the fits, this does not matter significantly.

## Functional forms

Following Kahler & Ling (2017), we consider six different candidates for fitting SEP light curves: the lognormal function, the Weibull function and the inverse Weibull function,

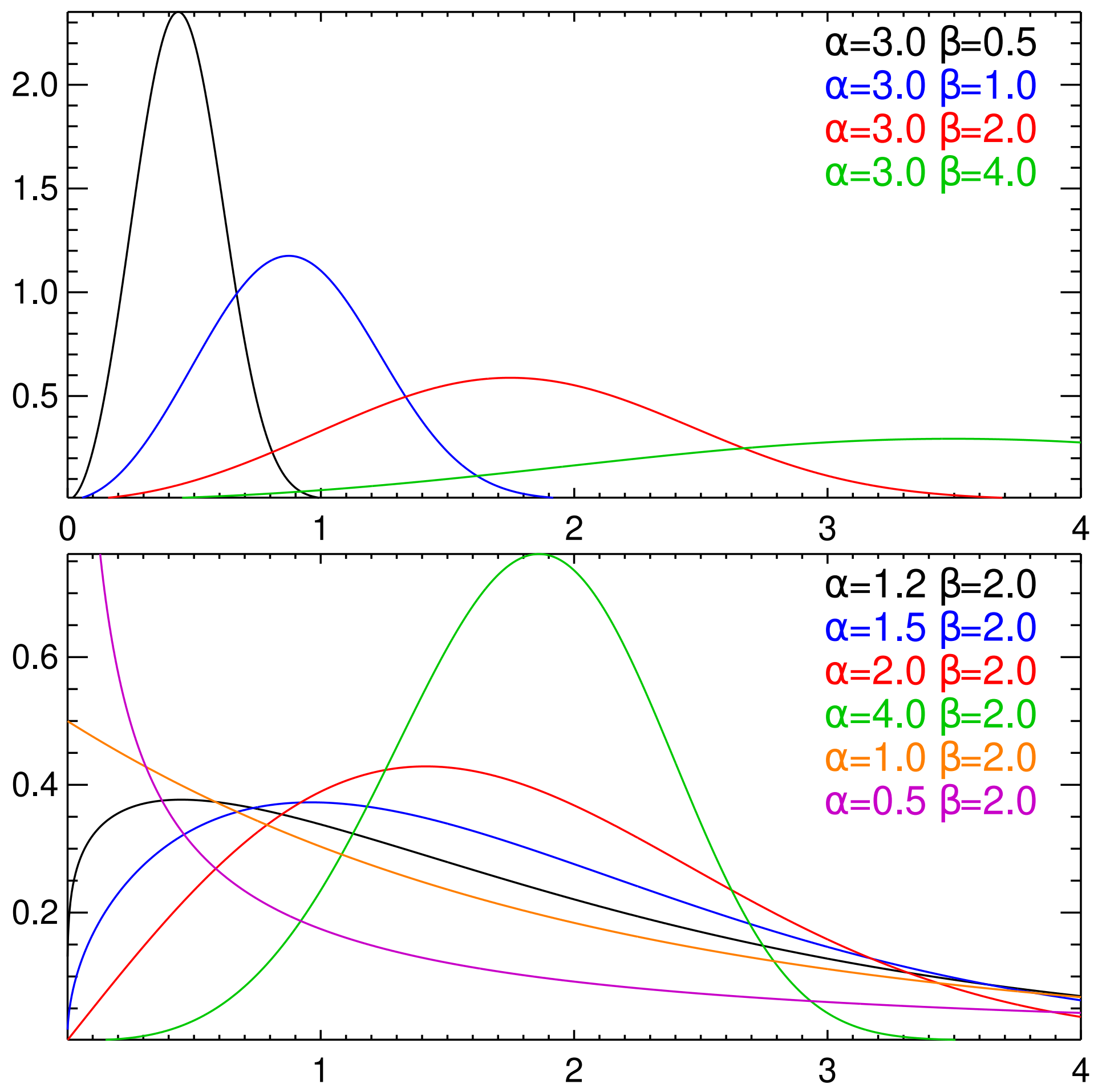

Fig. 3.— Changes in the shape of the Weibull function ($\alpha > 0$) as a function (top) of $\beta$ for fixed $\alpha$, and (bottom) as a function of $\alpha$ for fixed $\beta$.

and the cumulative (or fluence) versions of each. The basic functional forms are

- lognormal, with the 4 parameters $A$, $t_0$, $\sigma$ ($> 0$) and $\mu$:

$$\mathrm{Ln}(t) = \begin{cases} 0, & t \leq t_0 \\ \frac{A}{\sqrt{2\pi}\,\sigma\,(t-t_0)} \exp\left\{-\frac{(\ln(t-t_0)-\mu)^2}{2\,\sigma^2}\right\}, & t > t_0 \end{cases} \quad (1)$$

- Weibull, with the 4 parameters $A$, $t_0$, $\alpha > 1$ and $\beta$:

$$\mathrm{We}(t) = \begin{cases} 0, & t \leq t_0 \\ A\,\frac{\alpha}{\beta}\left(\frac{t-t_0}{\beta}\right)^{\alpha-1} \exp\left\{-\left(\frac{t-t_0}{\beta}\right)^{\alpha}\right\} & t > t_0 \end{cases} \quad (2)$$

- and inverse Weibull, with the same parameters but with $\alpha < 0$:

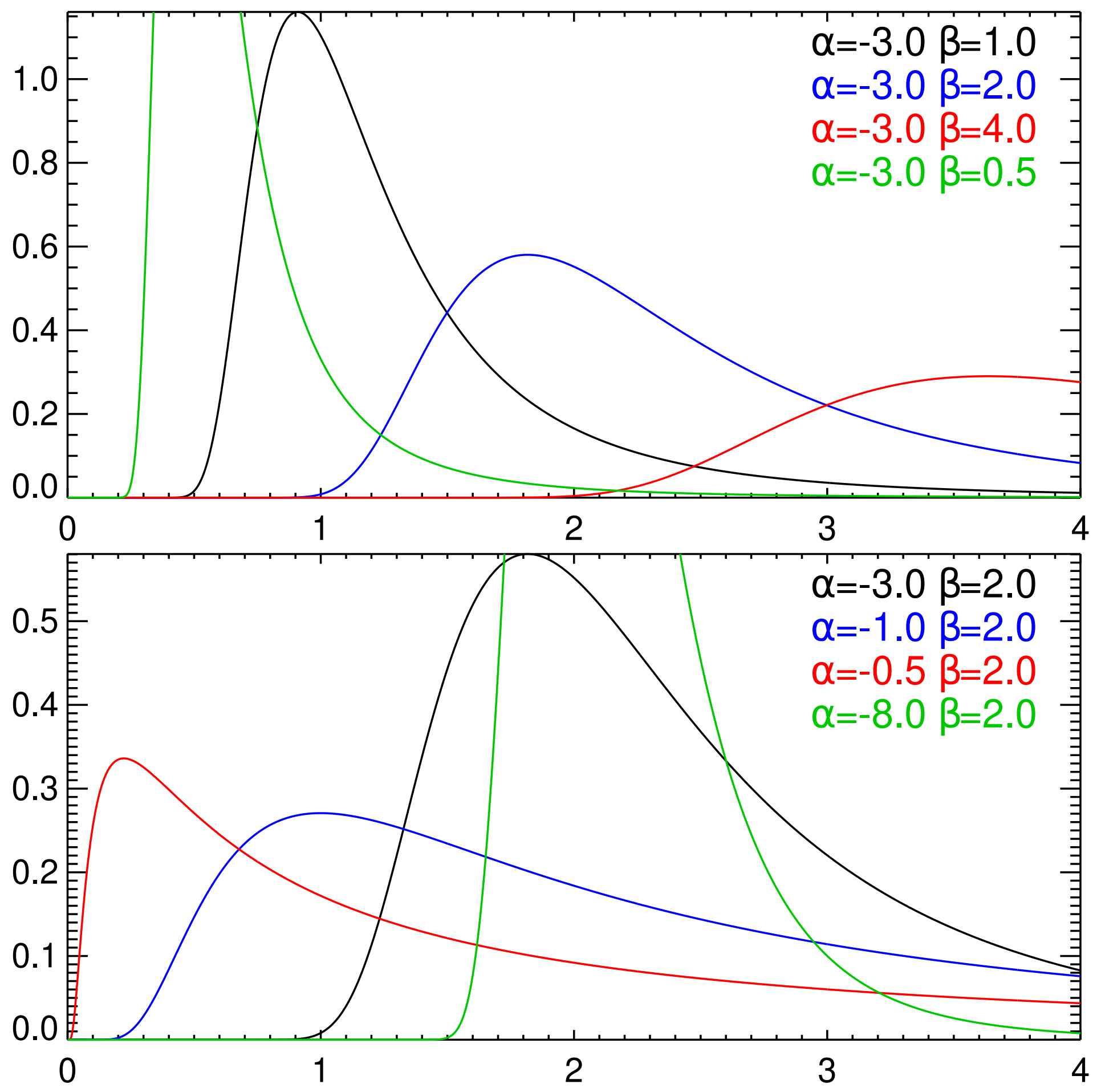

Fig. 4.— Changes in the shape of the inverse Weibull function ($\alpha < 0$) as a function (top) of $\beta$ for fixed $\alpha$, and (bottom) as a function of $\alpha$ for fixed $\beta$.

$$\mathrm{We}(t) \;=\; \begin{cases} 0, & t \le t_0 \\ A\,\frac{|\alpha|}{\beta}\,\left(\frac{t-t_0}{\beta}\right)^{\alpha-1} \exp\left\{-\left(\frac{t-t_0}{\beta}\right)^{\alpha}\right\} & t > t_0 \end{cases} \tag{3}$$

The lognormal and Weibull functions serve as distribution functions with 2 shape parameters. The forms used here need 2 more parameters: an amplitude, $A$, and a "zero" time, $t_0$, since the distribution functions do not exist for arguments less than zero. The reason for the distinction between Weibull and inverse Weibull functions is that the Weibull function is infinite at $x = 0$ for $0 < \alpha < 1$, resulting in a void in the allowed range for $\alpha$. Fitting can be achieved by the method described below without distinguishing the two cases, but we provide reasons for the distinction later. Note that the argument in the Weibull forms is always divided by $\beta$, so it has a clear role as a timescale parameter. In order to be suitable for fitting light curves that start from a low value, $\alpha > 1$ must be satisfied by the shape parameter in the Weibull function, and negative values of $\beta$ have no useful meaning. On the other hand $\mu$ in the lognormal function acts as a time offset, and can be positive or negative.

Plots of the distribution functions showing how the shape changes as a function of the two shape parameters are provided in Figures 2-4. In particular, note that while (for $\alpha > 1$) the Weibull function always rises promptly at the origin, the inverse Weibull function need not do so, and indeed in general does not. The rise of the inverse Weibull function does not start until about $x = \beta/2$. Similarly, the lower panel of Fig. 4 shows that the value of $\alpha$ also plays a major role in determining when the functional form starts to rise. This seems to leads to a strong co-dependence between $\alpha$ and $\beta$, whereby very different pairs of values can produce very similar shapes. We give examples of this further below. Due to this co-dependence, the fitted values for $\alpha$ and $\beta$ in the inverse Weibull function can change significantly depending on the allowed time ranges in the fits. This is not ideal for prescribing values of the shape parameters for the early phase of the fit.

The cumulative forms of these functions are given in the appendix: they are analytic functions for $t > t_0$. The reason for considering fits to the cumulative functions is that light curves seldom have the nice smooth shape that people often think of with logarithmic plots of SEP events in mind, and this complicates fits to the light curves: the cumulative light curves are smoother, and may provide fits that are less susceptible to fluctuations in the light curve. This will be demonstrated below. Figure 5 shows an event with fits for each of the 6 functional forms described above.

We note one convenient result of using probability distribution functions for fitting: by constructon they must integrate to unity. It follows that the fitted amplitude paramater $A$ is in fact just the fluence of the event[2], which is one of the properties of SEP events that is important for space-weather impacts and understanding radiation dosages that may be experienced.

## Analysis Methods for Events

As a prelude to providing initial estimates of parameters to enable fitting during the rise of an event, we first discuss fits to the full light curves of the selected events in order to assess the longitudinal dependence of the fitted parameters. The fitting procedure used for the results reported here was as follows:

- Each light-curve file was loaded; an iterative procedure was used to identify a linear background trend from the first 2.5 hours of data, then the point at which the >10

[2]with suitable adjustment for units: the proton flux unit (pfu) is 1 proton $cm^{-2}$ $s^{-1}$ $sr^{-1}$, whereas we are using days as our time variable.

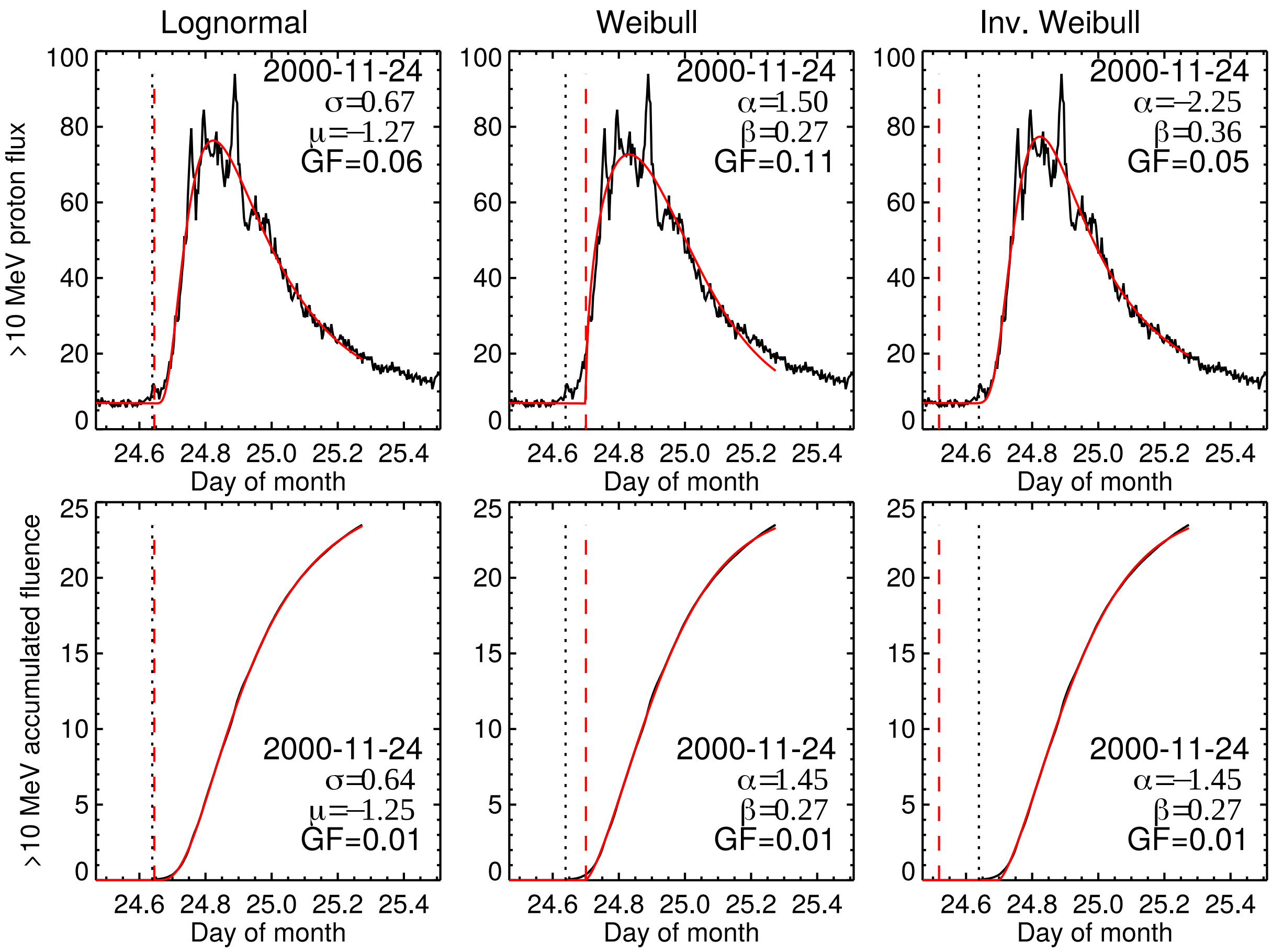

Fig. 5.— Functional fits to an SEP event for each of the 6 functional forms considered here. The fitting parameters are shown in each panel. The vertical red dashed line represents the fitted value of $t_0$ in each case, while the black dotted line marks the starting onset time $t_{on}$ determined from the data.

MeV count rate is 5 standard deviations and at least 0.6 sfu above this background is identified. This is the nominal onset time, $t_{on}$. The threshold of 0.6 sfu was found to be a convenient limit to avoid rise times longer than seem reasonable for some events, which would occur without that limit. The background is not permitted to increase with time, and is not allowed to be negative at any time.

- For the fits to full events, the data range considered extends to the point at which the flux has decayed to 20% of the peak value. Some events have to be truncated before this point is reached due to the onset of another event.

- The cumulative (fluence) datasets as a function of time for each event are determined by summing the flux in the background-subtracted light curve from $t_{on}$ forward to each subsequent proton measurement. They are set to zero at times prior to $t_{on}$.

- The background-subtracted data (forced to be $> 0$) are then fed to a Levenburg-Marquardt fitting routine[3] supplied with the functions described above: the functional forms permit negative values of the argument $t$ through setting the functions to zero for $t < t_0$ The fitting routine seems to work better if it does not have to work too hard to find $t_0$, so the time arguments are supplied with the nominal offset time $t_{on}$ subtracted, providing a guess at $t_0$.

- In order to optimize the speed of fitting, the parameters are constrained to lie within certain ranges: thus $0.1 < \sigma < 10$, $-10 < \mu < 10$ and $-0.1 < t_0 < 5$ for lognormal; $1 < \alpha < 10$, $0.01 < \beta < 10$ and $-5 < t_0 < 5$ for Weibull; and $-10 < \alpha < 0$, $0.01 < \beta < 10$ and $-9 < t_0 < 9$ for inverse Weibull.

- We note that $t_0$ is allowed large negative values for Weibull and inverse Weibull, but not for lognormal: we tested this restriction and found that the only events where a lognormal fit preferred a large negative value of $t_0$ were events that were fitted poorly, and widening the allowed range of $t_0$ did not significantly improve the quality of the fit in most of these cases. On the other hand, for reasons described earlier, allowing negative $t_0$ can significantly change the fits for the Weibull options.

- The four parameters for each functional form are permitted to float within reasonable ranges. The fitting routine does not always find an acceptable result, even when the light curve appears suitable: the reasons for such failures may be explored at a later point.

- For the resulting fits, a "goodness-of-fit" metric is calculated as the mean absolute fractional error relative to a 7-point smoothed version of the light curve, $\tilde{F}$, divided by the value at each point (with a minimum divisor of either the maximum of the noise level or 1 pfu), normalized over the number of points in the light curve after the event starts:

$$GF = \frac{1}{(n_f - n_0)} \sum_{i=n_0}^{n_f} \left| \frac{Y_{fit}(i) - \tilde{F}(i)}{\tilde{F}(i) > ((5 * RMS) > 1.0)} \right| \tag{4}$$

where $Y_{fit}$ is the functional fit, $n_0$ is the data point corresponding to the time $t_0$, $n_f$ is the last point in the fitted data range, and $RMS$ is the noise level in the background data prior to the start of the event. The resulting metric appears reasonable based on visual inspection, although possibly biased against events with very high count rates. Note that the goodness-of-fit measures for the light curves and the cumulative functions cannot be compared directly since they are not measuring the same quantity. We recognize that this measure is not statistically robust, and we use it only to select sub-samples of events that are well fit. We also determine the value of $\chi^2$ divided by the number of degrees of freedom returned by *mpfit*, but since our estimate of the noise in the data, particularly for the fluence, is uncertain, we are unsure of the value of this property as a measure of the success of the fit.

[3] "mpfit", as implemented in the Interactive Data Language (IDL) by Craig Markwardt; see `https://pages.physics.wisc.edu/~craigm/idl/`

As noted above, the variable $t_0$ that appears in the functional forms need not match the onset time $t_{on}$ of an event, for mathematical reasons. One can envisage that for real-time fitting there might be reason to enforce $t_0 = t_{on}$: to explore that issue, all the events were fitted with both cases: $t_0 = t_{on}$ ($t_0$ fixed, hence 3 free parameters to fit), and $t_0$ variable (floating; 4 free parameters to fit).

Figure 6 shows another set of events with lognormal and Weibull fits over-plotted. This particular sample of events has explicitly been chosen to illustrate an important point for this study, already mentioned above: the use of logarithmic axes for plots of SEP events tends to obscure the fact that most events do not actually have the nice regular shapes shown by the functional forms in Figs. 2-4, and in fact they are often very irregular, which will complicate the real-time fitting process.

Figure 7 illustrates a context in which such "irregular" SEP events may occur. A long-duration M7 flare located at heliographic coordinates S26E08 occurred at 1540 UT on May 08, about 12 hours before protons were first detectable above background. No flare of class C or greater was detected for another 10 days after this. Essentially all flares in the previous 3 days were produced by the region responsible for the M7 flare; another active region had gone over the west limb 3 days earlier. The linear plot seems to suggest that the SEP event starts at 20 UT on the 9th, whereas the log plot clearly shows protons arriving 16 hours earlier. The delay in first arrival compared to the timing of the M7 flare indicates that the SEP source was not magnetically well connected to the Earth. The observed light curve can be explained in 3 ways. Given the location of the flare, the SEP source may have been poorly connected until 20 UT on the 9th when it suddenly became well connected; an unseen flare on the 9th from the region 30° behind the west limb produced the SEPs; or the peak results from the arrival of the acceleration source at Earth. No matter the cause, fitting a single functional form cannot capture the two-part history of the event; but for space weather purposes, it is more important to capture the most intense period starting at 20 UT.

We note that in the third scenario, i.e., the passage of a shock locally conveying higher densities of trapped energetic particles, the resulting sharp increases in proton fluxes in the middle of an event are described as "energetic storm particles" (or ESP, e.g., Cohen 2006), since the accompanying shock passage at Earth often triggers a geomagnetic storm at the same time. ESP features are usually more prominent at energies below 10 MeV that cannot as easily escape the vicinity of a shock. In this analysis we make no attempt to separate out

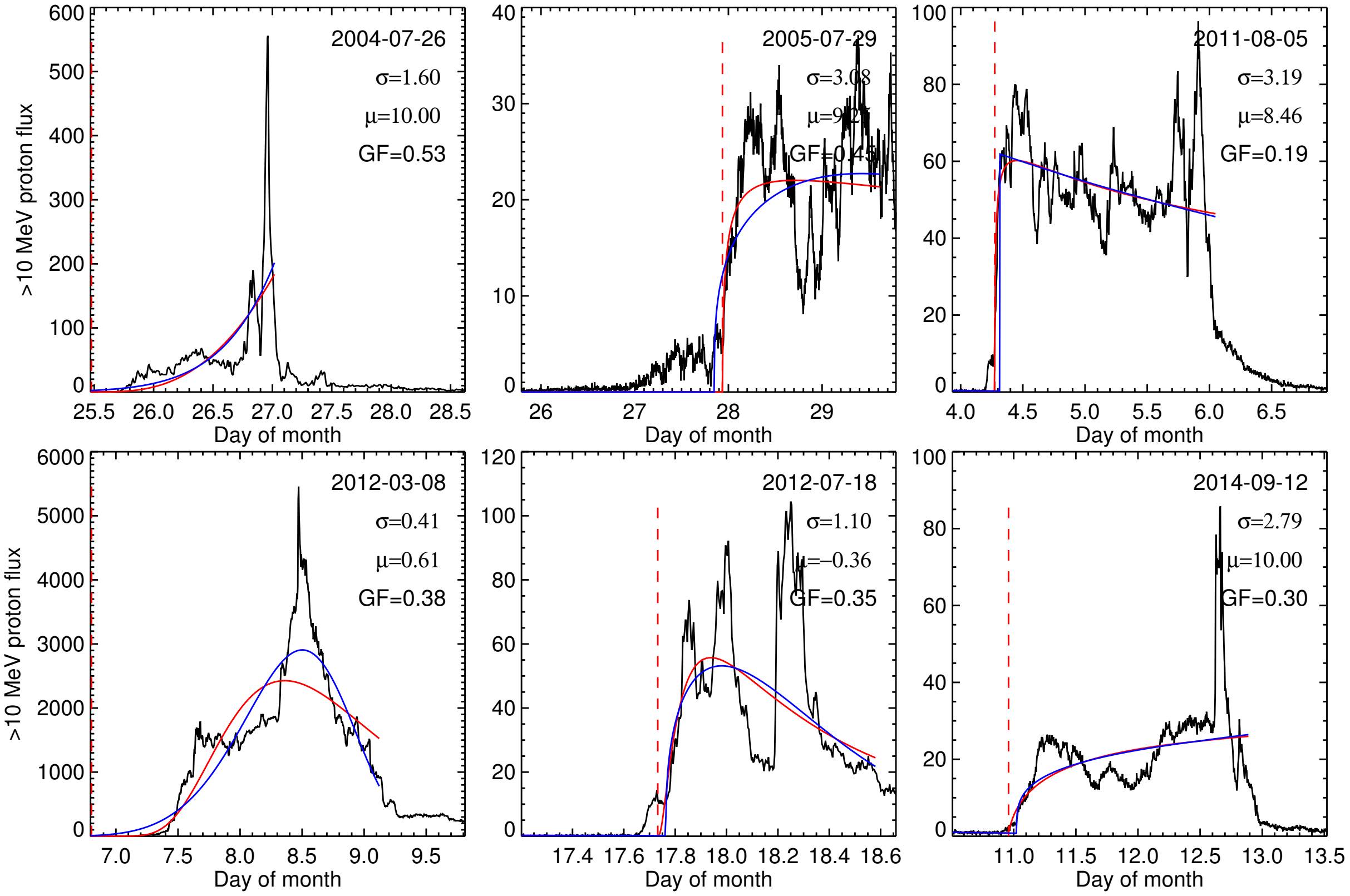

Fig. 6.— Six SEP events chosen to illustrate the fact that generally SEP events do not have the smooth regular behavior illustrated in the plots of the functional forms shown in Figs. 1-3. The red curve in each panel represents a lognormal fit to the light curve, with $\sigma$ and $\mu$ parameters, and goodness-of-fit GF, shown in each panel. The vertical dashed red line in each panel is the fitted $t_0$. The fitted Weibull function is shown by a blue line; in 5 of the 6 events, the fitting routine was unable to find solution for an inverse-Weibull fit so they are not plotted.

ESP peaks, since presently it is not possible during the early rise phase of an SEP event to be confident of knowing which CMEs will result in ESP features.

In any case, for our purpose here it makes sense to exclude highly irregular light curves from the process of determining a forecast procedure, so a set of "good" events has been identified to focus the analysis: these generally avoid periods of overlapping events and have forms that visually appear amenable to representation by functional forms (the events were selected by visual inspection). The list of 78 events, and plots of each together with "successful" fits, is provided in Appendix B.

For completeness we describe two further properties of the SEP sample that are relevant to the interpretation of our results.[4] Figure 8 shows the histogram of the rise times ($t_{on}$ to time of peak flux) of the datasets used here, comparing the "good" events to the full sample, and it is clear that events with a long rise time are under-represented in the "good" events.

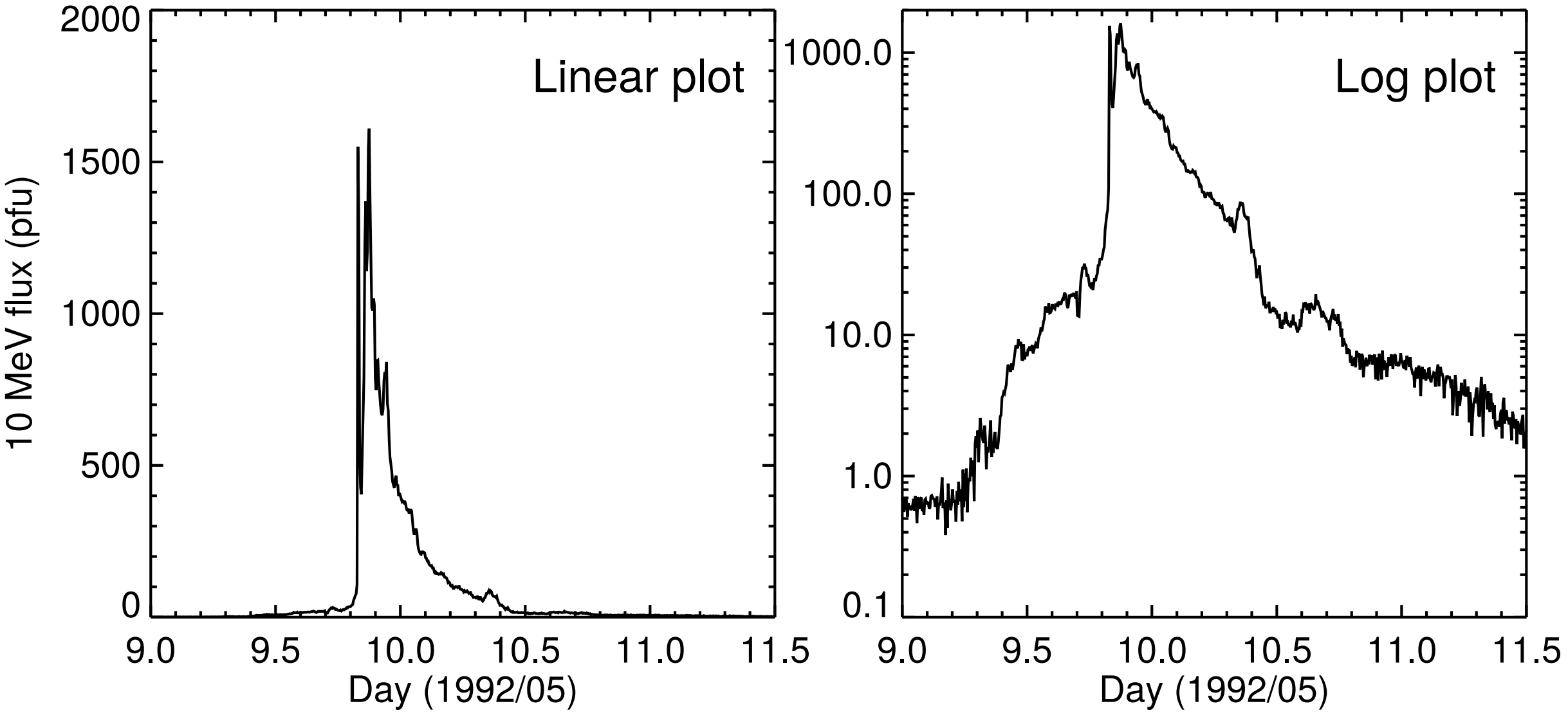

Fig. 7.— The light curve for an SEP event on 1992 May 09 plotted with (left panel) linear flux scale and (right panel) logarithmic flux scale, illustrating the fact that the apparent onset of the event depends on how it is represented.

Figure 9 plots the peak flux in each event against the rise time of the event. It is interesting that the largest peak fluxes tend to occur for events with rise times of 1 to 2 days, and there does seem to be a shortage of very large peak fluxes at long rise times. Figure 10 shows the variation of peak flux and fluence, respectively, versus longitude and addresses the question of whether poorly connected events have fluxes different from well-connected events: it is clear that eastern-hemisphere events do not have the very high fluxes seen from better-connected events, and the longer durations of east-limb events do not entirely compensate for this fact in the distribution of fluences.

## Robustness of parameter fits

With the constraints described above, successful results (here meaning a fitted maximum within 10% of the actual maximum) were achieved

[4]We note that the rise-time distribution depends on how the onset of each event is defined - this was described earlier. In particular, our onset time is intended to identify the first arrival of 10 MeV protons as part of the main phase of an event, needed for functional form fitting, and not the time at which the NOAA threshold of 10 pfu is breached.

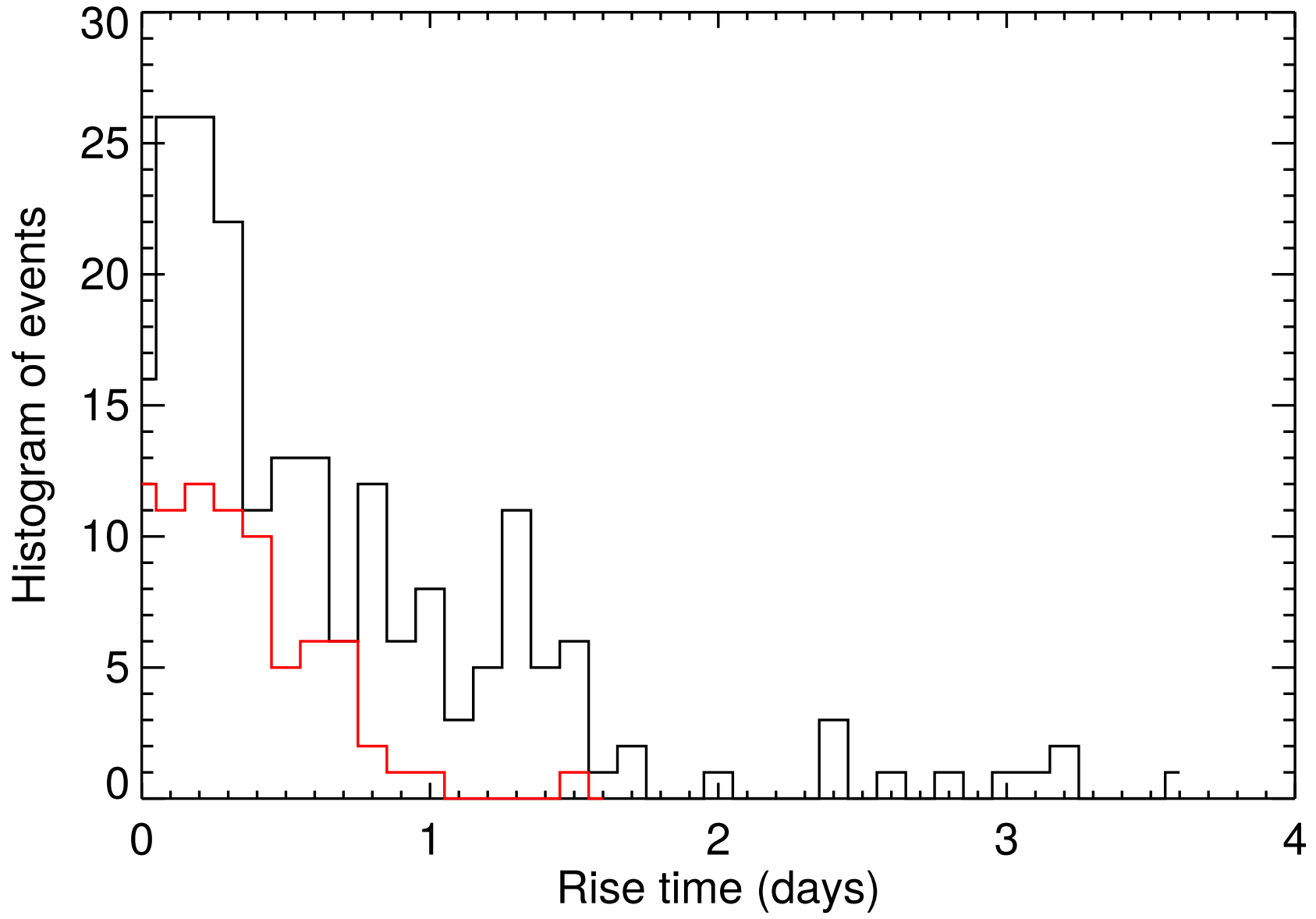

Fig. 8.— A histogram of the rise times of the 203 distinct SEP events used in this sample, binned into 0.1 day intervals. The black line shows the distribution of 203 events using the rise time from the lowest-numbered GOES satellite available, while the red line shows the distribution of the 78 "good" events listed in Appendix B.

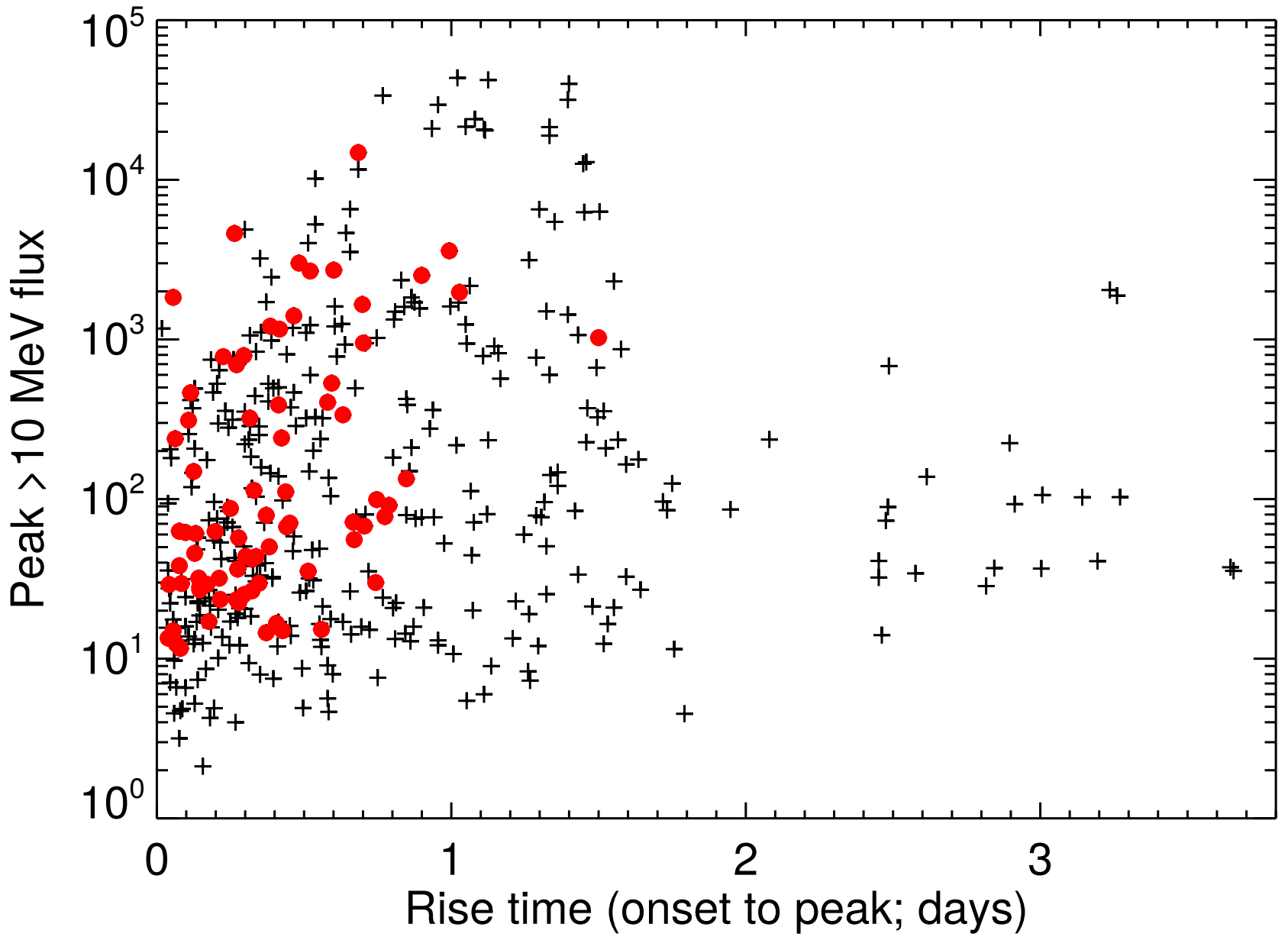

Fig. 9.— A plot of the peak flux versus rise time for the events used in this sample. Filled red circles represent the 78 "good" events, while the black plus symbols show all other events.

- in 410 (with floating $t_0$; 410 with fixed $t_0 = t_{on}$) of the lognormal fits to the light-curves;

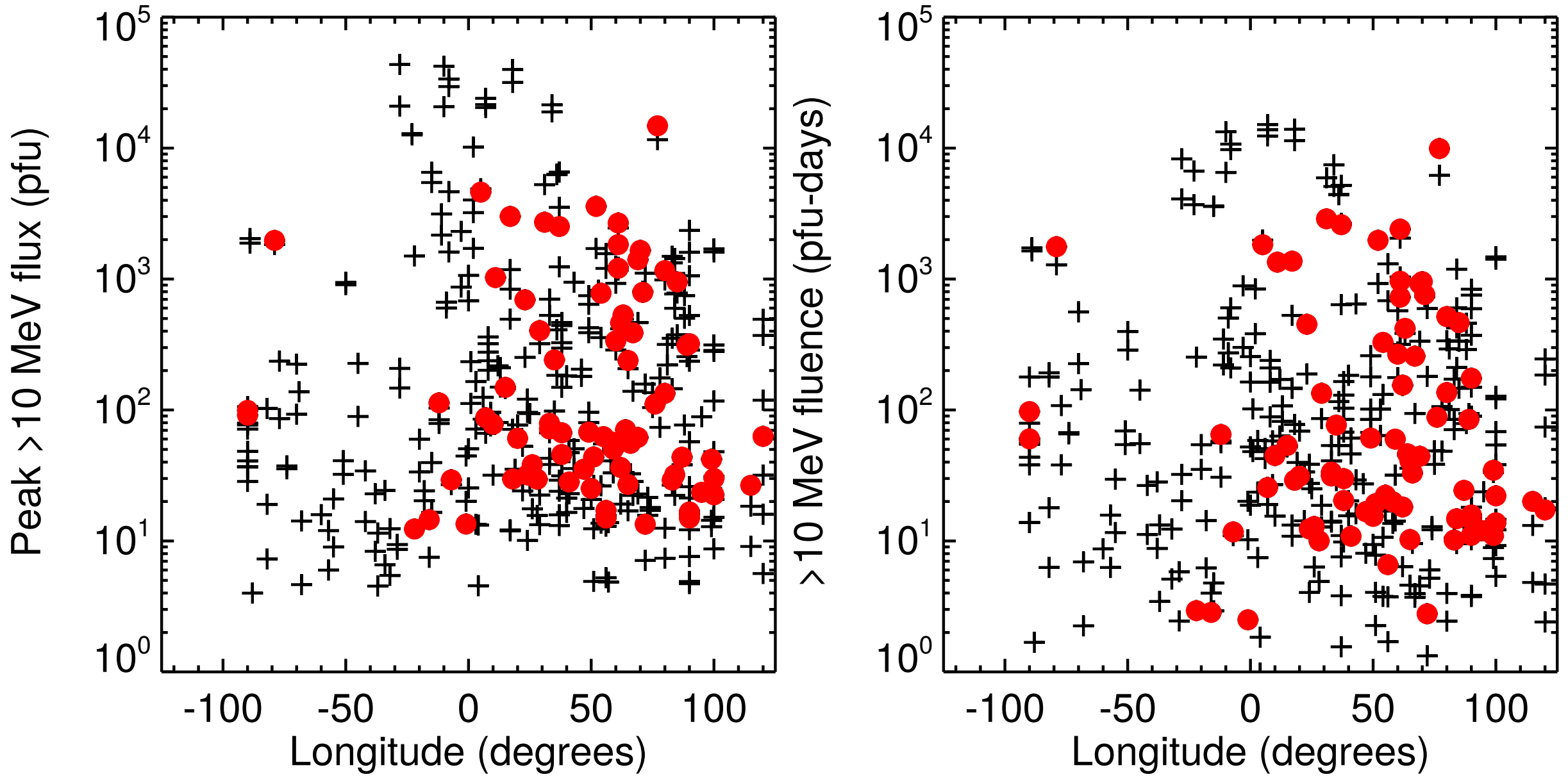

Fig. 10.— A plot of the peak flux (left panel) and event fluence (right panel) versus event longitude for the events used in this sample. Filled red circles represent the 78 "good" events, while the black plus symbols show all other events. Events beyond the west limb are limited to 120°.

- in 410 (410) of the lognormal fits to the cumulative fluence;
- in 408 (408) of the Weibull fits to the light-curves;
- in 410 (409) of the Weibull fits to the cumulative fluence;
- in 111 (196) of the inverse Weibull fits to the light-curves; and
- in 365 (405) of the inverse Weibull fits to the cumulative fluence.

It is clear from the numbers above that lognormal and Weibull functions will fit both most light curves and most fluence curves, and that the inverse Weibull function will fit most fluence curves but fails on an unacceptable fraction of light-curve fits.. It is likely that, e.g., a higher fraction of successful light-curve fits could be achieved for inverse Weibull functions with more tailored fitting efforts, but we choose not to pursue that here: our goal is to find a robust approach that can work for automated real-time fitting, and tailored fitting procedures that may require manual intervention are not suitable for that purpose. The increased rate of successful fits when $t_0$ is not allowed to vary presumably reflects the ability of the particular fitting routine being used to find a solution when there are 3 free parameters rather than 4. In many of the additional successful solutions for the Weibull function when $t_0$ is fixed, the resulting values of $\alpha$ and $\beta$ can be extreme since, e.g., in the case of a gradual rise before a sharp peak, $t_{on}$ may be well separated from the main period of the proton flux.

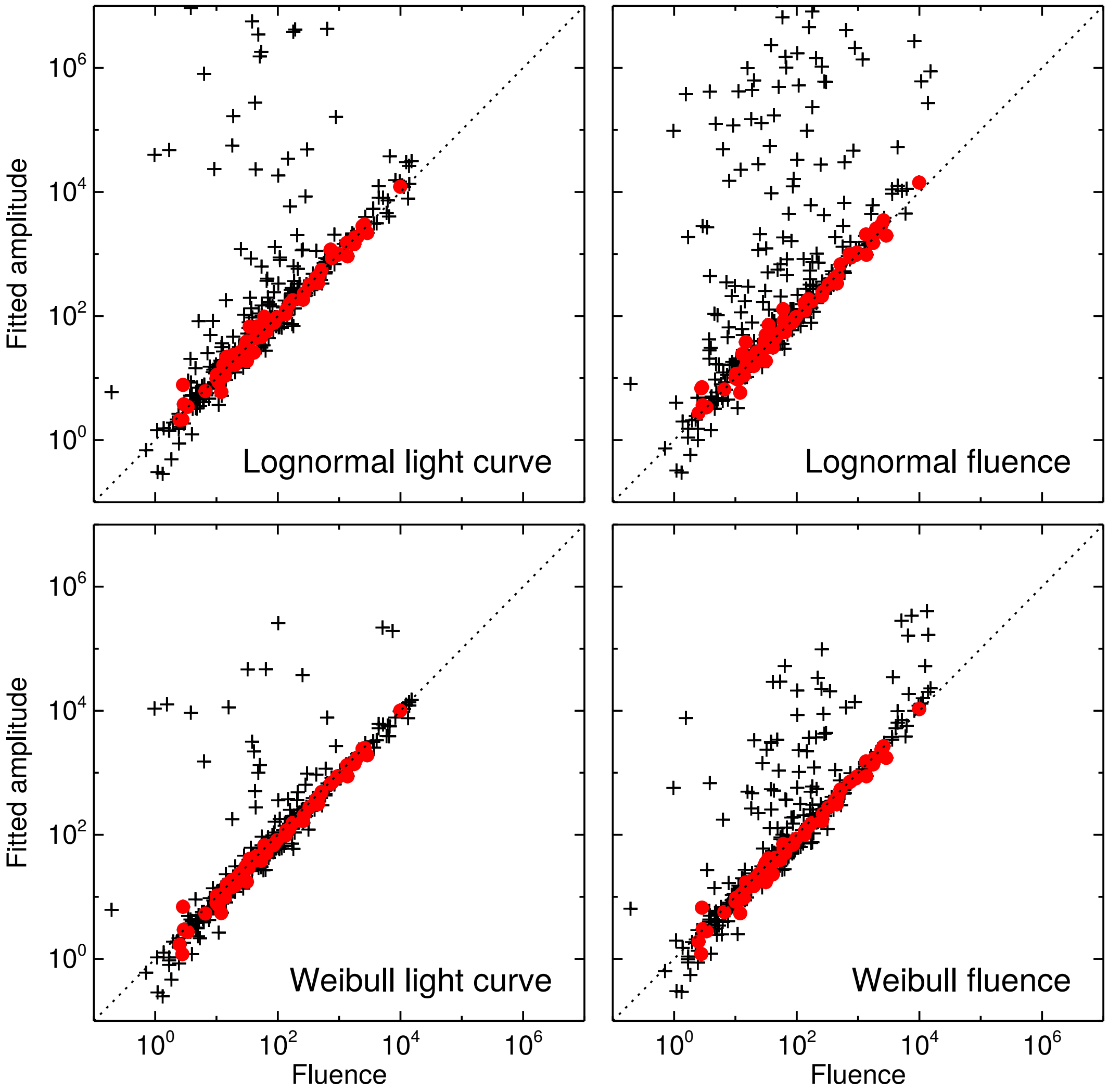

Fig. 11.— A comparison of the fluence for each event (calculated by integrating the background-subtracted light curve to the end of the fitted period) with the fitted amplitude of the functional form, which should represent the fluence (here in units of pfu-days). Filled red circles show results for the "good" events listed in App. B, while plus symbols represent all other events. Results are shown with the fitted amplitude determined by lognormal and Weibull fits to light curve and fluence, respectively, as labelled.

This results in significantly larger values for $\alpha$ for the same event, while $\beta$ tends to change less.

It is relevant to question how robust the parameter fits for a given event are. Since SEP events do not generally have the nice smooth shapes of the functional forms, the fits depend very much on the exact time range used for the fit, since the addition or removal of any irregular data will change the fit. One aspect of this issue, as already discussed, is whether or not $t_0$ is fixed, and if so, where. Another aspect is whether the actual light curve or the

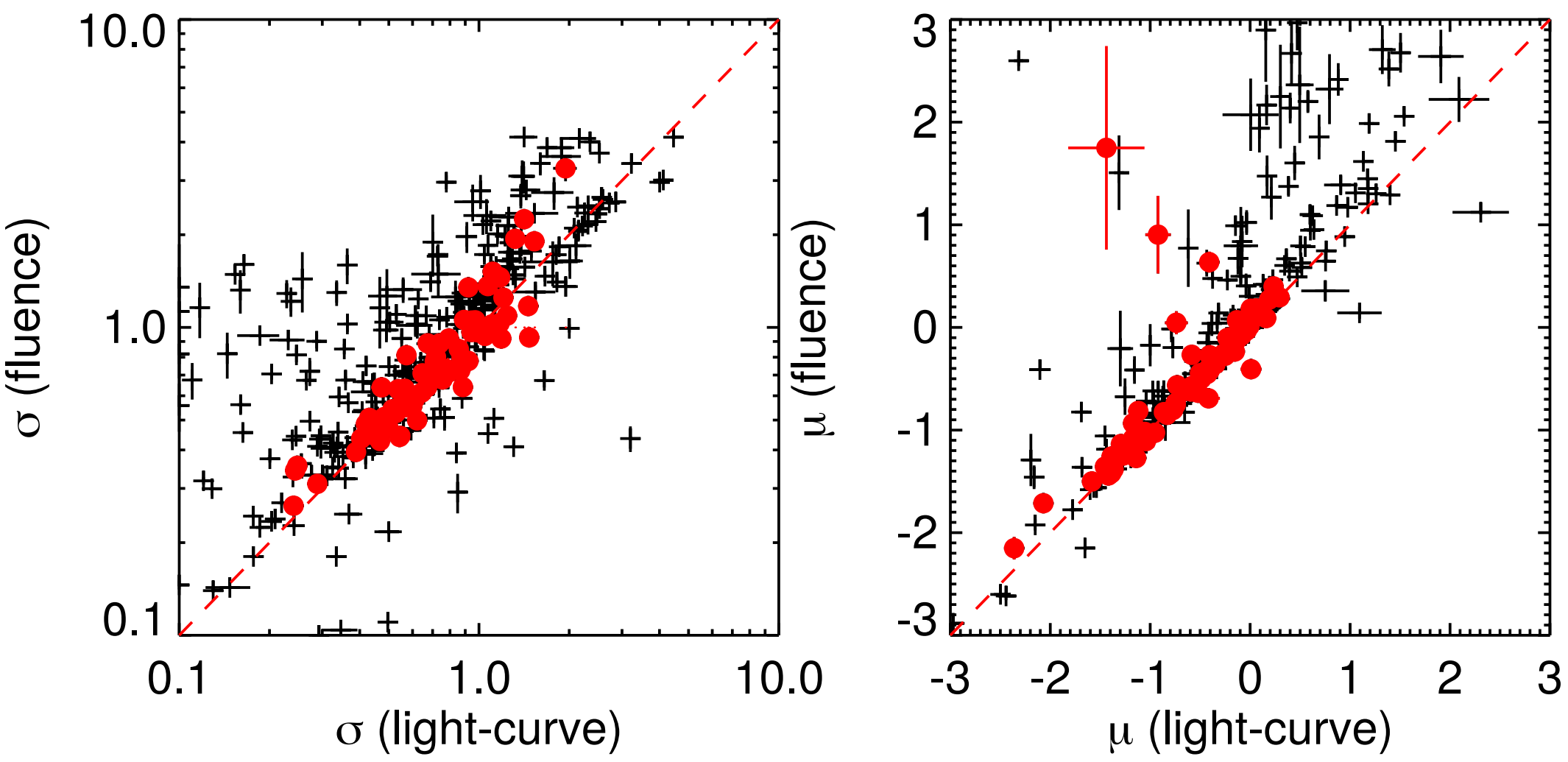

Fig. 12.— A comparison of the fitted shape parameters $\sigma$ (left panel) and $\mu$ (right panel) for lognormal fits (with floating $t_0$) to the light-curve (horizontal axis) and cumulative fluence (vertical axis) for common events. Filled red circles show results for the "good" events listed in App. B, while plus symbols represent all other events.

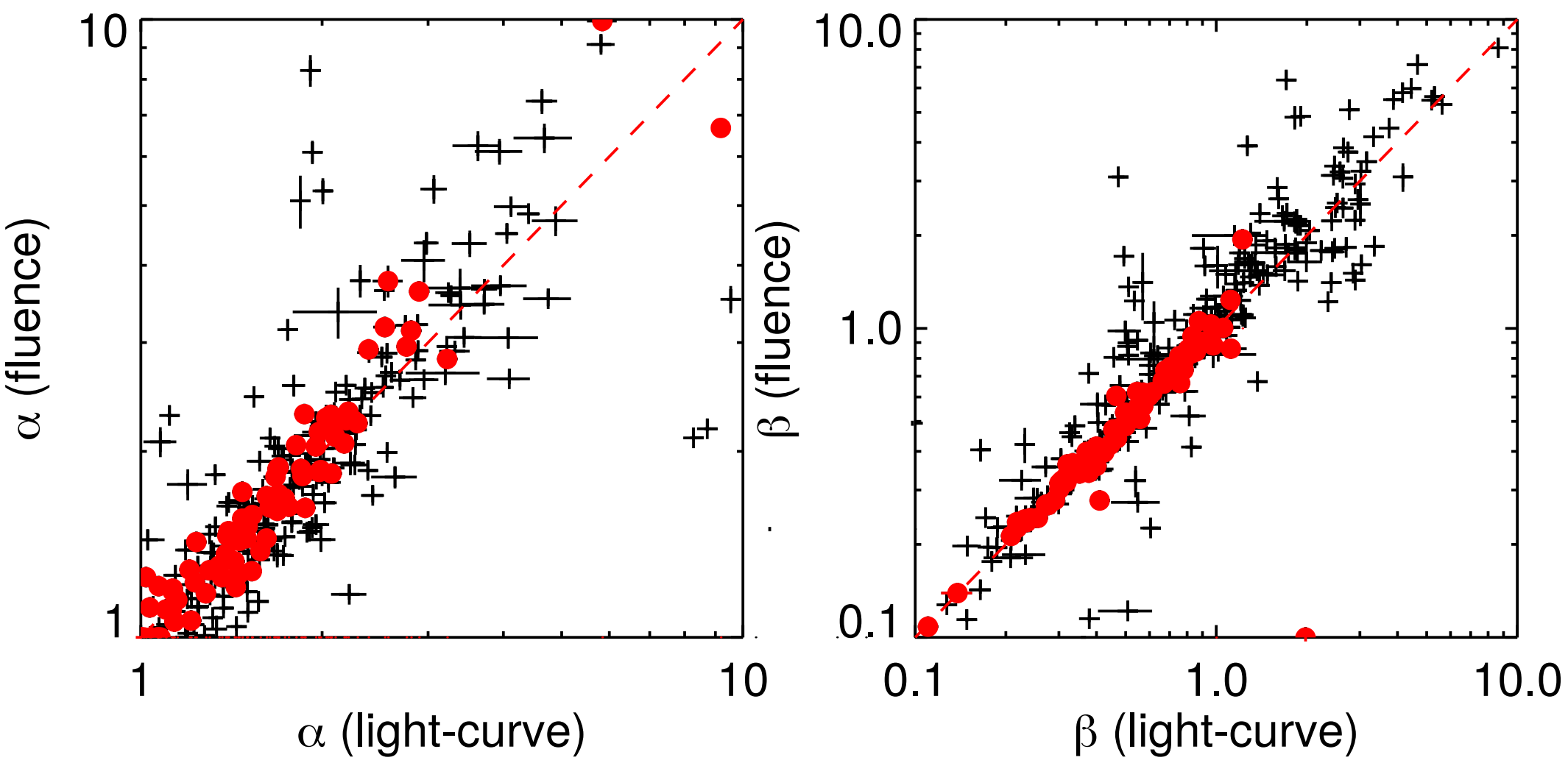

Fig. 13.— A comparison of the fitted shape parameters $\alpha$ (left panel) and $\beta$ (right panel) for Weibull fits (with floating $t_0$) to the light-curve (horizontal axis) and cumulative fluence (vertical axis) for common events. Filled red circles show results for the "good" events listed in App. B, while plus symbols represent all other events.

cumulative fluence is used for the fit, since, e.g. in the case of inverse Weibull fits, this can make a big difference in the success of the fitting procedure. We note that by construction the cumulative fluence does not include the same data as the light curve: all values prior to $t_{on}$ are set to zero in the fluence, whereas in the light curve those data are present, ideally

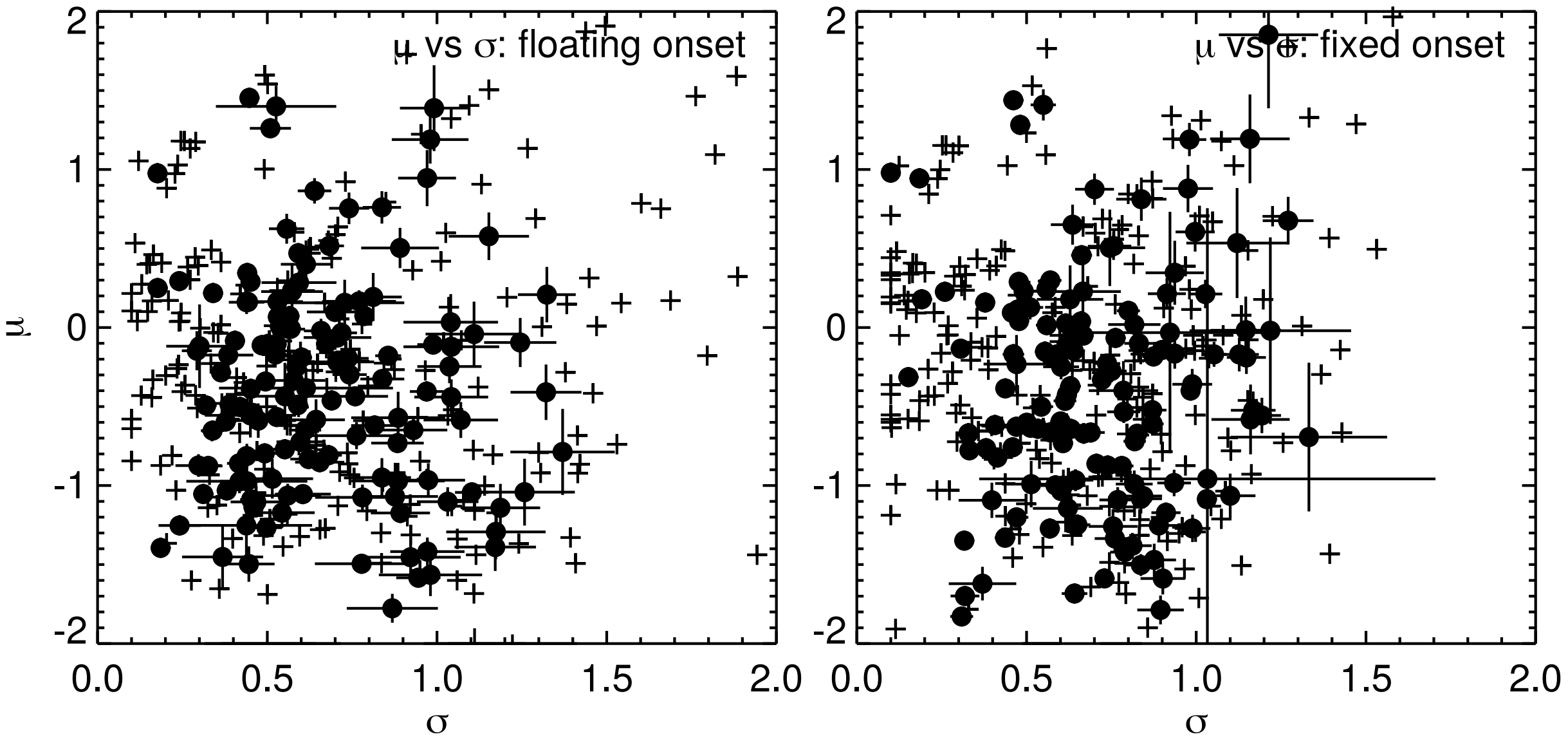

Fig. 14.— A plot of the fitted shape parameters for lognormal fits to the SEP light curves: $\sigma$ is represented on the horizontal axis while $\mu$ is on the vertical axis. Filled circles represent the well-fit events used to determine the variation with longitude discussed in the next section, while plus symbols show all other events. The left panel shows the fits in which $t_0$ is allowed to vary, while the right panel shows the fits when $t_0$ is fixed at $t_{on}$.

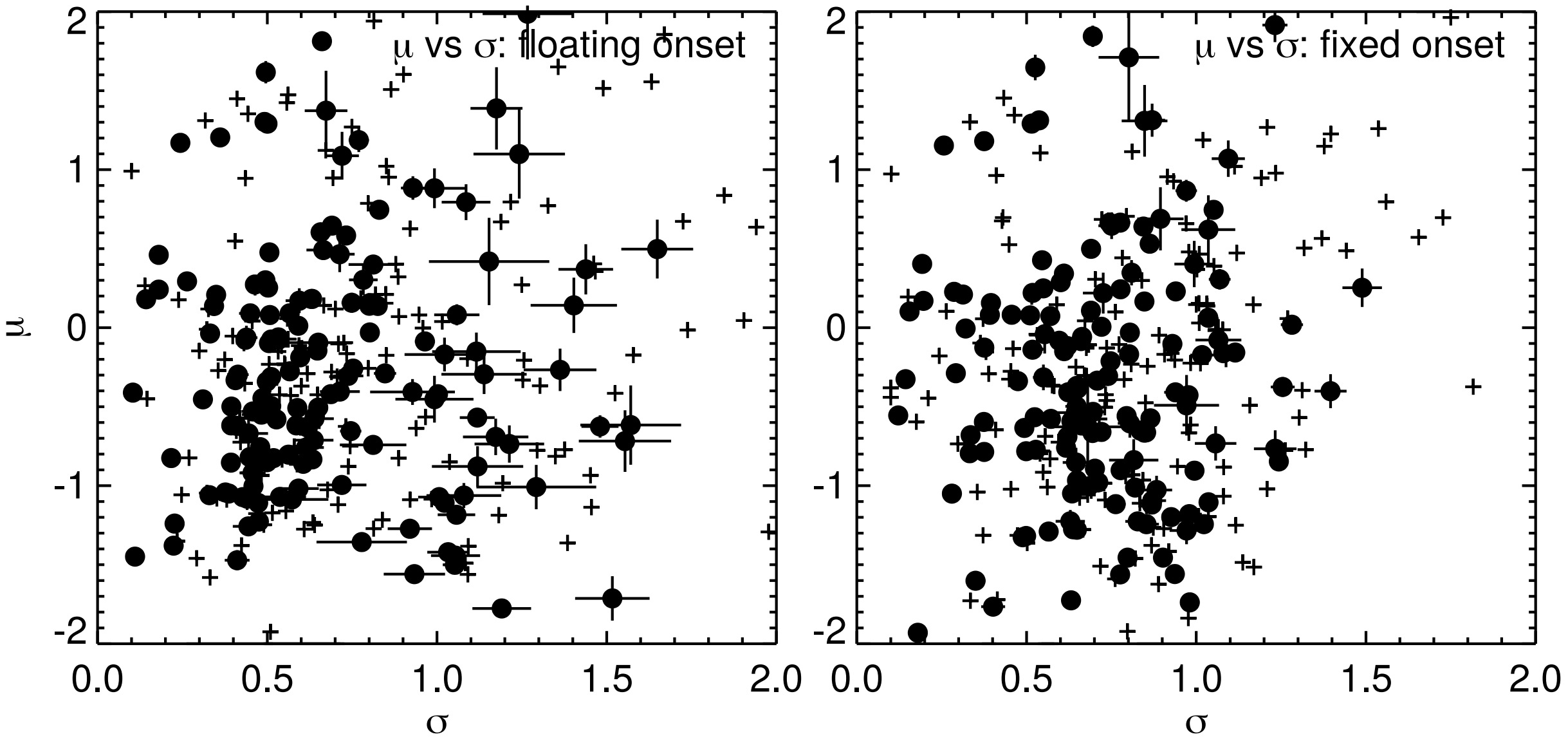

Fig. 15.— A plot of the fitted shape parameters for lognormal fits to the SEP accumulated fluence curves, with the same format as Fig. 14.

as noise, but they can affect the fits.

In discussing the functional forms earlier we noted that the fitted amplitude should represent the fluence of the event, since the functional forms are probability distributions that integrate to unity. To confirm this, Figure 11 compares the fluence with the fitted value of the amplitude for four of the functional fits. For the "good" events the match is excellent. However, many of the other events have fitted amplitudes that are much larger than the observed fluence: we believe that this is because many events decline faster after they peak than is characteristic of the standard picture of SEP events that decay more gradually than they rise. There are several such examples in Fig. 6.

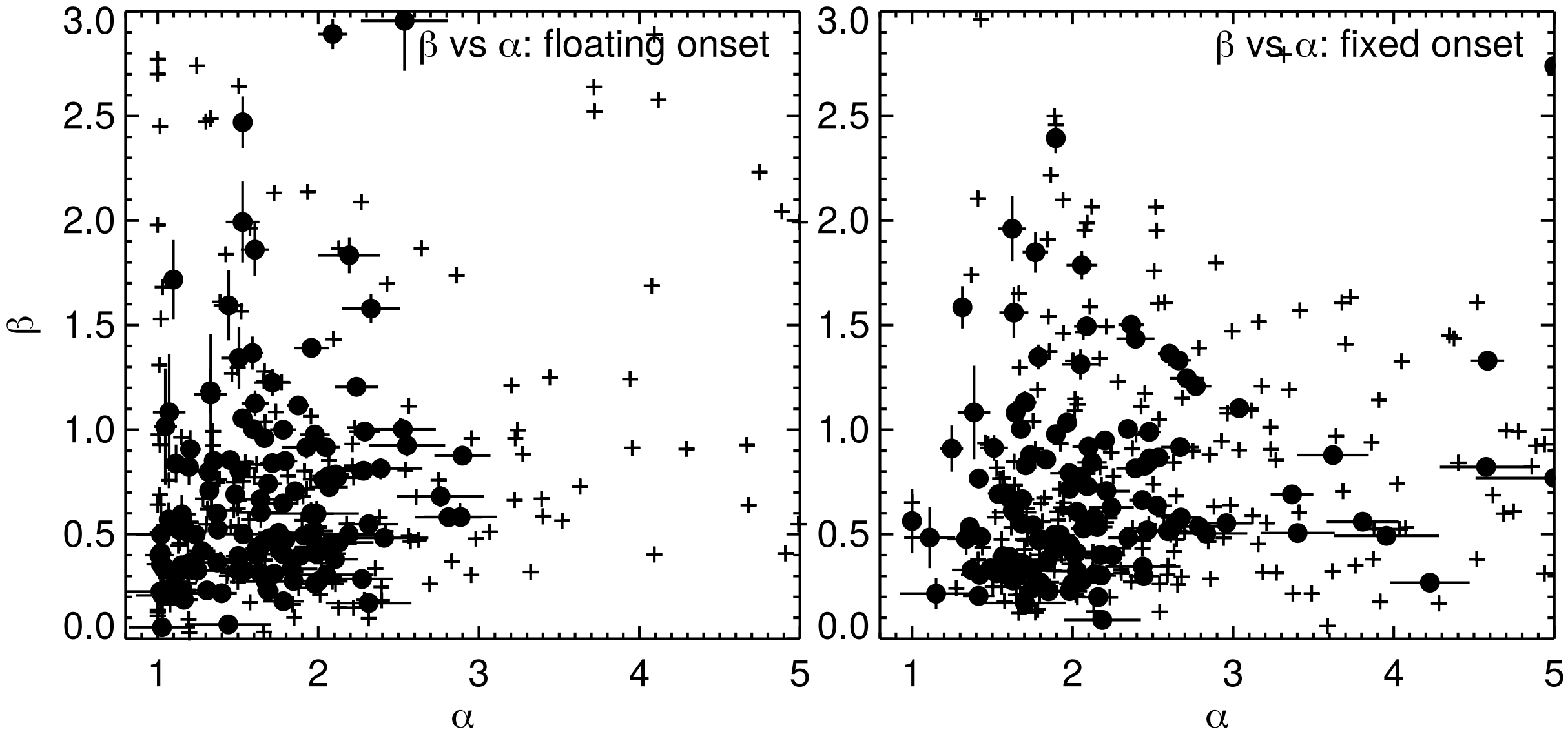

Fig. 16.— A plot of the fitted shape parameters $\alpha$ and $\beta$ of Weibull function fits to the SEP light curves, with the same format as Fig. 14.

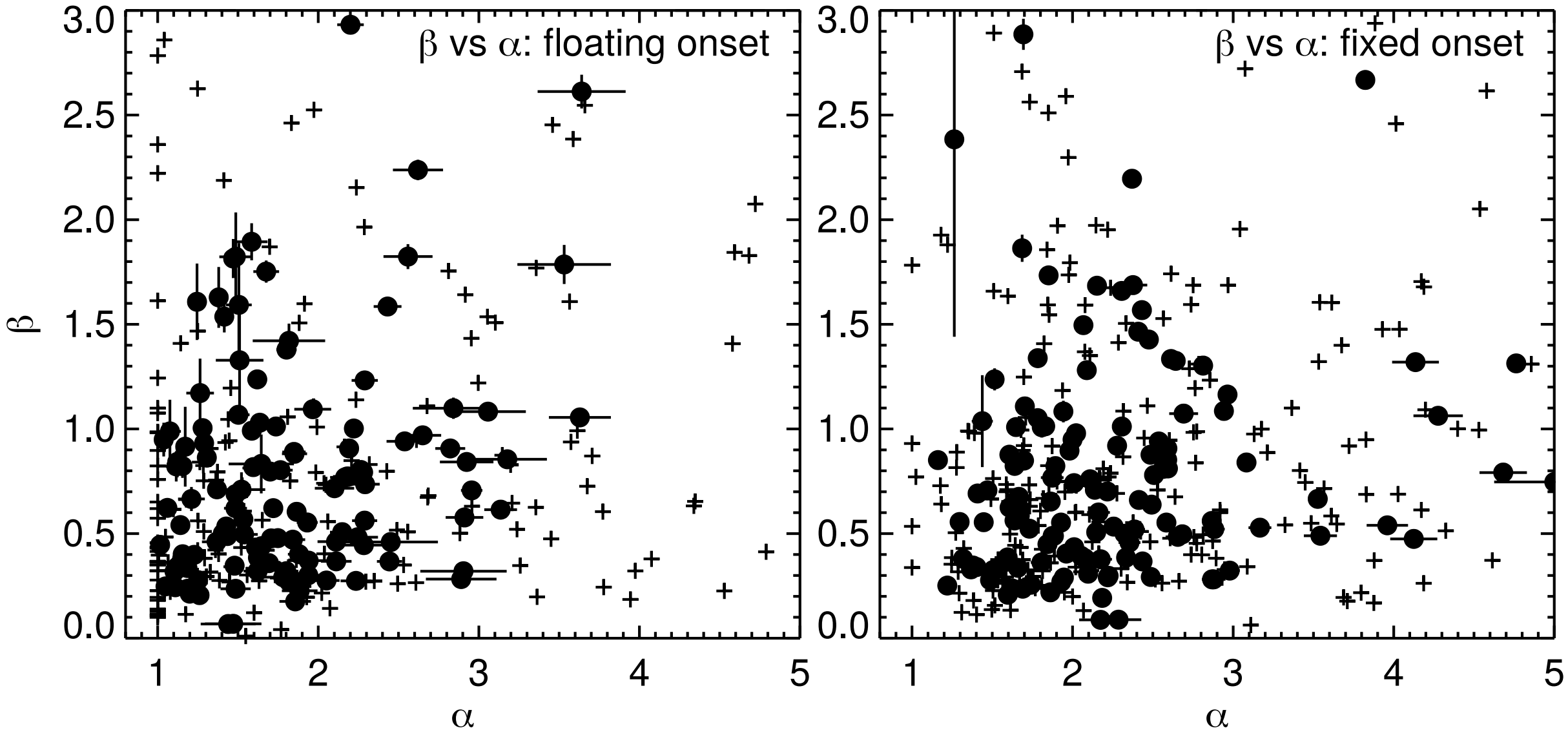

Fig. 17.— A plot of the fitted shape parameters for Weibull function fits to the SEP accumulated fluence curves, with the same format as Fig. 14.

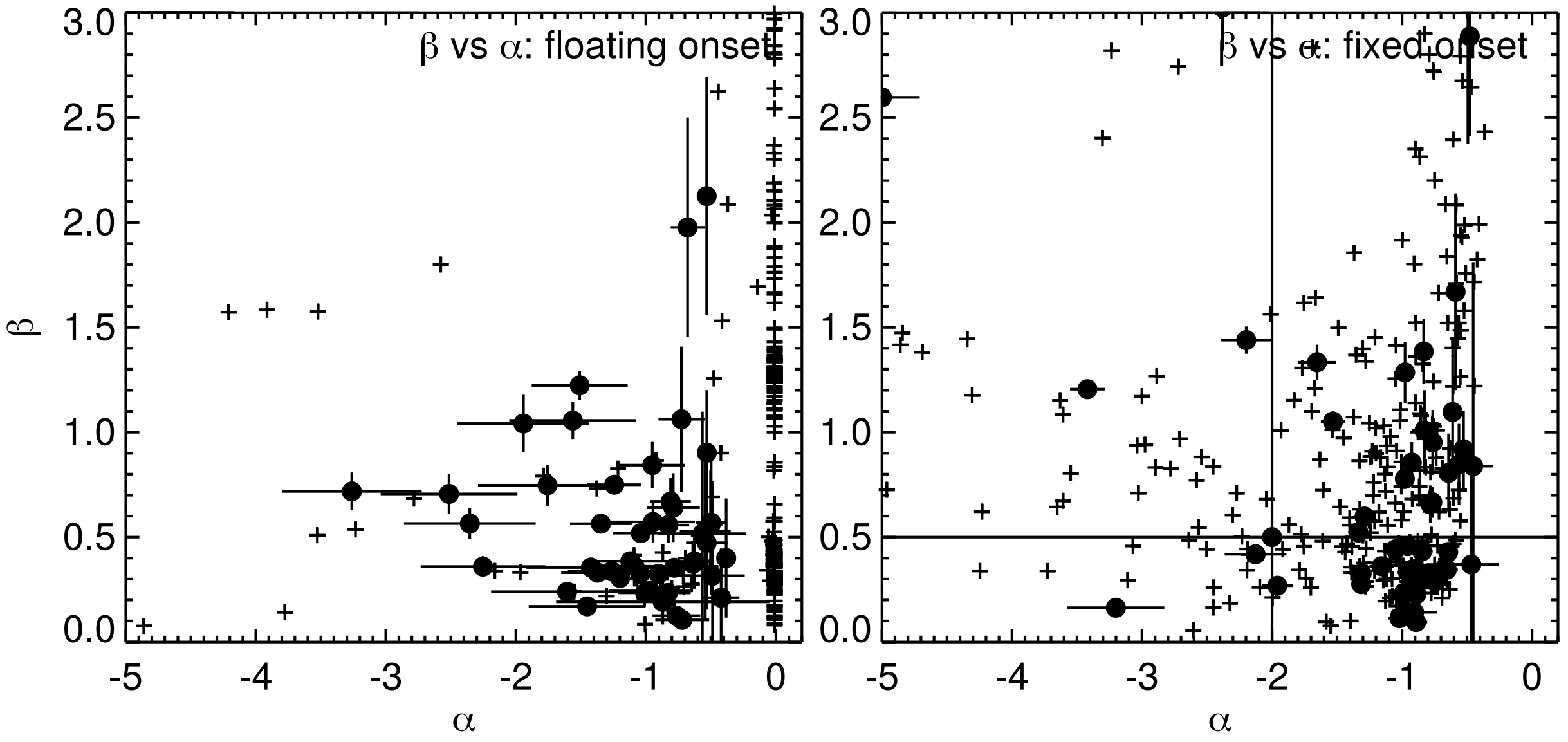

Fig. 18.— A plot of the fitted shape parameters $\alpha$ and $\beta$ of inverse Weibull function fits to the SEP light curves, with the same format as Fig. 14. The paucity of points and the pile-up of poorly-fit events at $\alpha = 0$ when $t_0$ is not fixed is due to the fact that the number of fits in this case was relatively low, whereas inverse-Weibull fits to the cumulative fluence had a much higher success rate: see Fig. 19 below.

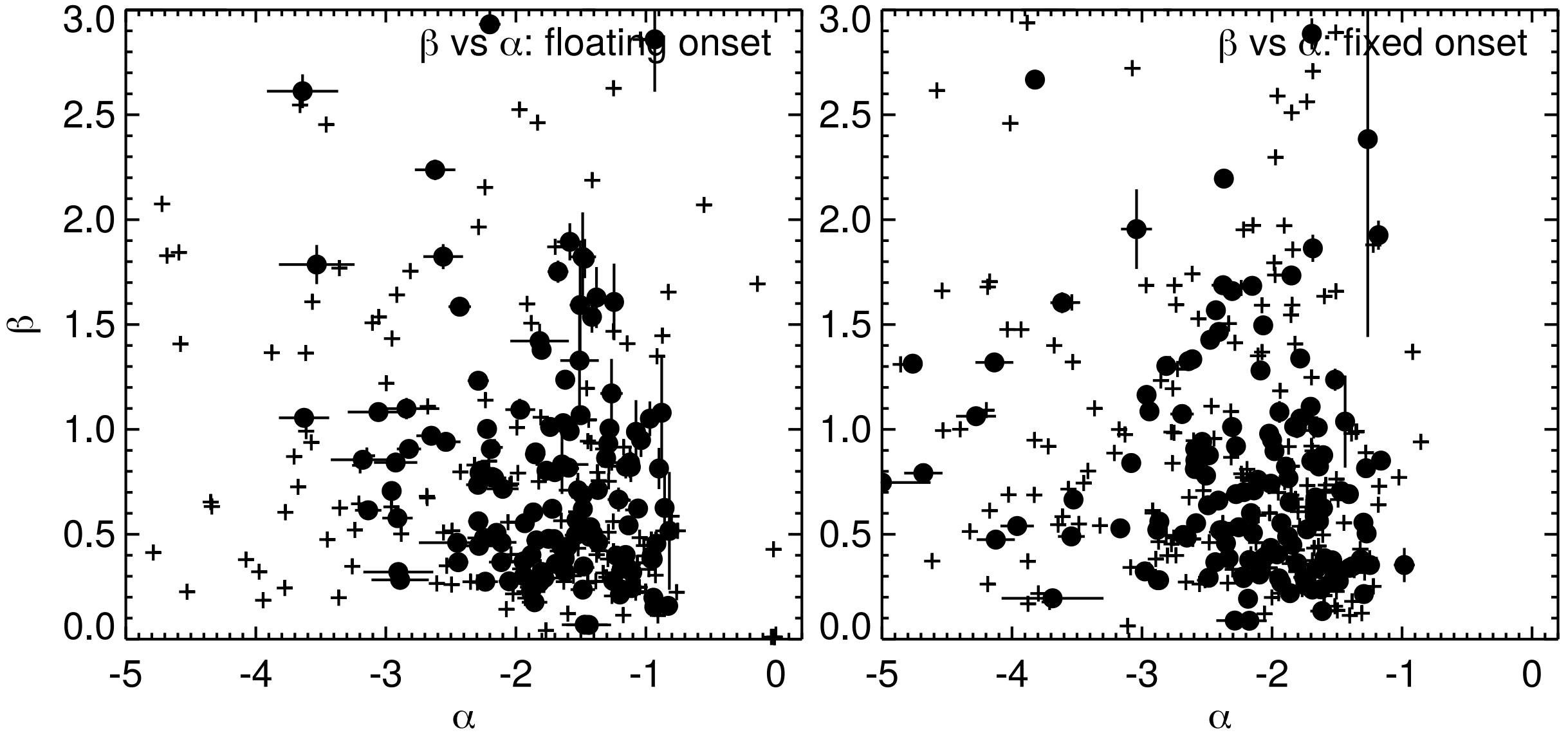

Fig. 19.— A plot of the fitted shape parameters for inverse Weibull function fits to the SEP accumulated fluence curves, with the same format as Fig. 14.

Addressing the latter point, Figure 5 shows an example of a well-behaved event where the fits to the light-curve and cumulative fluence produced very similar parameters for the lognormal and Weibull functions, and similar $\beta$ but seemingly different $\alpha$ in the case of inverse Weibull. Note that $t_0$ was a fitted parameter in these plots, and is quite different for the different functional forms. To explore the difference between fits to the light curve and fluence further, Figures 12 and 13 compare the fits for (respectively) $\sigma$ and $\mu$ values and $\alpha$ and $\beta$ values. The filled circles in these figures represent the better-behaved events expected to be well fitted. In general there is a good match between the values derived from the light curves and the fluence, although it is interesting that $\mu$ (lognormal) and $\beta$ (Weibull) show a tighter correlation than $\sigma$ or $\alpha$. The corresponding figure for inverse Weibull had relatively few good fits to the event light curves, as noted above, and is provided in Appendix C for completeness (Fig. 39).

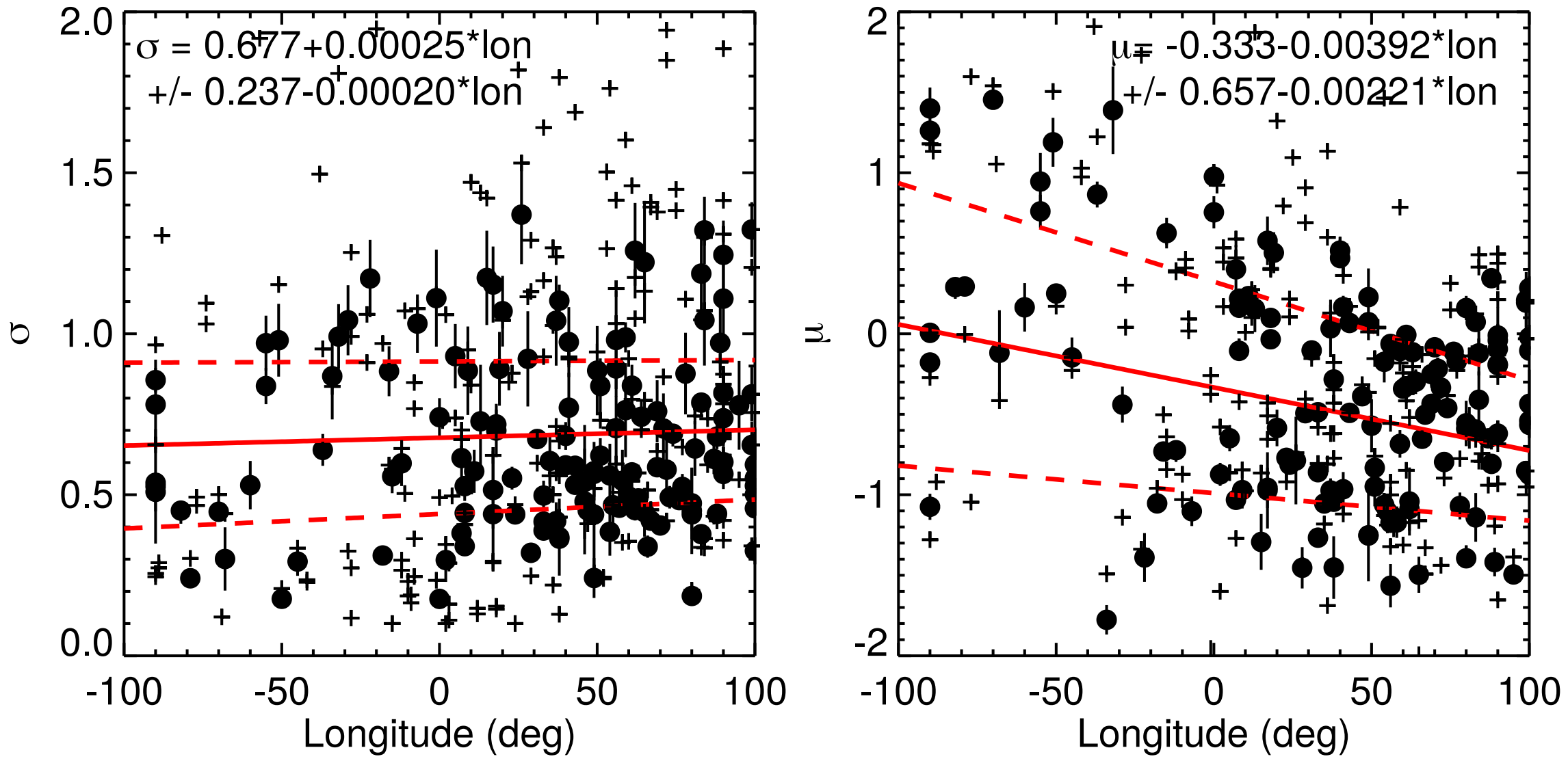

Fig. 20.— The longitudinal dependence of $\sigma$ and $\mu$ from lognormal fits to 139 SEP light curves, chosen to be unique events with good fits to the light curves (filled circles; error bars are plotted but are mostly small). In these fits $t_0$ is allowed to float. Small crosses (plotted without error bars, for clarity in the figure) represent fits to the remaining datasets. A linear polynomial is fitted to the filled circles, and shown by the solid red lines and reported in the upper right of each panel. A fit to the residuals as a function of longitude provides the dashed lines, which represent the uncertainty in the polynomial fit: the uncertainties in the fitted parameters are shown below the polynomial.

If there is a correlation between shape parameters for common events, that could reduce the number of degrees of freedom that need to be fitted for an event. To explore this possibility, Figs. 14-19 show the correlation between the fitted shape parameters for the 6 types of fit (lognormal, Weibull, and inverse Weibull fits to both light curves and cumulative fluence, with both floating and fixed $t_0$). It is clear from these figures that there is no strong

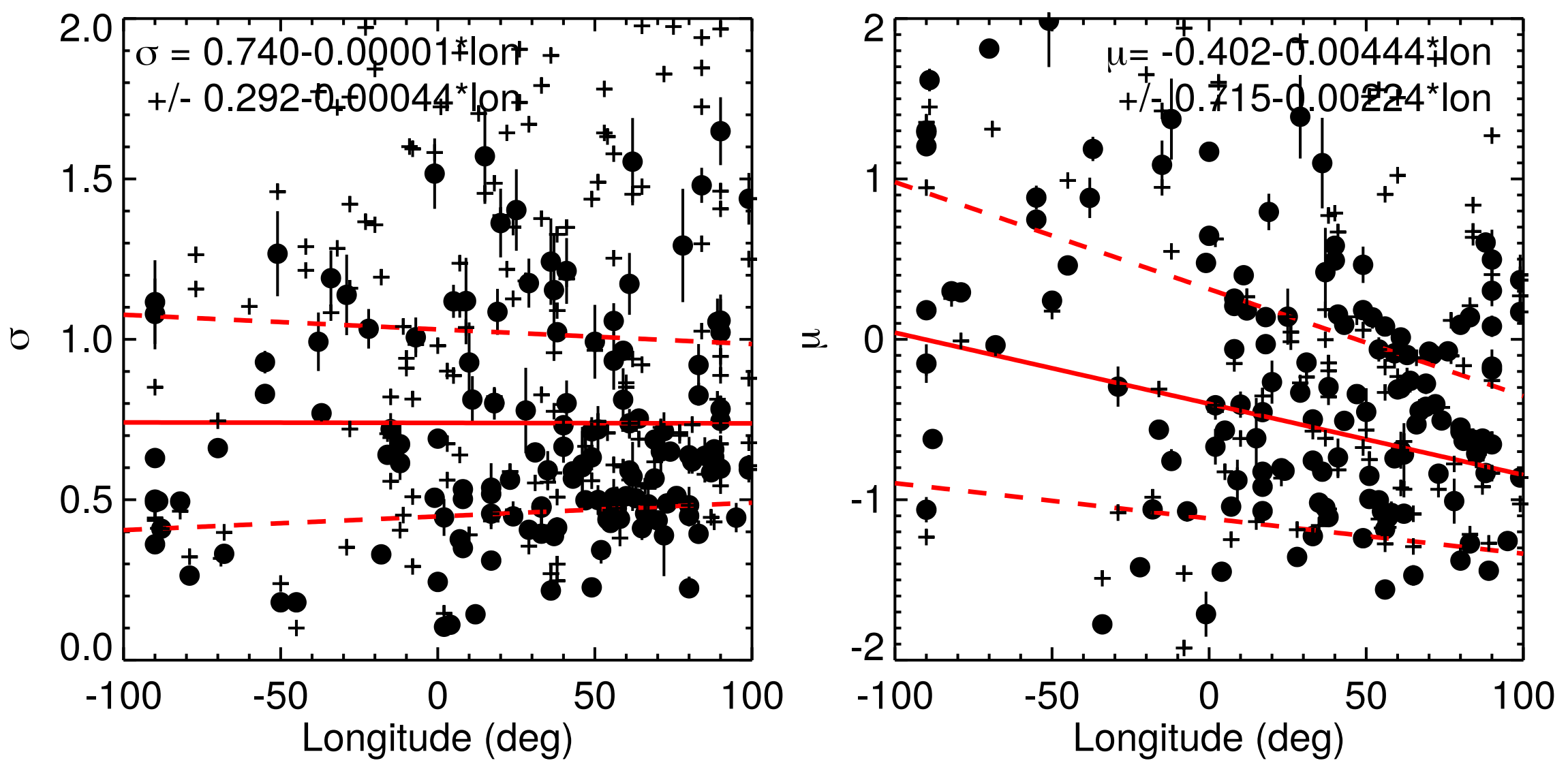

Fig. 21.— The longitudinal dependence of $\sigma$ and $\mu$ from lognormal fits to 141 SEP cumulative fluences, chosen to be unique events with good fits to the light curves (filled circles). In these fits $t_0$ is allowed to float. Small crosses represent fits to the remaining datasets. Other features are described in the caption to Fig. 20.

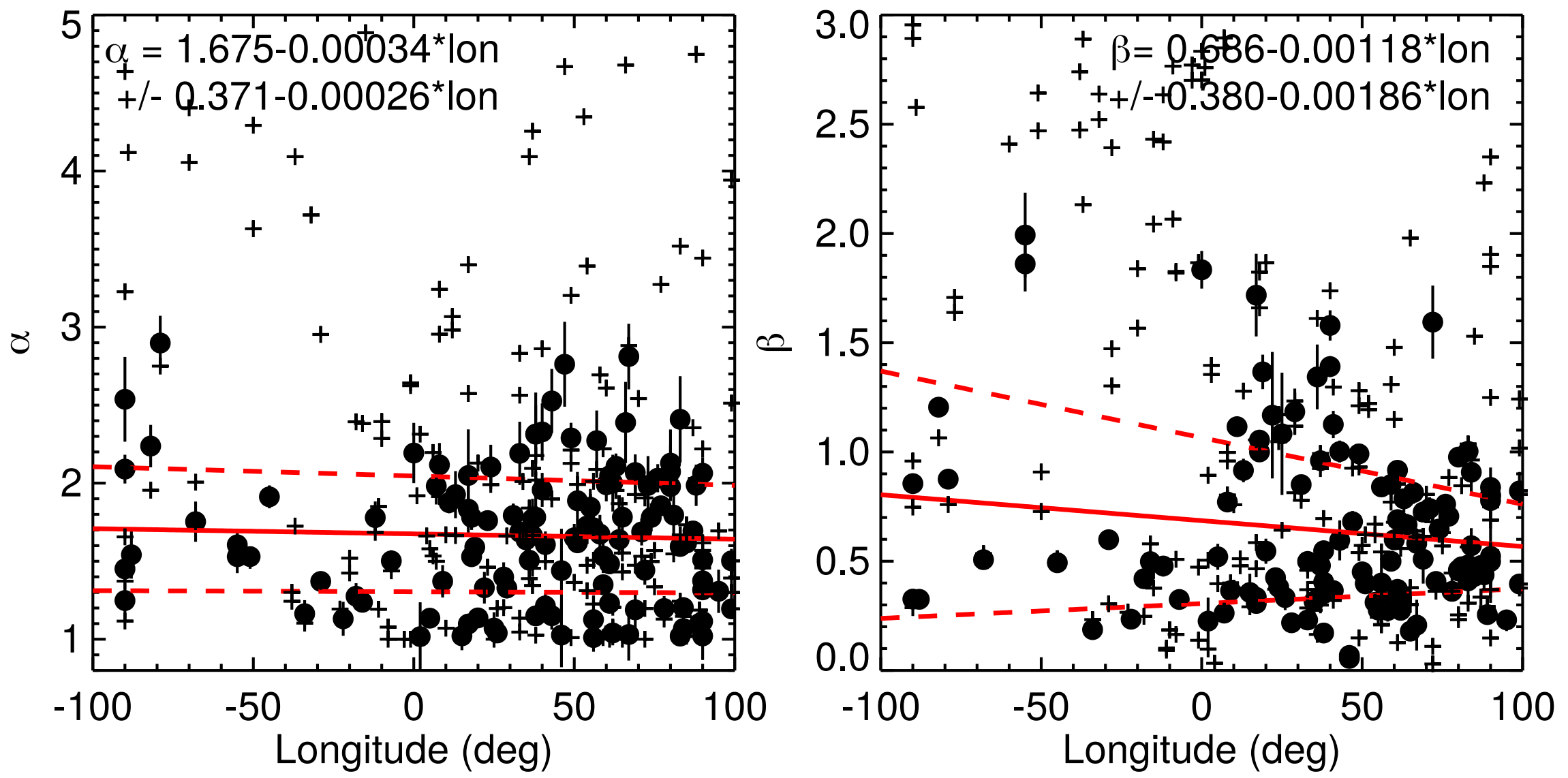

Fig. 22.— The longitudinal dependence of $\alpha$ and $\beta$ from Weibull fits to 118 SEP light curves. The format matches the description of Fig. 20.

relationship between the shape parameters that can be exploited in the fitting process.

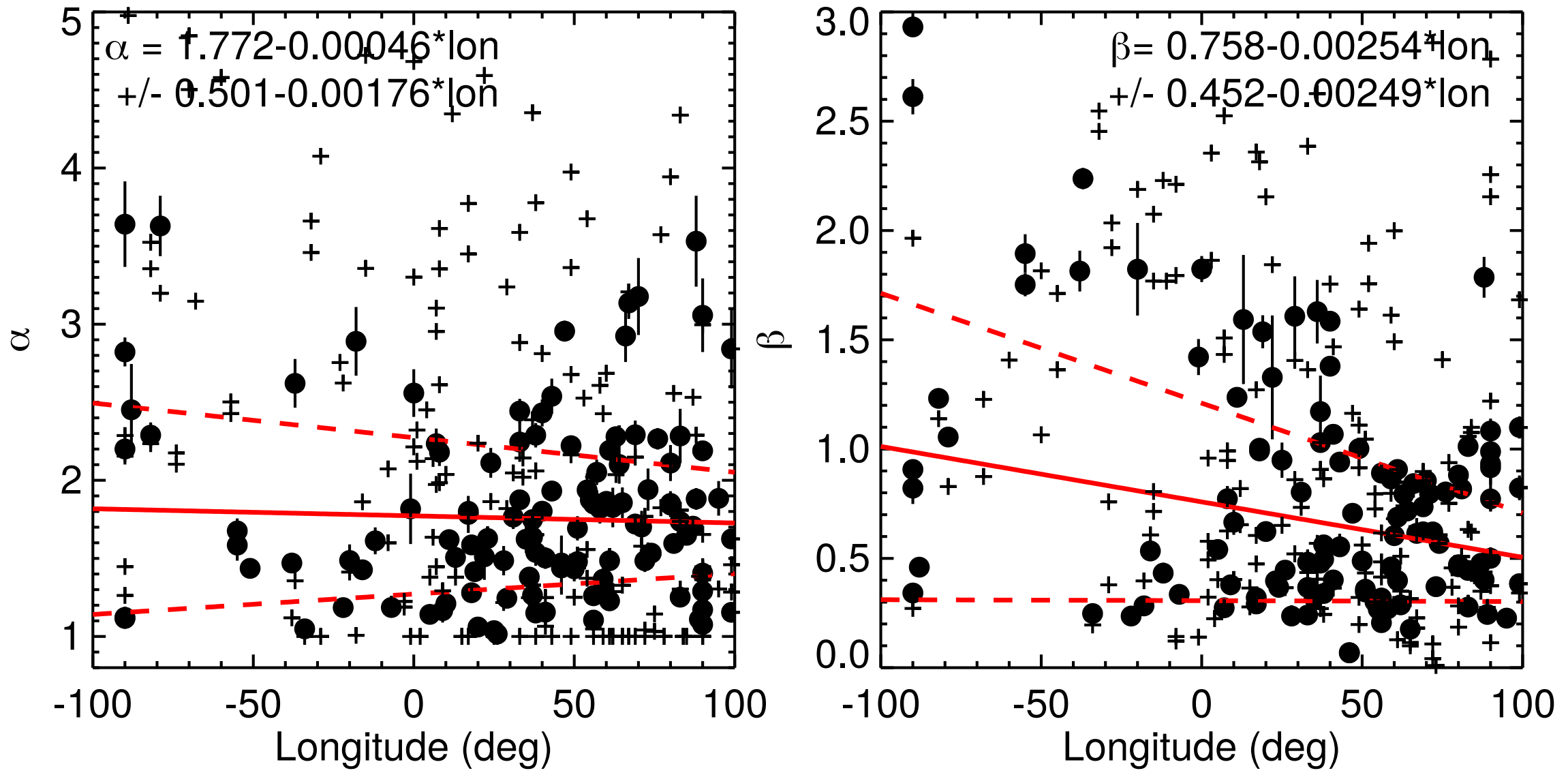

Fig. 23.— The longitudinal dependence of $\alpha$ and $\beta$ from Weibull fits to 116 SEP fluences. The format matches the description of Fig. 20.

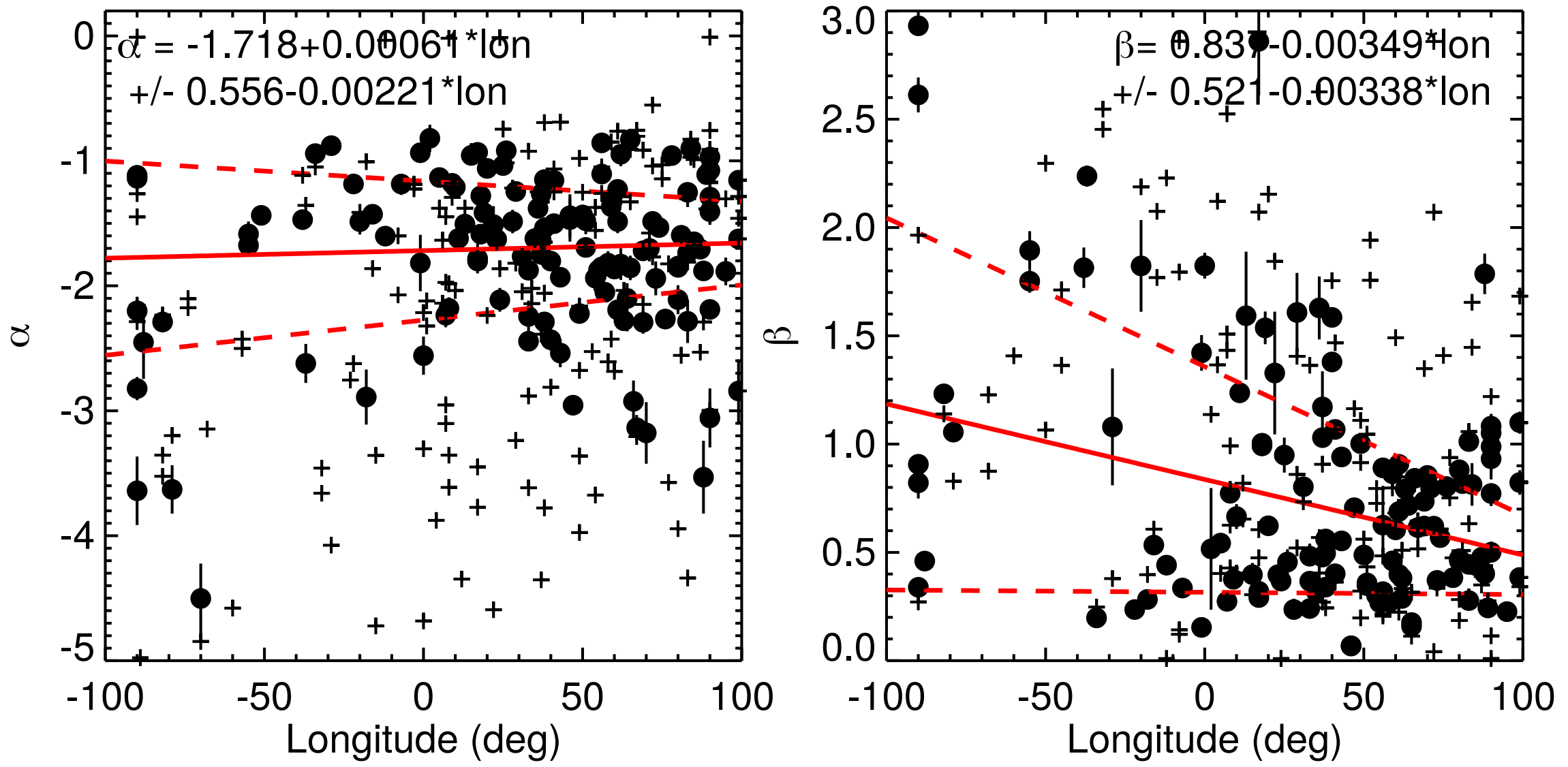

Fig. 24.— The longitudinal dependence of $\alpha$ and $\beta$ from inverse Weibull fits to 126 SEP fluences. The format matches the description of Fig. 20.

### Variation with longitude

In this section we investigate the dependence of the functional fitting parameters on longitude, in order to be able to specify plausible guesses at the shape parameters early in the rise phase of an SEP event rather than try to fit 4 free parameters with a small number of data points. 340 of the 410 data files have event longitudes based on associations with solar events listed in the SEP catalogs. Since many of these are redundant fits, and many do not represent good fits to the data, we down-selected to a sample of order 100 unique events (different for each set of fits, see captions for exact numbers) with known longitudes, better goodness-of-fit metrics and small uncertainties in the fit parameters[5]. The resulting variation with longitude is shown for lognormal fits in Figs. 20 and 21, where the filled circles represent the points used in fitting the longitude dependence, and the plus symbols represent all other events with associated longitudes. Figs. 22 and 23 show the corresponding behavior of the shape parameters in the Weibull fits, while Figure 24 shows the results for inverse Weibull fits to the fluence (Fig. 40 in App. C shows the results for inverse Weibull fits to the light curves, which are generally poor). These are the fits made with $t_0$ being fitted; the results when $t_0$ is fixed at the onset time, $t_{on}$, are shown in Figs. 41 - 46 in Appendix C.

A linear variation with longitude was fitted for each parameter in each set of fits, and these are shown in the figures. In order to determine confidence limits for the fitted lines, we made linear polynomial fits to the residuals about the fit and that produces the dashed lines shown in the figures.

For lognormal fits to the light curves, there is a clear decrease in $\mu$ as longitude moves from east to west, and a less significant increase in $\sigma$. For $\mu$, the spread is larger at eastern longitudes where there are fewer events. A similar pattern is seen in the lognormal fits to the cumulative fluences. For Weibull fits, both $\alpha$ and $\beta$ show a small decrease with increasing longitude, and again the uncertainty in the fit for $\beta$ is larger at eastern longitudes. The inverse-Weibull fits to the cumulative fluence find that $\alpha$ values increase with longitude, while $\beta$ values decrease but have high uncertainty.

Table 1 presents a summary of the fitting results, in the form of the coefficients for the linear fits to the longitudinal dependence of the shape parameters for each functional form, with uncertainties.

---

[5]This sample is different from that provided in Appendix B, since the latter has relatively few eastern hemisphere events and thus negative longitudes are not as well represented as in the sample shown in Fig. 20.

Table 1. Longitude dependence of fitted shape parameters

| Parameter | Floating $t_0$ Constant | Floating $t_0$ Slope | Fixed $t_0$ Constant | Fixed $t_0$ Slope |
|---|---|---|---|---|
| Log. l-c: $\sigma$ | 0.677 ± 0.237 | 0.00025 ± 0.00020 | 0.686 ± 0.205 | 0.00108 ± 0.00014 |
| Log. flu: $\sigma$ | 0.740 ± 0.292 | -0.00001 ± 0.00044 | 0.735 ± 0.220 | 0.00113 ± 0.00012 |
| Log. l-c: $\mu$ | -0.333 ± 0.657 | -0.00392 ± 0.00221 | -0.313 ± 0.697 | -0.00501 ± 0.00244 |
| Log. flu: $\mu$ | -0.402 ± 0.715 | -0.00444 ± 0.00224 | -0.440 ± 0.754 | -0.00461 ± 0.00237 |
| Wei. l-c: $\alpha$ | 1.675 ± 0.371 | -0.00034 ± 0.00026 | 2.184 ± 0.524 | -0.00377 ± 0.00228 |
| Wei. flu: $\alpha$ | 1.772 ± 0.501 | -0.00046 ± 0.00176 | 2.158 ± 0.502 | -0.00265 ± 0.00180 |
| Wei. l-c: $\beta$ | 0.686 ± 0.380 | -0.00118 ± 0.00186 | 0.799 ± 0.416 | -0.00298 ± 0.00253 |
| Wei. flu: $\beta$ | 0.758 ± 0.452 | -0.00254 ± 0.00249 | 0.805 ± 0.465 | -0.00330 ± 0.00296 |
| Inv. l-c: $\alpha$ | ··· | ··· | -1.146 ± 0.411 | 0.00428 ± 0.00208 |
| Inv. flu: $\alpha$ | -1.718 ± 0.556 | 0.00061 ± 0.00221 | -2.161 ± 0.529 | 0.00299 ± 0.00174 |
| Inv. l-c: $\beta$ | ··· | ··· | 0.786 ± 0.626 | -0.00037 ± 0.00211 |
| Inv. flu: $\beta$ | 0.837 ± 0.521 | -0.00349 ± 0.00338 | 0.816 ± 0.476 | -0.00319 ± 0.00287 |

Note. — The table presents the coefficients from linear fits to the longitudinal dependence of the parameters, in the form Parameter = Constant + Slope*Longitude where longitude is in degrees. Abbreviations: "Log." = lognormal function, "Wei." = Weibull function, "Inv." = inverse Weibull function, "l-c" = light curve fits, "flu" = fluence fits. The empty cells correspond to unreliable fits to light curves with the inverse Weibull function allowing $t_0$ to vary.

## Analysis Methods for the SEP Rise Phase

The fitting procedure used for the results reported here was as follows:

- Each light-curve file was loaded; an iterative procedure was used to identify a linear background trend from the first 2.5 hours of data, then the point at which the >10 MeV count rate is 5 standard deviations above this background is identified (the nominal onset time, $t_{on}$). The background is required to be level or falling in time, and cannot be negative.
- Once 6 points on the rise of the event have been obtained (30 minutes), the background-subtracted data (forced to be $> 0$) are then fed to a Levenburg-Marquardt fitting routine supplied with the functions described above: the functional forms permit negative values of the argument $t$ through setting the functions to zero for $t < t_0$ The fitting routine seems to work better if it does not have to work too hard to find $t_0$, so the time arguments are supplied with the nominal offset time $t_{on}$ subtracted, providing a starting guess that $t_0 = 0$.
- Constraints consistent with the functional forms are enforced in the fitting: $\sigma > 0$ and $t_0 > t_{on} - 0.1$ for lognormal, $\alpha > 1$, $\beta > 0$, $t_0 > t_{on} - 5$ for Weibull, and $\alpha < 0$, $\beta > 0$, $t_0 > t_{on} - 5$ for inverse Weibull.
- The fit is repeated and the results updated as each subsequent 5-minute measurement is added to the data being fitted.
- For the resulting fits, a "goodness-of-fit" metric is calculated as described earlier.

This process is carried out for the following 8 combinations of constraints, e.g., for lognormal:

- $\sigma$, $\mu$ fixed, $t_0$, $A$ *fitted*;
- $\sigma$, $\mu$, $t_0$ fixed, $A$ *fitted*;
- $\sigma$ fixed, $\mu$, $t_0$, $A$ *fitted*;
- $\sigma$, $t_0$ fixed, $\mu$, $A$ *fitted*;
- $\mu$ fixed, $\sigma$, $t_0$, $A$ *fitted*;
- $\mu$, $t_0$ fixed, $\sigma$, $A$ *fitted*;
- $\sigma$, $\mu$, $t_0$, $A$ *fitted*;

- $t_0$ fixed, *σ, μ, A fitted*;

The analogous 8 combinations of constraints were also carried out for the Weibull function. In view of the poorer performance of inverse Weibull fits in the automated fitting process, it was not used to forecast maxima. Fitting each of the functional forms for both light-curve fitting and cumulative fitting with 8 options over 410 datasets every 5 minutes requires of order 6 million separate fits.

Where the shape variables $(\sigma, \mu; \alpha, \beta)$ are fixed in the above combinations, they are specified based on the event longitude using the fits in Table 1 derived for floating $t_0$. For events with no known longitude, the fit corresponding to a longitude of 50° is used, since this is the most common longitude for SEP events, but this choice has little impact since the slopes in Figs. 20 - 23 are generally small. The choice of the floating $t_0$ fits in Table 1 even when we are fitting with $t_0$ fixed is because in real-time fitting, it provides more flexibility with irregularly shaped profiles. For lognormal fits there is little difference between the floating and fixed $t_0$ results; there is more difference for Weibull and inverse Weibull, but the fixed $t_0$ fits in those cases have larger uncertainties than the floating $t_0$ fits.

Figure 25 shows the results for lognormal fits to the event maximum at 6 different times with $\sigma$ and $\mu$ fixed. In this and subsequent figures, the filled circles represent the 78 "good" events of Appendix B, while + symbols show the remaining 332 datasets. The long-dashed line in each panel is a fit to the good events with unity slope enforced, and the short-dashed lines show upper and lower uncertainty bounds. The dotted line corresponds to a correct prediction of the maximum. The second row of text in each panel gives the average multiplicative correction factor needed to convert the forecast maximum into the actual maximum; the uncertainty in the correction factor as a multiplicative factor is in the third row of text, and it results in the upper and lower bounds to the multiplicative correction provided in parentheses in the second row. This uncertainty represents the standard deviation of the points about the fitted line, calculated in log space, with outliers removed (points more than $\pm 3\sigma$ outside the initial fit, in log space, to all good events). In each panel, blue-filled circles show "good" events with rise times shorter than the time labelled in the panel, while red-filled circles are "good" events with longer rise times.

Figure 26 shows the corresponding results for Weibull fits to light curves with the two shape parameters fixed and $t_0$ floating. Figures 27 and 28 show the results for lognormal and Weibull fits to the cumulative fluence, respectively, with the two shape parameters fixed and with $t_0$ fixed at $t_{on}$. Figure 29 shows the lognormal fits to the fluence with $\mu$ and $t_0$ fixed but floating $\sigma$. For completeness, the other 27 sets of fits are shown in Appendix C (Figs. 47 - 73).

These plots reveal that none of the techniques tested with floating $t_0$ provided a fit in the first hour after the event start. The fitting methods fail and simply return the initial guess at the parameters being fitted: this can be seen in the first panel of Figs. 25 - 26, where 10 pfu is the initial guess provided for the fitting routine. With $t_0$ fixed, one less parameter is being fitted, and this makes the fitting with a small number of data points early in the event more likely to succeed.

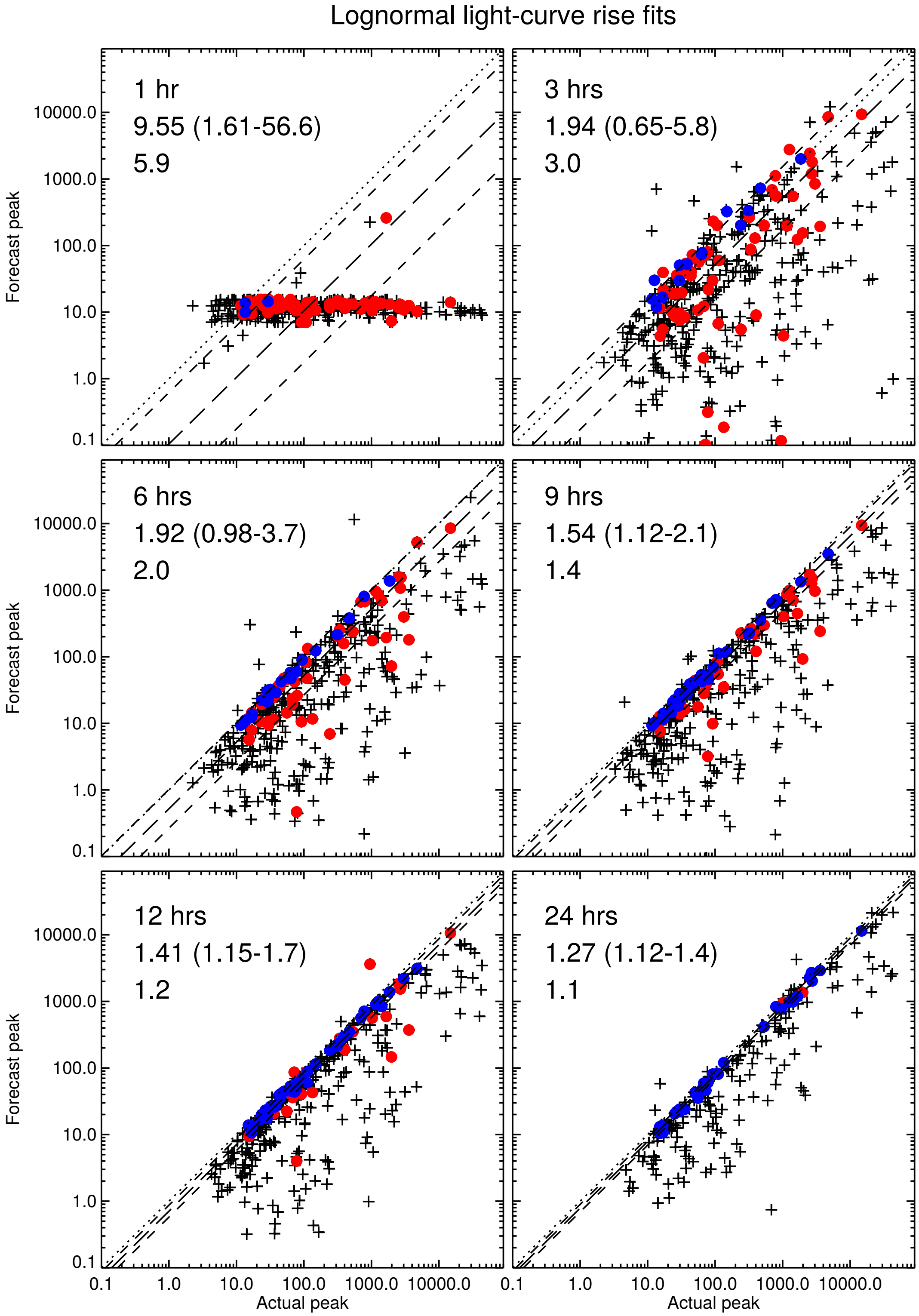

Fig. 25.— Comparison of the forecast peak flux (vertical axis) versus the actual peak flux (horizontal axis) at 6 different times after event onset for lognormal fits to light curves with fixed $\sigma$ and $\mu$ and floating $t_0$. See text for further details.

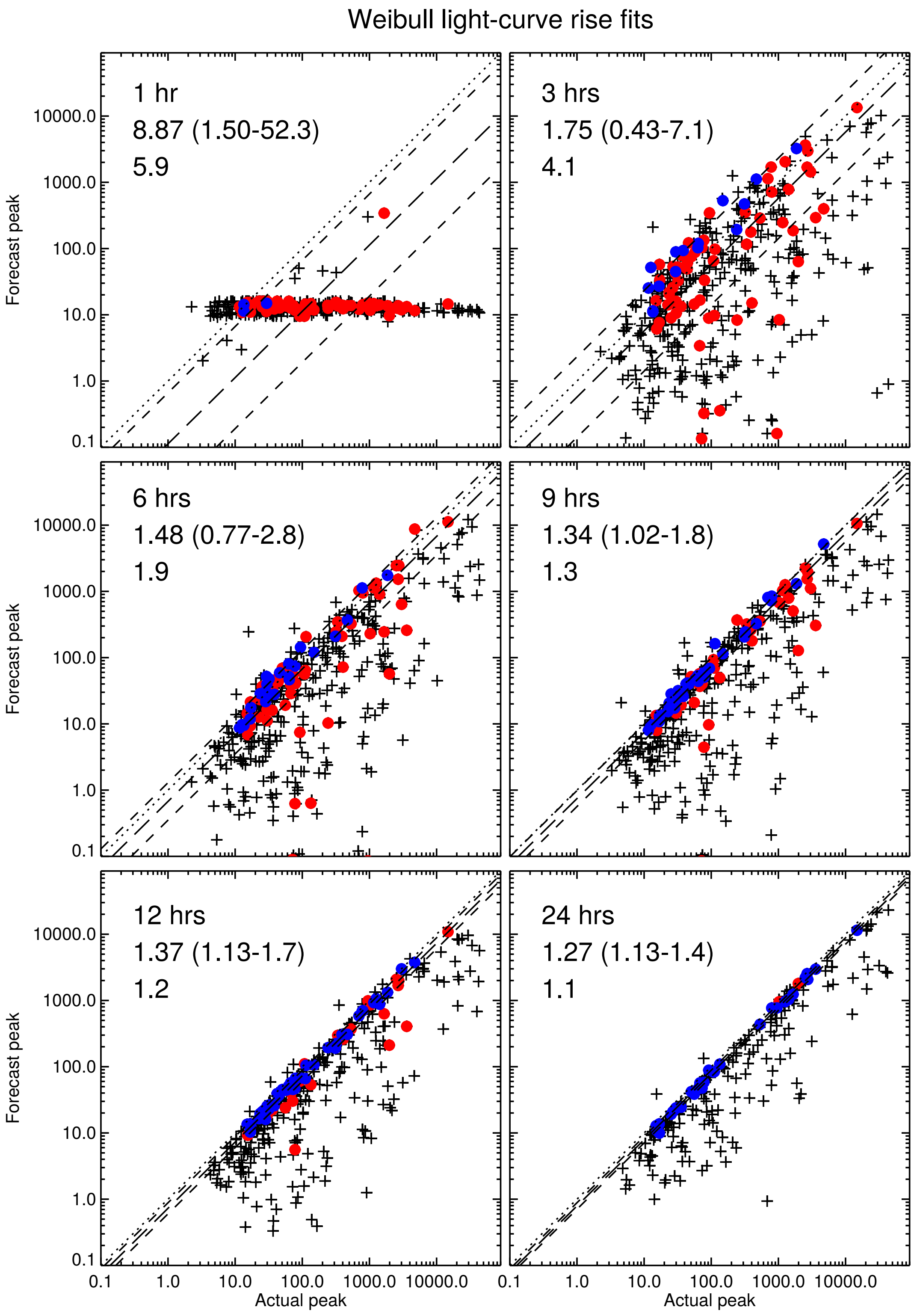

Fig. 26.— Comparison of the forecast maximum flux versus the actual peak flux for Weibull fits to light curves with fixed $\alpha$ and $\beta$ and floating $t_0$.

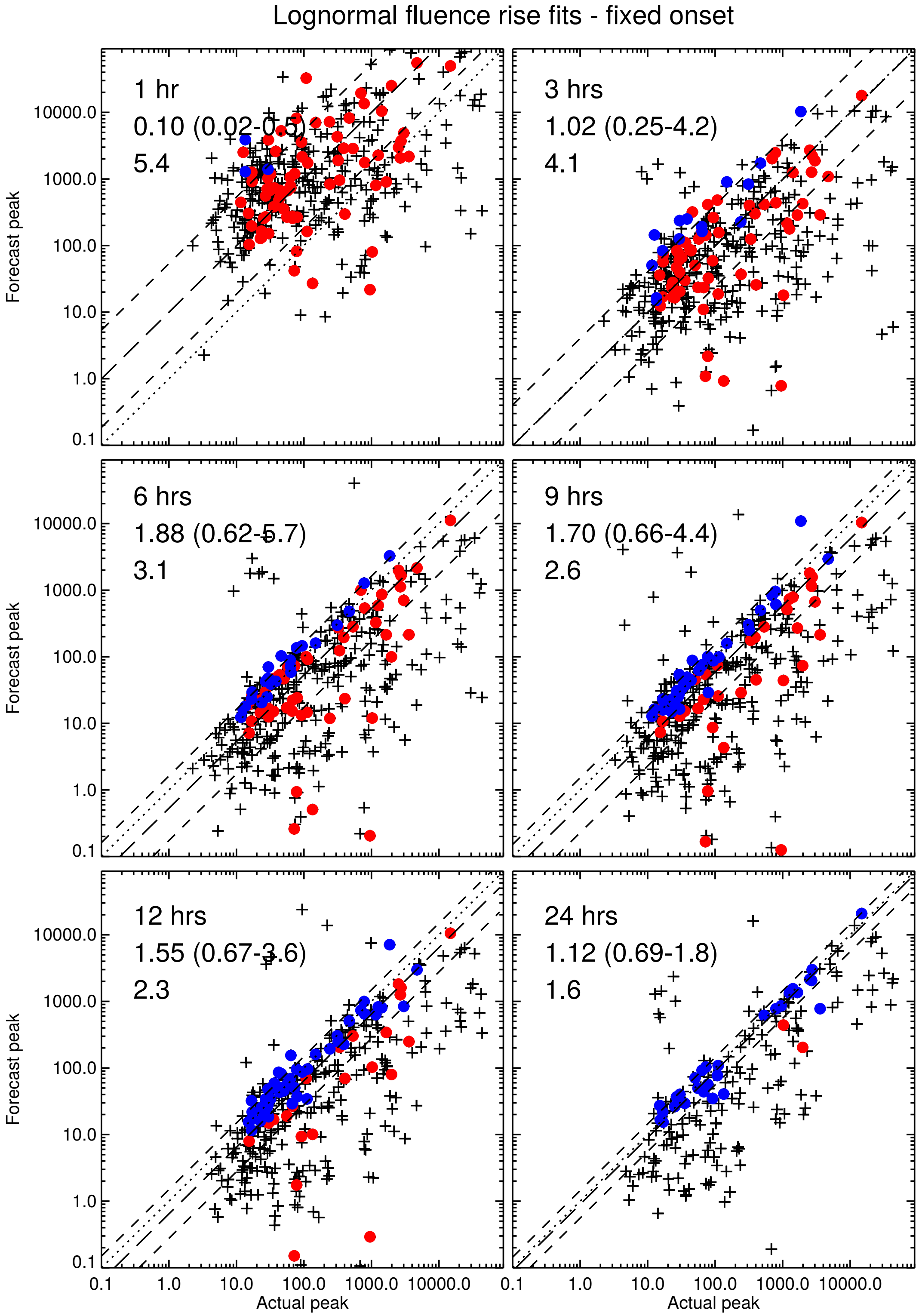

Fig. 27.— Comparison of the forecast maximum flux versus the actual peak flux for lognormal fits to cumulative fluence with fixed $\sigma$, $\mu$ and $t_0$.

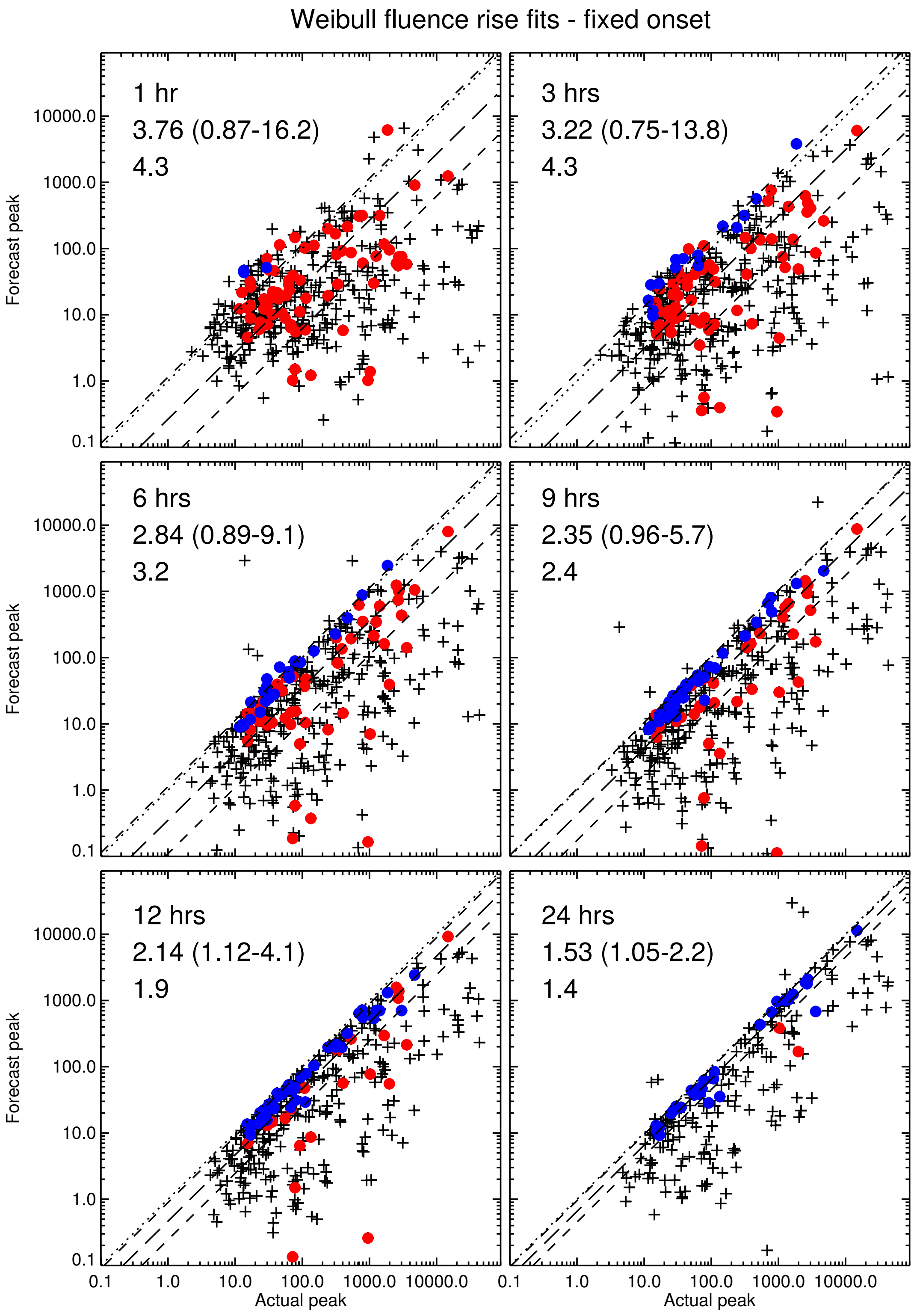

Fig. 28.— Comparison of the forecast maximum flux versus the actual peak flux for Weibull fits to cumulative fluence with fixed $\alpha$, $\beta$ and $t_0$.

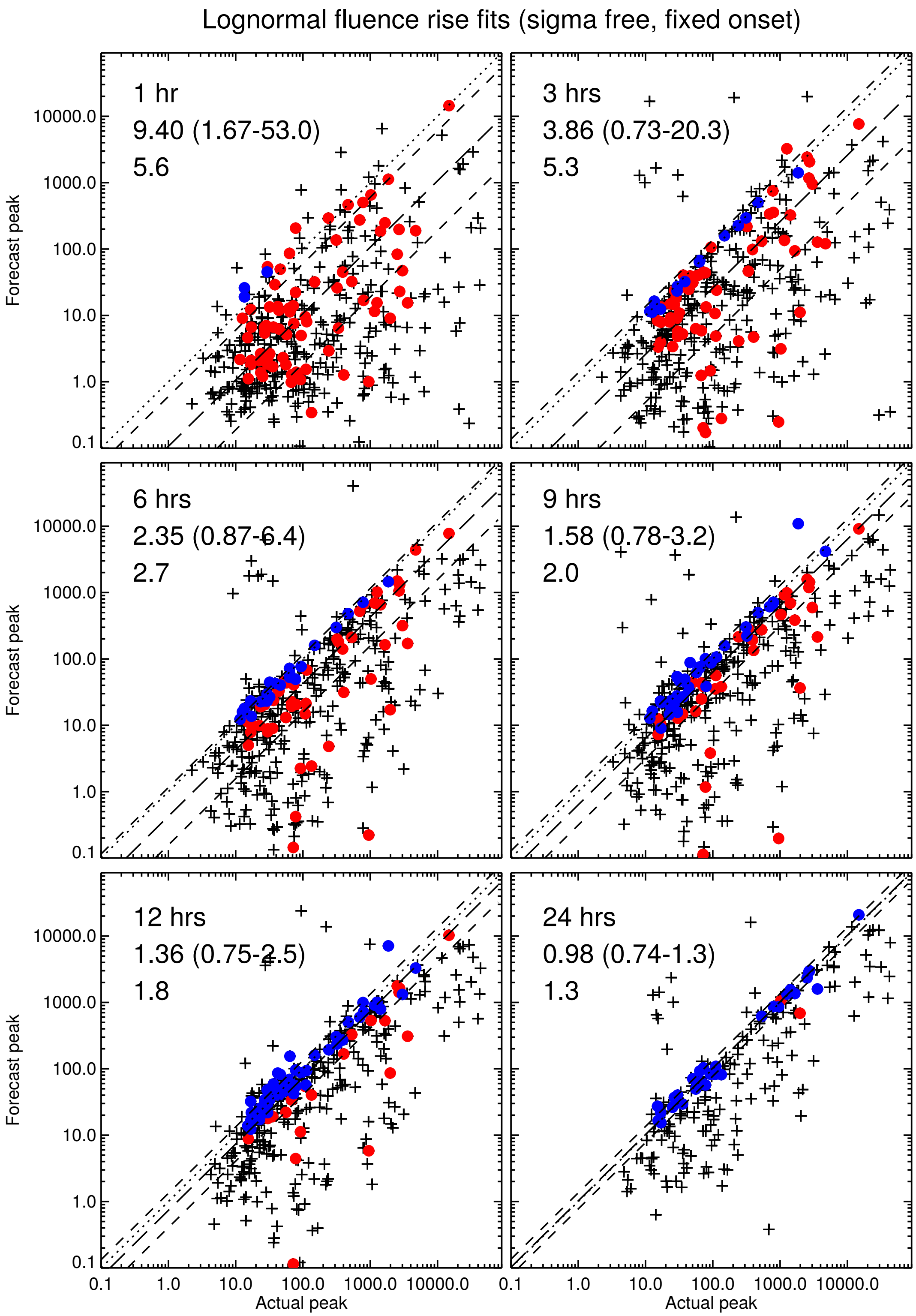

Fig. 29.— Comparison of the forecast maximum flux versus the actual peak flux for lognormal fits to cumulative fluence with fixed $\mu$ and $t_0$ and floating $\sigma$.

It is also apparent in Figures 25 - 48 that the forecast maximum is almost always below the actual maximum: this is because the fitting process generally finds that the peak of the event is "not far" (in some vague sense) above the current flux being fitted. This is true of all functional forms being tried when $t_0$ is floating. Curiously, it is not true at early times in some cases when $t_0$ is fixed: e.g., in Fig. 27 where on average the predicted maxmum is too high and the mean correction factor is actually less than unity.

It is apparent from the figures that the well-fit events tend to be those with short rise times (thus, the blue points in each panel), while most of the spread in the correction factors results from the red points, corresponding to events that have yet to peak. This result simply emphasizes the fact that generally there is insufficient information during the rise of an event to provide a robust estimation of the full shape of the event. The correction factors shown in the panels represent a compromise between these two groups: it is then relevant to ask if the correction factor should be based only on the "better-fit" events, or if it should include all "good" events. If the poorer fits are predominantly events with longer rise times, then their peak values will initially be under-predicted, but more time is available to correct the predictions.

From 6 hours and onwards the predictions are often quite good and the correction factor required is not too far from unity: this probably reflects the fact that we have enough data to do a reasonable job of getting the fit correct despite not fitting for the shape parameters. Once the peak of an event has been reached and the flux starts to decline, it makes sense to fit all 4 functional parameters in order to provide a better estimate of the longer-term behavior, and of the total fluence to be expected. This is implemented in the fitting process used for these figures after the flux decays to 50% of the peak, and so at some of the later times, when the predicted maximum is doing very well, it is the case that all 4 parameters are being fitted correctly, producing better results. The results from fitting all 4 parameters from the beginning are shown in Figs. 66 - 69. The spread in results at early times beyond 1 hour is generally much wider than when only 2 parameters $(A, t_0)$ are being fitted.

The most important parameter in the fitting process used here from the forecasting perspective is the uncertainty in the correction factor, since it represents the reliability of the forecast. The uncertainty is the spread about the fitted line in log-log space, and thus is a multiplicative uncertainty, i.e., one multiplies the correction factor by the uncertainty to get an upper bound, and divides by the multiplicative uncertainty to get the lower bound. Figure 30 plots the fitted multiplicative uncertainty versus time from event onset for each of the 32 fitting assumptions. Each panel has 4 curves corresponding to floating (solid lines) and fixed (dashed lines) $t_0$ for fits to the light curves (red) and fluence (blue); the eight panels correspond to all 4 possible combinations of the pairs of shape parameters being fixed or floating, for lognormal (left column) and Weibull (right column) functional fits, respectively. As noted above, the early fits with floating $t_0$ are unsuccessful, and this can be seen in the delayed start in the solid lines in each panel.

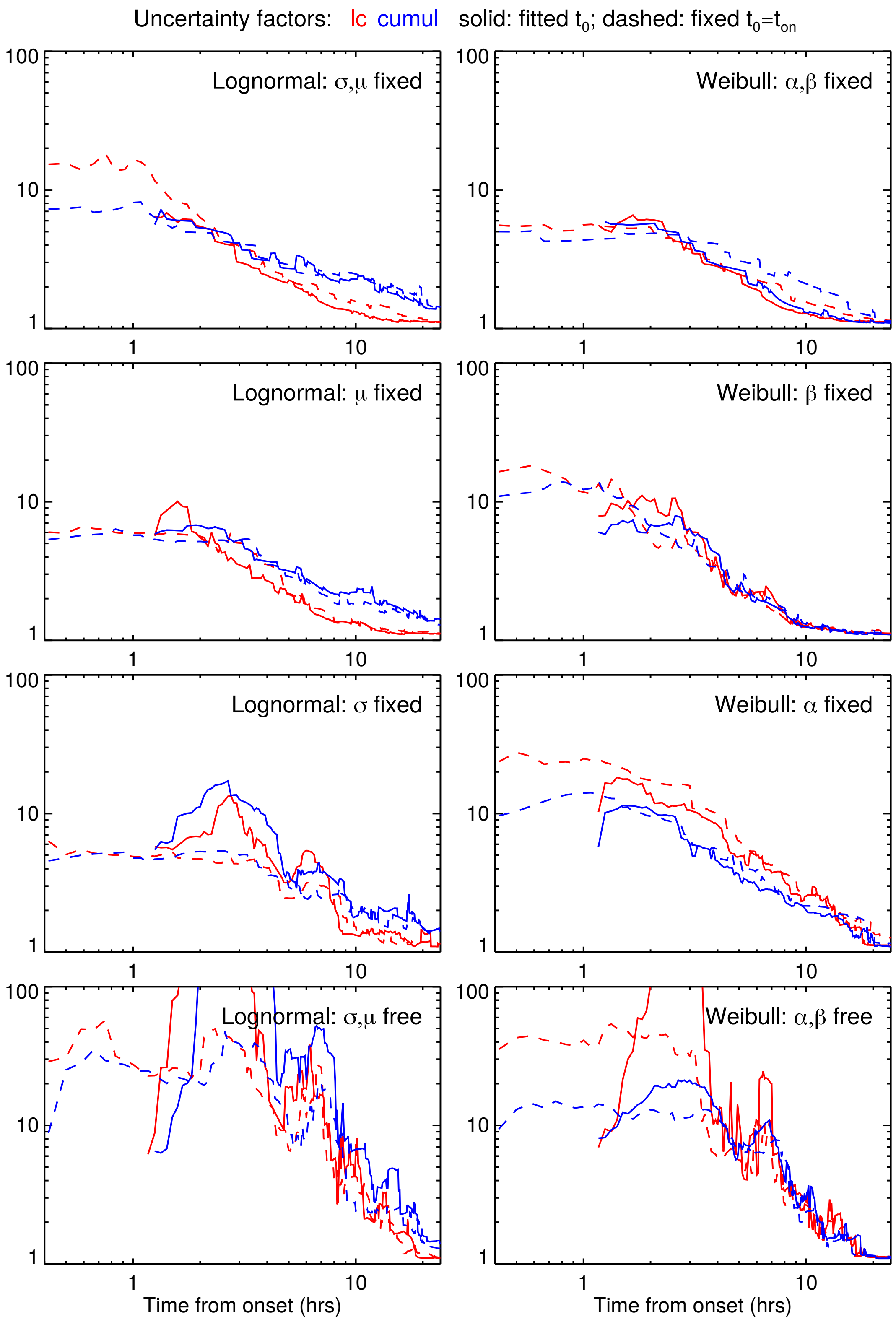

Fig. 30.— The multiplicative uncertainty in the correction factor versus time for the 32 types of fit carried out.

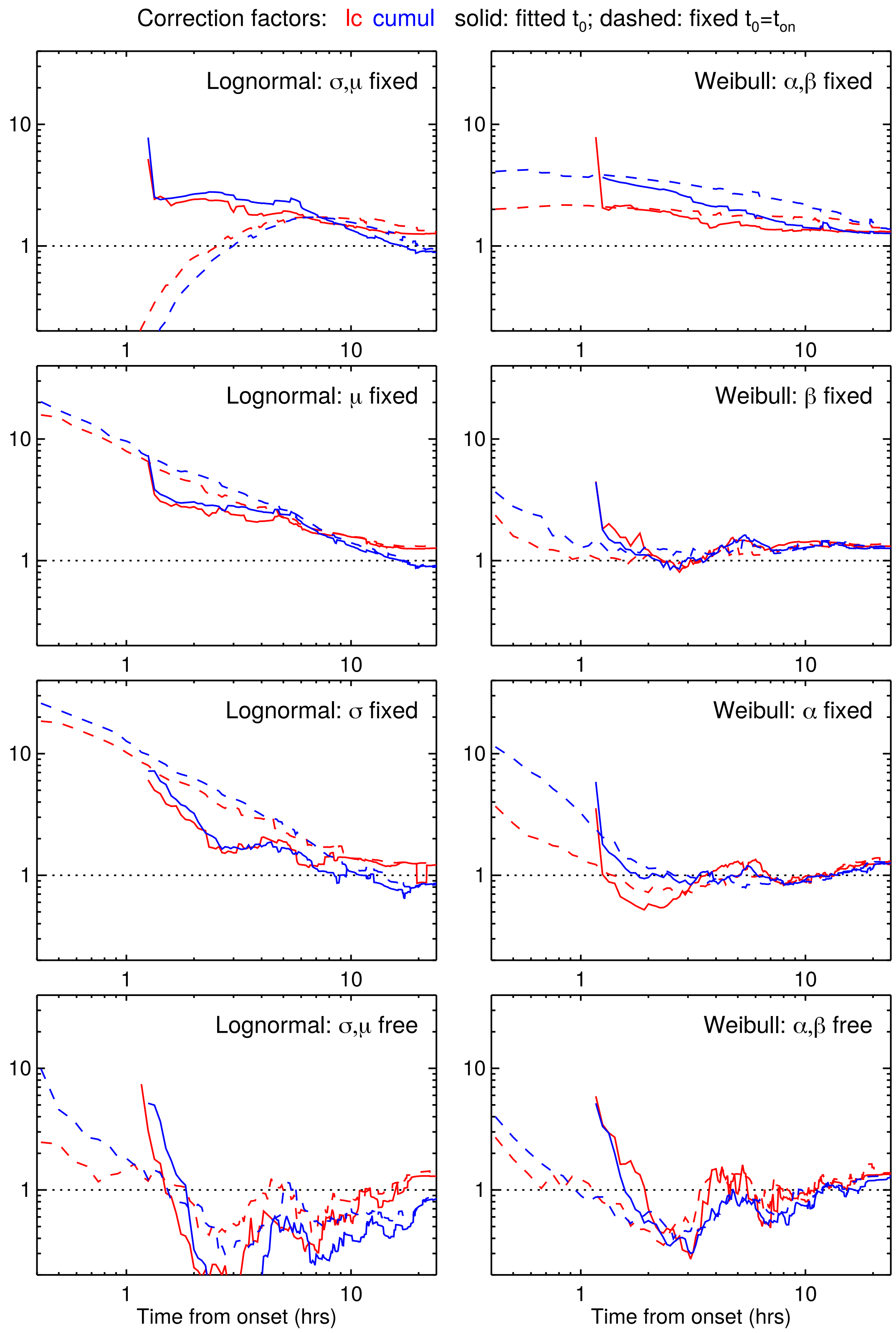

Fig. 31.— The correction factor versus time for the 32 types of fit carried out.

The lowest uncertainties in the first hour are provided by the Weibull function with fixed $\alpha$, $\beta$ and $t_0$ (Fig. 28), showing similar results for both light curve and cumulative fluence fitting, and the lognormal fit to the fluence with fixed $t_0$ and either $\sigma$ (Fig. 63) or $\mu$ (Fig. 29) fixed, but not both. For intervals between 3 and 12 hours, either lognormal or Weibull fits to the light curves with fixed $\alpha$ and $\beta$, and floating $t_0$, appear to give the best results.

The corresponding behavior of the correction factor itself as a function of time is shown in Figure 31. A smooth variation with time is desirable for forecasting purposes, and several of the fitting options can be ruled out from this perspective, but the options mentioned above appear to be satisfactory. The uncertainty in the correction factors is compared directly for four of the options in Figure 32. For the first 2 hours of an event, Weibull fits to the fluence appear to provide the lowest uncertainty. From 3 to 6 hours after onset, lognormal fits to the light curves (with shape parameters fixed but $t_0$ variable) do best, while after 6 hours both lognormal and Weibull fits to the light curves do equally well.

The corresponding correction factors for these options are shown in Figure 33. The time dependence of the correction factor is sufficiently smooth in all cases that it is not a negative factor in choosing a fitting process.

## A First Proposal for Operational Fitting

The results of the previous section suggest the following modifications to the fitting process described at the beginning of that section:

- Prior to about 90 minutes after the SEP event start, fits with variable $t_0$ are likely to fail. Therefore for this initial phase fixed $t_0 = t_{on}$ should be enforced, together with fixed shape parameters. Weibull fluence fits give the best performance out to 2 hours after event onset.
- Beyond 2 hours, both Weibull and lognormal light-curve fits with fixed shape parameters (based on event longitude if available) and floating $t_0$ work well, but with lognormal fits being slightly better from 2 to 6 hours.
- By 6 hours some events have already reached their peak, and better results may be achieved by fitting the shape parameters as well as $A$ and $t_0$. However, in real time one cannot know if a local peak will prove to be the actual peak of the event (e.g., see Fig. 6). Therefore one needs to be able to revert back to fixed-shape-parameter fitting if the rise begins again after a local peak. This is different from the process used above, where the peak value was known in advance.

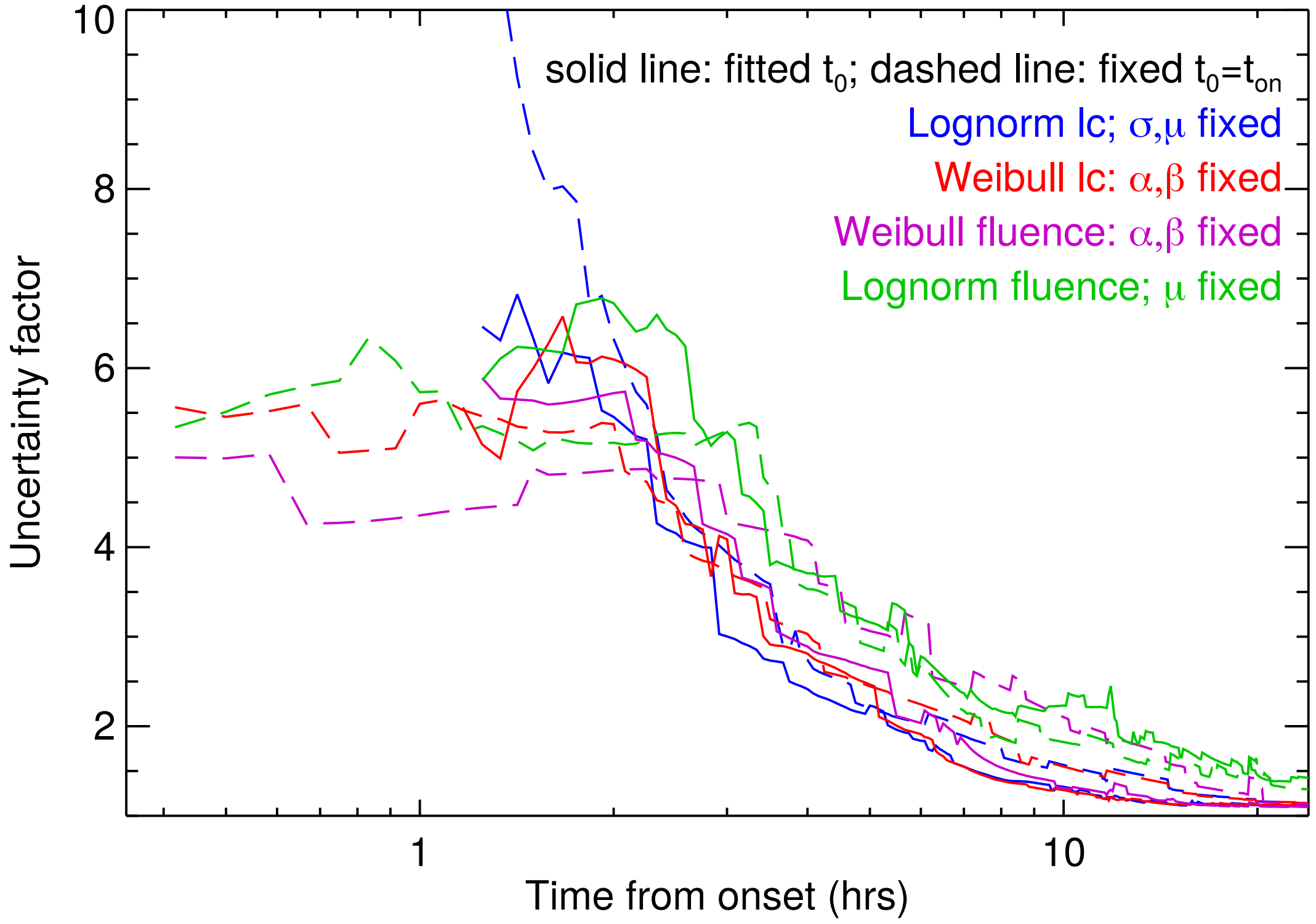

Fig. 32.— The multiplicative uncertainty in the correction factor versus time for four of the better fitting options: lognormal (blue) and Weibull (red) fits to the light-curves and Weibull fits to the fluence (purple) with both shape parameters fixed, and lognormal fits to the fluence with $\mu$ fixed (green). Solid lines represent fits with floating $t_0$, while dashed lines show fits with fixed $t_0 = t_{on}$.

This algorithm has been implemented as a test and applied to the event light curves, in both Weibull and lognorm forms for comparison. One parameter not addressed in the discussion above is how far down on the decay of the peak to wait before starting to fit the shape parameters. The issue is that the correct shape parameters are needed in order to provide the best estimate of duration and therefore total fluence, and it is desirable to provide this as soon as possible. In the version described here, the shape parameters are fitted when the decay reaches 80% of the most recent peak.

However, this version in which both shape parameters are fitted on the decline of local peaks introduces a lot of variability, and does not do as well as seen in Fig. 32. By trial and error, the following process was found to perform better:

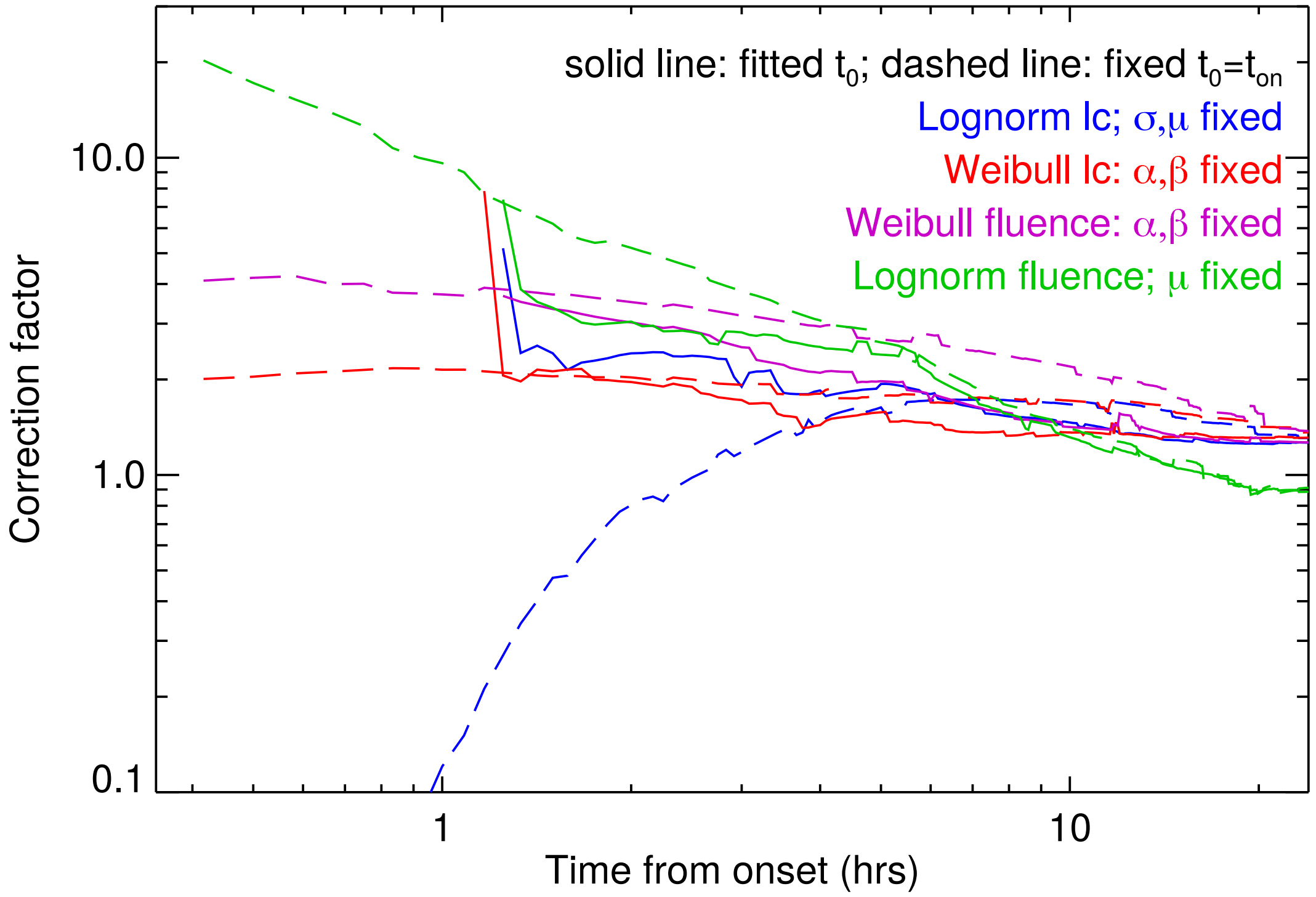

Fig. 33.— The correction factor versus time for fitting options presented in Fig. 32.

- Weibull fluence fits with fixed $t_0 = t_{on}$ and fixed shape parameters are used for the first 2 hours.
- From 2 to 6 hours, either Weibull and lognormal light-curve fits are made with fixed shape parameters (based on event longitude when available) and floating $t_0$.
- These fits are continued from 6 hours onwards, until a point 2 hours beyond the most recent peak as long as the flux has dropped to below 80% of the peak value. This condition is intended to avoid interpreting short local fluctuations as the main peak, at the cost of waiting 2 hours beyond the peak for a better fit. Once the condition is satisfied, one of the shape parameters is permitted to float: $\mu$ for lognormal fits and $\alpha$ for Weibull fits.
- In order to minimize extremes, when fitted the shape parameters are only permitted to lie in the ranges identified by the upper and lower red dashed lines in Figs. 20 and 22 for lognormal and Weibull light-curve fits, respectively.

There is in fact little difference in the results of the last stage resulting from the choice of fitting one or other shape parameter or both. The fitting distributions for all events at 6 different times resulting from the process described above are shown in Figures 34 and 35.

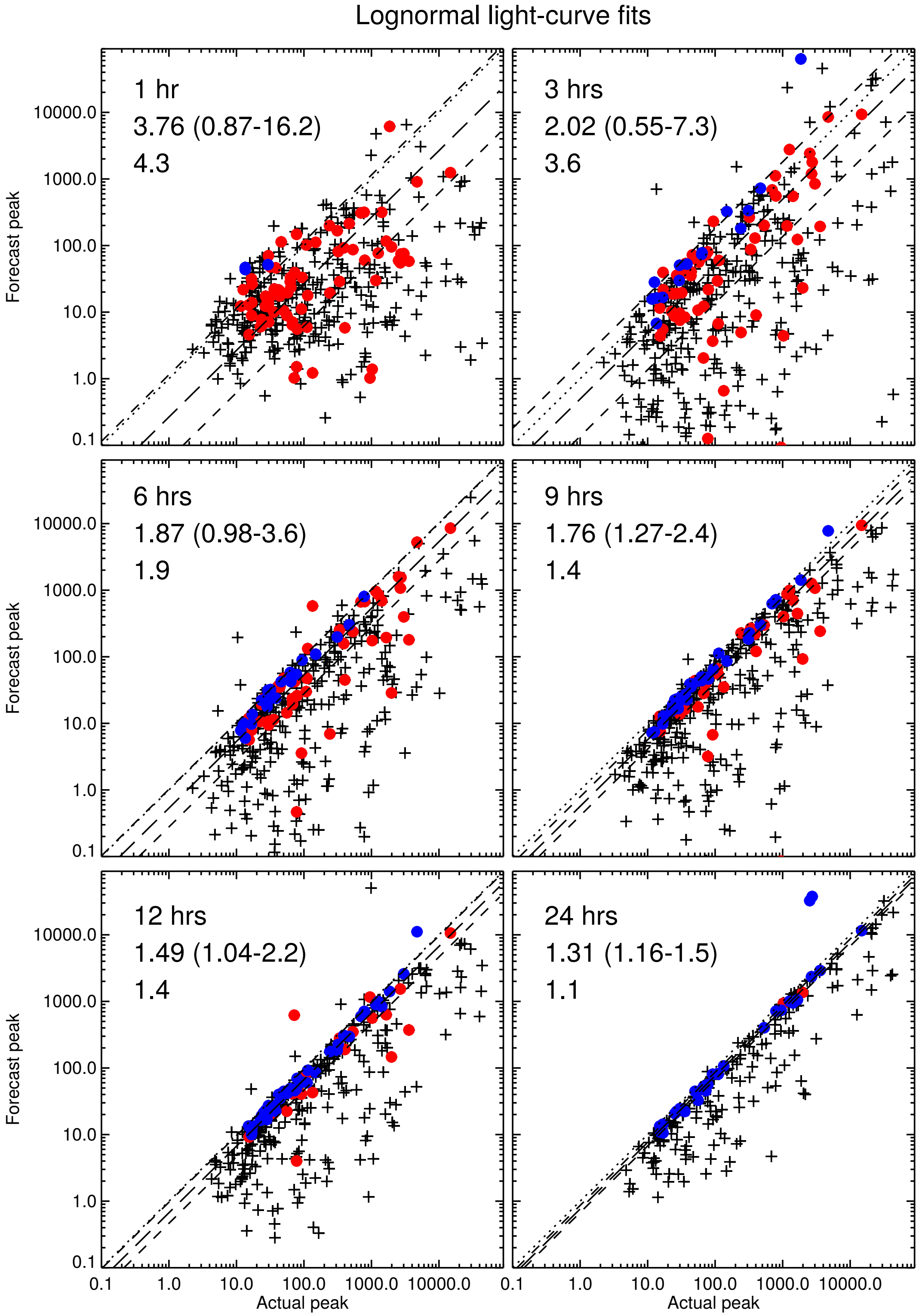

Fig. 34.— Comparison of the forecast peak flux versus the actual peak flux at 6 times after event onset for the test forecasting process with Weibull fits to the fluence for the first 2 hours, followed by lognormal fits to light curves. See text for further details.

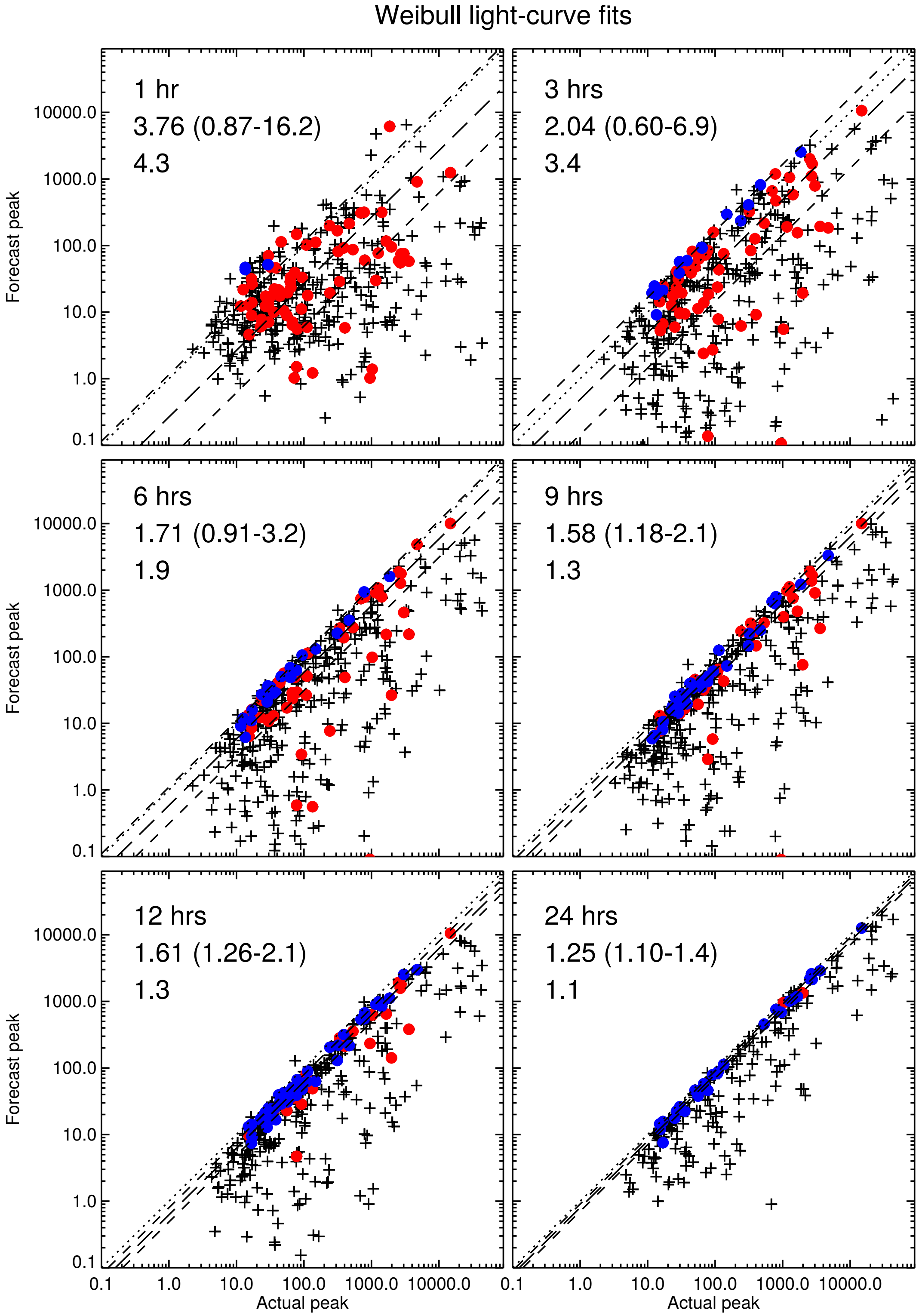

Fig. 35.— Comparison of the forecast peak flux versus the actual peak flux at 6 times after event onset for the test forecasting process with Weibull fits to the fluence for the first 2 hours, followed by Weibull fits to light curves. See text for further details.

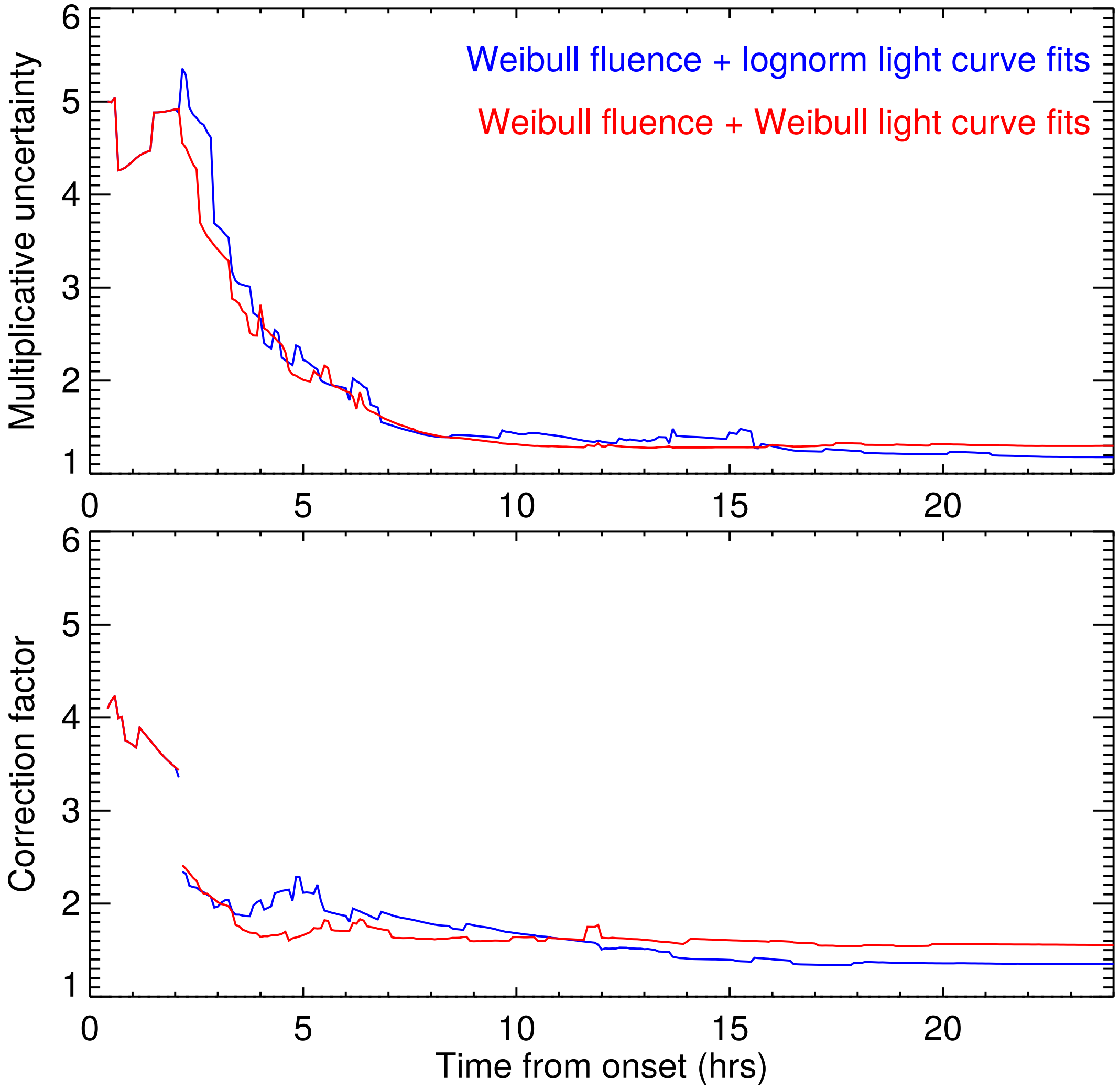

Fig. 36.— The multiplicative uncertainty in the correction factor (upper panel) and the corresponding correction factor (lower panel) versus time for the lognormal (blue) and Weibull (red) versions of the test light-curve forecasting process. The correction factor has a discontinuity at 2 hours when the transition from fitting fluence to fitting light curves occurs.

The corresponding multiplicative uncertainty and correction factors are shown as a function of time in Figure 36. We recognize that for an arbitrary event, these values are unrealistic, based as they are on carefully-selected "well-behaved" events with outliers removed. However, for initial steps it is appropriate to focus on such events where functional forms with just 2 shape parameters might be expected to represent the light-curve forms, rather than events such as those in Fig. 6 which are unlikely to be matched by simple functional forms.

In order to employ the corrections presented in Fig. 36, we need analytic representations of the curves shown. Note that for the correction factor itself there is a jump at the 2 hour break where we transition from fitting fluence to fitting light curves. Convenient fits for the first 24 hours are as follows:

- Correction factor, where time $t$ is measured in hours from the onset of the event (with onset identified by the protocol defined earlier):

$$\text{Lognorm correction}(t) \;=\; \begin{cases} 4.28 \;-\; 0.412 * t, & t \;\leq\; 2\text{hrs} \\ -0.217 \;+\; 0.047 * t \;+\; \frac{509}{t^{2.13} + 225}\;, & 2 \;<\; t \;<\; 24\text{hrs} \end{cases} \tag{5}$$

$$\text{Weibull correction}(t) \;=\; \begin{cases} 4.28 \;-\; 0.412 * t, & t \;\leq\; 2\text{hrs} \\ 1.711 \;-\; 0.007 * t \;+\; \frac{1305}{t^{7.32} + 1544}\;, & 2 \;<\; t \;<\; 24\text{hrs} \end{cases} \tag{6}$$

- Multiplicative uncertainty factor:

$$\text{Lognorm uncertainty}(t) \;=\; 1.674 - 0.023 * t + \frac{4200}{t^{5.58} \;+\; 1320}\;, \qquad 0 \;<\; t \;<\; 24\text{hrs} \tag{7}$$

$$\text{Weibull uncertainty}(t) \;=\; 1.274 + 0.0007 * t + \frac{342}{t^{3.65} \;+\; 96.5}\;, \qquad 0 \;<\; t \;<\; 24\text{hrs} \tag{8}$$

These fits are compared to the actual curves in Figure 74. Implementation of the correction factors in the fitting process described above produces the results shown in Figures 37 and 38. For the "good" events the effect is more or less as expected: the mean predictions, excluding extreme outliers, are centered about the actual maxima to within the corresponding uncertainty. There remain anomalies: the spread at 3 hrs for lognormal fits is larger than at 1 hr, whereas the Weibull fits at 3 hrs are better than at 1 hr. Some of these anomalies can be due to the process of removing outliers: a wider spread makes it less likely that extreme cases will be removed, so there is some arbitrariness in the process. At 24 hours, the formal spread is small but in both cases the mean predictions are over-estimates by an amount larger than the nominal spread, indicating that it should be larger. The analytic expressions in (5) and (6) do not capture deviations from a smooth curve evident in Fig. 36, although these are small after 16 hours. In general the non-"good" events are under-estimated by this fitting process, defined as it is by the "good" events..

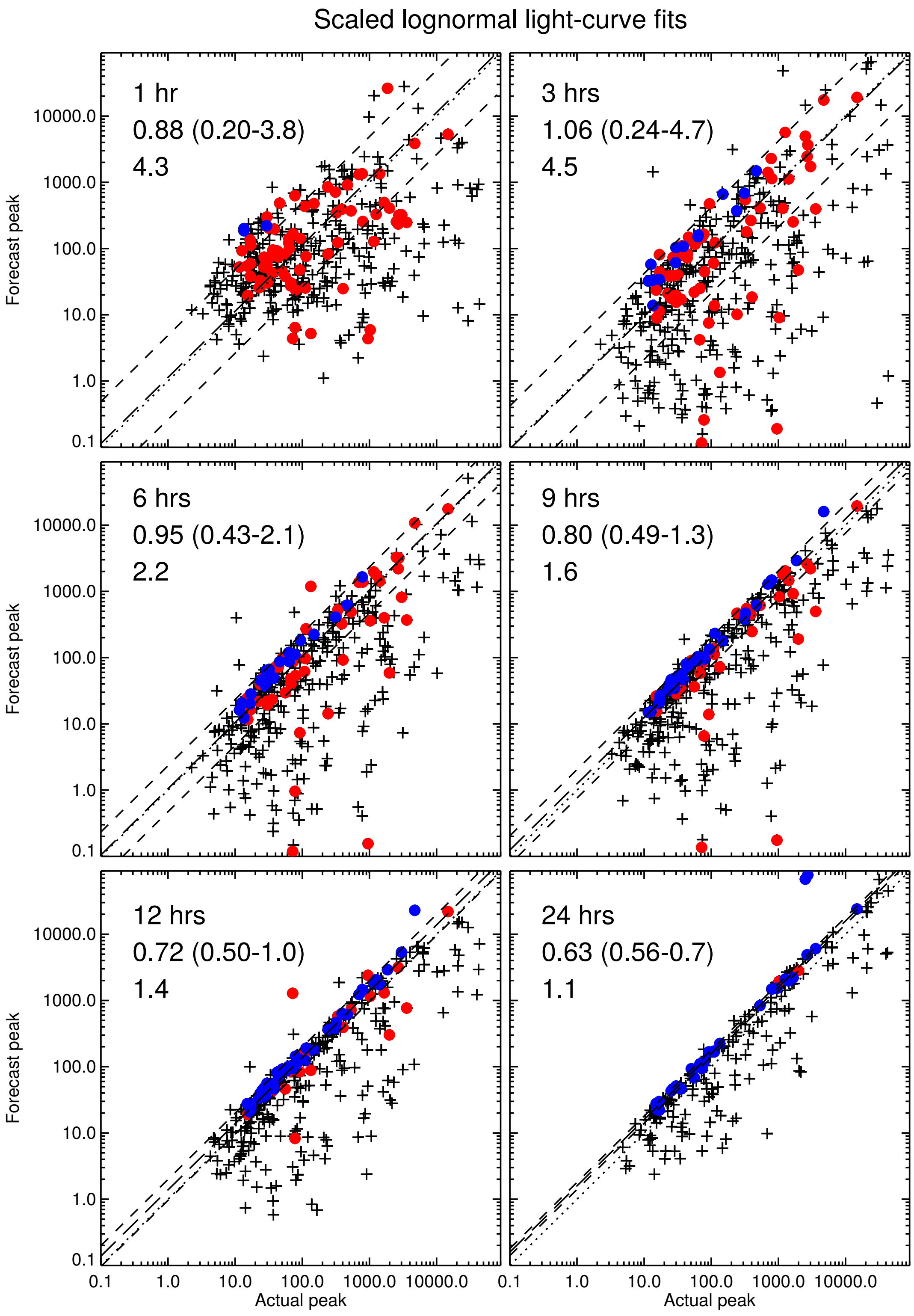

Fig. 37.— The result of applying the correction factor (5) in quasi-"real-time" to the fitting process for the lognormal light-curve forecasting option.

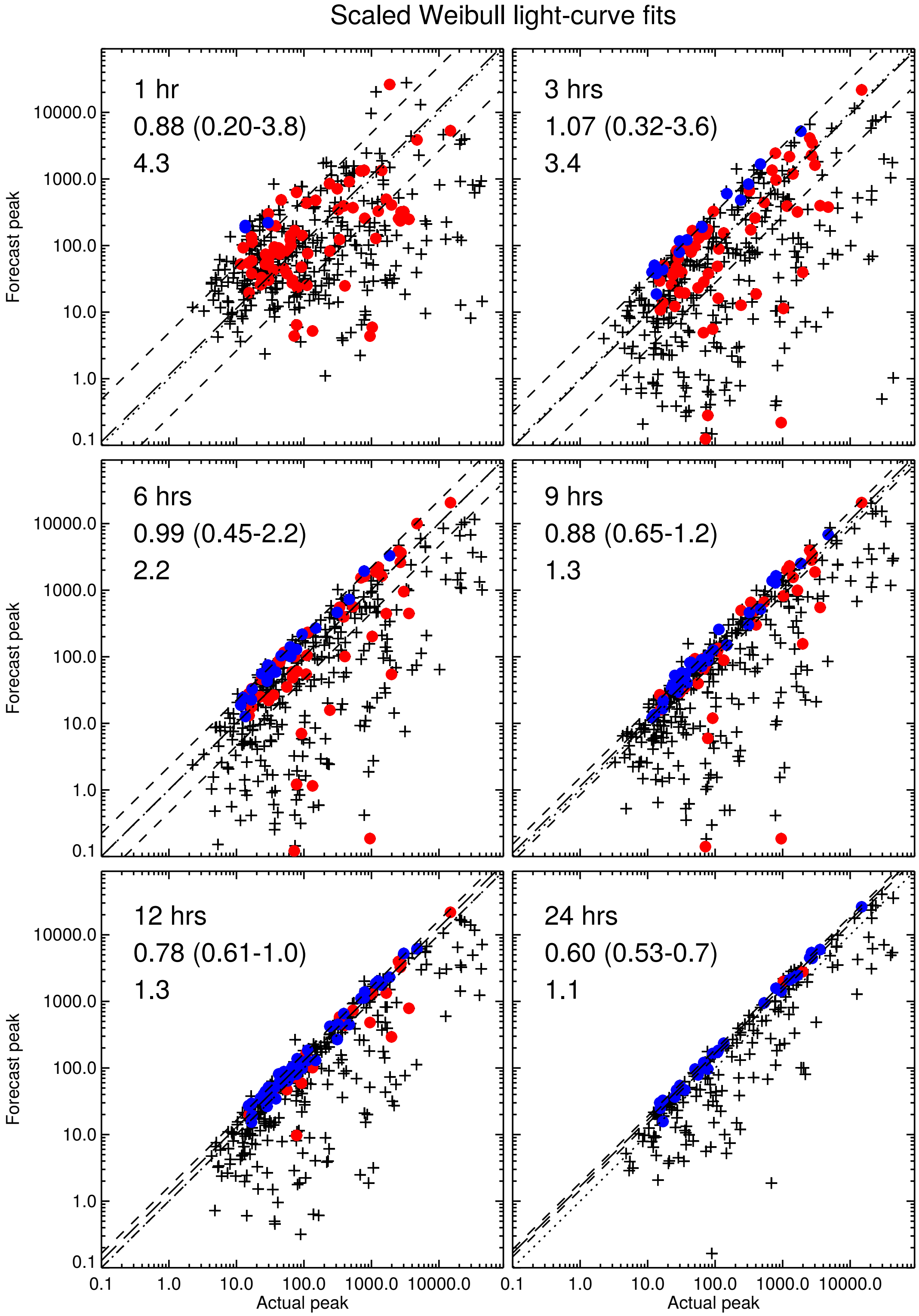

Fig. 38.— The result of applying the correction factor (6) in quasi-“real-time” to the fitting process for the Weibull light-curve forecasting option.

## Summary

Based on the results and discussion above, a process for a prototype forecasting method is as follows:

- The 5-minute GOES 10 MeV proton fluxes are continually monitored in real time ( the latency in the availability of the measurements at present seems to be of order 5-10 minutes);
- a time-varying background trend is fitted to the last, say, 3 hours of data, and the variation around that trend is measured;
- when the data rise $5\sigma$ (and at least 0.6 sfu) above the trend, the fitting process is triggered: an onset time is determined by extrapolation backwards to the trend line, and 30 minutes (6 data points) after that time the first fit of the cumulative fluence data to a Weibull function is carried out, after subtracting the background trend. In the 410 datasets investigated, this trigger works reliably. The two shape parameters are fixed at the values determined from the longitude of the associated flare event, which should be available by then; if not, default values are used. It is found that the results are not very sensitive to this choice, since the range of shape parameters found in the archival events is not that large.
- For the first 2 hrs the "zero" time in the functional form is fixed at the onset time, and after 2 hrs it is to be fitted. This has the advantage that we do not need to try to keep track of overlapping events: if a new event starts and takes over from a previous event, the zero time will automatically shift as the flux from the new event starts to dominate the light curve.
- After 6 hours we allow one or more shape parameters to vary in fitting the light curves, if the event is 2 hours beyond a main peak.
- In line with Figure 36, we have a pre-determined correction factor as a function of time from the onset: the maximum determined from the fit at the current time is multiplied by the appropriate correction factor, and this represents the forecast. The uncertainty in this forecast is similarly pre-determined: it is necessarily large for the first few hours.
- The fixed shape parameters mean that the timescale of the event is fixed during the rise phase: as noted above, improved knowledge of the timescale can be determined once the shape parameters can also be fitted.

We refer to this algorithm as the Air Force Dynamic Energetic Particle Tool, or ADEPT. It is being implemented for testing in AFRL's SEP Forecasting System (SFS). We emphasize again that getting the onset time "right" is important in the fitting process: for most of this effort the $5\sigma$ criterion alone was used, but it tended to result in onsets significantly before the main peak, which drew out the rise phase of the events and correspondingly produced much larger uncertainties than can be seen in Figs. 37 and 38.

There are numerous potential modifications that will likely improve outcomes. Some features that will require further investigation are:

- The threshold of 0.6 sfu with $5\sigma$ used here to define event onset is somewhat arbitrary: for a focus on forecasting large events, a larger threshold may be more effective.
- The effective uncertainty at early times is dominated by the spread in forecasts relative to the actual peak, but at later times can be dominated by the application of the correction factor: the quoted uncertainty does not yet represent this accurately. Similarly, the question of whether the uncertainty should be defined by the "good" events needs to be considered.
- Fitting $t_0$ allows flexibility in fitting events with complexity, or overlapping events, but complicates the application of the correction factor which, in this version, relies on knowing an appropriate value for $t_{on}$. It is possible that correction relative to $t_0$ rather than $t_{on}$ may be more appropriate if it better represents the main onset, but as noted in the discussion of the functional forms, $t_0$ is not always a good representation of the onset for Weibull fits.
- In addition, there is the need for a protocol to identify the end of an event, which has not been addressed here.
- An analysis of the performance of the fluence forecasts is needed.

---

## Appendix A: Functional forms

The three functional forms being used for SEP curve fitting are the lognormal function, the Weibull function and the inverse Weibull function.

### Lognormal function

The lognormal distribution function takes the following form:

$$f(x) = \frac{1}{\sqrt{2\pi}\,\sigma\,x}\ \exp\{-\frac{(\ln(x)-\mu)^2}{2\,\sigma^2}\}$$

This function can take on forms that represent the conventional shapes of SEP light curves, and in particular both the impulsive shapes of well-connected west limb events as well as the more gradual east-limb events. The lognormal distribution has just the two parameters:

- $\mu$ is the log of the median of the function and thus influential in determining where the peak occurs;
- and $\sigma$ (always positive), which affects the width of the peak.

Variations in the shape of the resulting form as a function of $\sigma$ and $\mu$ are shown in Figure 2.

The form shown in (1) represents a probability distribution and thus integrates to unity. For light curve fitting, we need to add an amplitude parameter in order to match the peak count rate observed. In addition, $f(x)$ defined above does not exist for $x < 0$, and this is an inconvenient constraint for fitting a light curve. We therefore extend $f(x)$ to negative arguments by setting it to zero at $x < 0$. This should work well in principle for background-subtracted light curves that are dominated by noise prior to the rise of the event. For events superimposed on the decay of a preceding event, this necessitates that we carry out background subtraction before fitting in order to facilitate the match of the data to the functional form. The location of the $x = 0$ point in the light curve provides a fourth parameter that needs to be fitted.

Thus the actual function used in the light-curve fit is

$$\mathrm{Ln}(t) \;=\; \begin{cases} 0, & t \,\leq\, t_0 \\ \frac{A}{\sqrt{2\pi}\,\sigma\,(t-t_0)}\ \exp\Big\{-\frac{(\ln(t-t_0)-\mu)^2}{2\,\sigma^2}\Big\}, & t \,>\, t_0 \end{cases} \tag{9}$$

with the 4 parameters $A$, $t_0$, $\sigma$ and $\mu$.

Once $\sigma$ and $\mu$ are known, the following properties can be determined:

- the location of the peak is at $x = \exp(\mu - \sigma^2)$, providing a rise-time estimate;
- the value of $f(x)$ at the peak is $A \exp(\sigma^2/2 - \mu)/\sqrt{2\pi}/\sigma$;
- two-thirds of the distribution lies between $\exp(\mu - \sigma)$ and $\exp(\mu + \sigma)$, thus providing a width estimate.

The "cumulative lognormal" function takes the following analytic form:

$$\int_{t_0}^{t} \mathrm{Ln}(t)\, dt \;=\; \frac{A}{2}\, \mathrm{erfc}\Big\{-\frac{\ln(t - t_0) - \mu}{\sigma\sqrt{2}}\Big\} \qquad t > t_0 \tag{10}$$

where erfc is the complementary error function.

**Weibull function**

The Weibulll probability density function also has two parameters, labelled $\alpha$ and $\beta$:

$$f(x) = \frac{\alpha}{\beta}\left(\frac{x}{\beta}\right)^{\alpha-1} \exp\Big\{-\left(\frac{x}{\beta}\right)^{\alpha}\Big\} \qquad x > 0$$

In this case $\beta$ is more explicitly a scaling factor, appearing in the divisor wherever the independent variable occurs. Examples of the shape of the Weibull function depending on the values of the two parameters are shown in Figure 3. In order to be suitable for fitting light curves that start from a low value, $\alpha > 1$ must be satisfied by the shape parameter, and negative values of $\beta$ have no useful meaning.

For fitting light curves, we again need an amplitude $A$ and an onset-time parameter $t_0$: the fitted function is thus

$$\mathrm{We}(t) \;=\; \begin{cases} 0, & t \le t_0 \\ A\,\frac{\alpha}{\beta}\left(\frac{t-t_0}{\beta}\right)^{\alpha-1} \exp\Big\{-\left(\frac{t-t_0}{\beta}\right)^{\alpha}\Big\} & t > t_0 \end{cases} \tag{11}$$

The cumulative function for the Weibull function takes a particularly convenient form:

$$\int_{t_0}^{t} \mathrm{We}(t)\, dt \;=\; A\Big(1 \;-\; \exp\Big\{-\Big(\frac{t - t_0}{\beta}\Big)^{\alpha}\Big\}\Big) \tag{12}$$

Other relevant properties of the Weibull function are:

- The maximum occurs at $x = \beta\left(\frac{\alpha-1}{\alpha}\right)^{1/\alpha}$, which provides the rise time.

- The value of the maximum is $A\,\frac{\alpha}{\beta}\,\left(\frac{\alpha-1}{\alpha}\right)^{\frac{\alpha-1}{\alpha}}\ \exp\Big\{-\ \frac{\alpha-1}{\alpha}\Big\}$

**"Inverse" Weibull function**

When used as a distribution, the $\alpha$ parameter in the Weibull function needs to be non-negative. However, mathematically, negative $\alpha$ produces perfectly acceptable functional forms with appropriate modifications:

$$\mathrm{We}(t) \;=\; \begin{cases} 0, & t\,\le\,t_0 \\ A\,\frac{|\alpha|}{\beta}\,\left(\frac{t-t_0}{\beta}\right)^{\alpha-1}\,\exp\Big\{-\left(\frac{t-t_0}{\beta}\right)^{\alpha}\Big\} & t\,>\,t_0 \end{cases} \tag{13}$$

$$\int_{t_0}^{t}\,\mathrm{We}(t)\;dt \;=\; A\Big(1\;-\;\exp\Big\{-\Big(\frac{t\,-\,t_0}{\beta}\Big)^{|\alpha|}\Big\}\Big) \tag{14}$$

Figure 4 shows examples of the resulting functional forms for varying $\alpha$ and $\beta$. Kahler & Ling (2017) found that in many cases the inverse Weibull function produced better fits to SEP light curves than the regular Weibull function.

However, we note some features of the inverse Weibull function that may make it less satisfactory for fitting the rise of a light curve in real time. Unlike the Weibull function with $\alpha\,>\,1$, which starts its rise at the origin (x=0), effectively the inverse Weibull function does not start to rise until well after the origin: in the examples shown in the upper panel of Figure 4, the rise of the light curve does not start until about $x\,=\,\beta/2$. Similarly, the lower panel shows that the value of $\alpha$ also plays a major role in determining when the functional form starts to rise. This seems to leads to a strong co-dependence between $\alpha$ and $\beta$, whereby very different pairs of values can produce very similar shapes. The fitted values for $\alpha$ and $\beta$ depend on the allowed ranges in the fits. This is less satisfactory for prescribing values of $\alpha$ and $\beta$ for the early phase of the fit, as we do with the regular Weibull function using the longitude dependence of a sample of fits.

## Appendix B: Examples of SEP light curves

For completeness, in the following figures we show 78 SEP events chosen for reasonably well-behaved light curves likely to be receptive to functional-form fitting (the "good" event list). These are the events shown as filled circles in many earlier plots. This selection is somewhat arbitrary, but is used to help evaluate the results of fitting for well-behaved events; for poorly-behaved events, we cannot expect good results. The events are listed in the following table, providing the onset date and time, the rise time (days) and the longitude of the flare event identified with the SEP event, the peak flux (sfu) and the time of the peak, and the end of the period used for the functional fits (also apparent in the figures).

In the following figures the red curve is the lognormal fit, the blue curve is the Weibull fit, and the green curve is the inverse Weibull fit, when successful. For reference, suitable lognormal fits were achieved in all 78 events, Weibull fits for 77 events and inverse Weibull for 36 events. Note that frequently the inverse Weibull and lognormal fits appear to lie directly on top of each other while the Weibull fit is significantly different, which is a curious feature of the fits.

Table 2. “Good” SEP event list

| Onset time | Rise (days) | Longitude (degrees) | Peak time | 10 MeV peak flux (pfu) | Fitting end | Fitting duration (days) | GOES |
|---|---|---|---|---|---|---|---|
| 1986-02-07 14:40 | 0.13 | 20 | 1986-02-07 17:50 | 78 | 1986-02-08 13:30 | 0.95 | 06 |
| 1986-02-14 10:40 | 0.44 | 76 | 1986-02-14 21:10 | 111 | 1986-02-15 19:45 | 1.38 | 06 |
| 1986-03-06 17:55 | 0.04 | -1 | 1986-03-06 18:50 | 14 | 1986-03-07 00:05 | 0.26 | 06 |
| 1988-01-02 23:15 | 0.34 | 18 | 1988-01-03 07:30 | 30 | 1988-01-05 16:45 | 2.73 | 06 |
| 1988-06-30 10:00 | 0.07 | -22 | 1988-06-30 11:35 | 13 | 1988-06-30 20:45 | 0.45 | 07 |
| 1989-03-23 20:20 | 0.08 | 28 | 1989-03-23 22:20 | 30 | 1989-03-24 05:40 | 0.39 | 07 |
| 1989-07-25 08:55 | 0.14 | 84 | 1989-07-25 12:20 | 32 | 1989-07-26 04:25 | 0.81 | 07 |
| 1989-09-04 00:25 | 0.37 | -16 | 1989-09-04 09:20 | 17 | 1989-09-05 05:50 | 1.23 | 07 |
| 1989-10-22 16:50 | 0.60 | 31 | 1989-10-23 07:10 | 2740 | 1989-10-24 00:35 | 1.32 | 06 |
| 1989-10-24 16:50 | 0.39 | 61 | 1989-10-25 02:05 | 1260 | 1989-10-25 22:45 | 1.25 | 06 |
| 1989-11-15 07:15 | 0.08 | 26 | 1989-11-15 09:05 | 38 | 1989-11-15 19:20 | 0.50 | 07 |
| 1989-12-30 13:40 | 0.99 | 52 | 1989-12-31 13:30 | 3590 | 1989-12-31 20:35 | 1.29 | 07 |
| 1990-04-28 06:50 | 0.44 | 38 | 1990-04-28 17:25 | 67 | 1990-04-29 02:50 | 0.83 | 06 |
| 1990-06-12 08:05 | 0.37 | 33 | 1990-06-12 17:00 | 79 | 1990-06-13 06:30 | 0.93 | 07 |
| 1991-01-31 06:10 | 0.42 | 35 | 1991-01-31 16:20 | 242 | 1991-02-01 01:10 | 0.79 | 07 |
| 1991-06-11 02:45 | 0.48 | 17 | 1991-06-11 14:20 | 3010 | 1991-06-12 00:15 | 0.90 | 07 |
| 1991-06-15 08:45 | 0.47 | 69 | 1991-06-15 19:55 | 1420 | 1991-06-16 13:50 | 1.21 | 07 |
| 1992-02-06 16:40 | 0.77 | 10 | 1992-02-07 11:15 | 78 | 1992-02-08 08:15 | 1.65 | 07 |
| 1992-06-25 20:20 | 0.41 | 67 | 1992-06-26 06:15 | 389 | 1992-06-26 17:05 | 0.86 | 07 |
| 1992-10-30 18:40 | 0.52 | 61 | 1992-10-31 07:10 | 2680 | 1992-11-01 06:40 | 1.50 | 07 |
| 1993-03-04 13:20 | 0.12 | 56 | 1993-03-04 16:15 | 17 | 1993-03-05 05:40 | 0.68 | 07 |
| 1993-03-12 18:40 | 0.30 | 51 | 1993-03-13 01:55 | 44 | 1993-03-13 12:40 | 0.75 | 07 |
| 1994-10-19 22:30 | 0.21 | 24 | 1994-10-20 03:35 | 32 | 1994-10-20 14:40 | 0.67 | 07 |
| 1995-10-20 07:25 | 0.20 | 55 | 1995-10-20 12:10 | 63 | 1995-10-20 19:55 | 0.52 | 08 |
| 1997-11-03 19:20 | 0.67 | 33 | 1997-11-04 11:20 | 72 | 1997-11-04 17:50 | 0.94 | 09 |
| 1997-11-06 12:45 | 0.59 | 63 | 1997-11-07 03:00 | 532 | 1997-11-07 21:35 | 1.37 | 08 |
| 1998-05-02 13:55 | 0.12 | 15 | 1998-05-02 16:55 | 149 | 1998-05-03 03:45 | 0.58 | 08 |
| 1998-05-06 08:20 | 0.06 | 65 | 1998-05-06 09:45 | 239 | 1998-05-06 13:55 | 0.23 | 08 |
| 1998-10-30 14:25 | 0.42 | 80 | 1998-10-31 00:25 | 1160 | 1998-10-31 08:20 | 0.75 | 08 |
| 1999-06-04 08:35 | 0.10 | 69 | 1999-06-04 10:55 | 64 | 1999-06-05 11:25 | 1.12 | 08 |
| 2000-04-04 17:25 | 0.67 | 66 | 2000-04-05 09:30 | 56 | 2000-04-05 22:30 | 1.21 | 08 |
| 2000-06-10 17:40 | 0.13 | 38 | 2000-06-10 20:45 | 46 | 2000-06-11 10:35 | 0.70 | 08 |
| 2000-07-22 12:30 | 0.06 | 56 | 2000-07-22 13:50 | 17 | 2000-07-23 01:50 | 0.56 | 11 |
| 2000-10-16 08:25 | 0.43 | 90 | 2000-10-16 18:40 | 15 | 2000-10-17 19:25 | 1.46 | 08 |
| 2000-10-25 14:15 | 0.56 | 90 | 2000-10-26 03:40 | 15 | 2000-10-26 20:15 | 1.25 | 08 |
| 2000-11-08 23:30 | 0.68 | 77 | 2000-11-09 15:55 | 14800 | 2000-11-10 01:25 | 1.08 | 08 |
| 2000-11-24 15:20 | 0.25 | 7 | 2000-11-24 21:20 | 94 | 2000-11-25 06:40 | 0.64 | 08 |
| 2001-04-15 02:30 | 0.70 | 85 | 2001-04-15 19:15 | 951 | 2001-04-16 08:45 | 1.26 | 08 |
| 2001-04-18 03:10 | 0.32 | 90 | 2001-04-18 10:45 | 321 | 2001-04-19 02:45 | 0.98 | 08 |
| 2001-05-07 14:05 | 0.74 | 100 | 2001-05-08 07:55 | 30 | 2001-05-08 21:45 | 1.32 | 08 |
| 2001-06-15 16:20 | 0.32 | 115 | 2001-06-16 00:05 | 27 | 2001-06-16 23:55 | 1.32 | 08 |
| 2001-09-15 13:00 | 0.08 | -999 | 2001-09-15 14:55 | 12 | 2001-09-15 23:15 | 0.43 | 08 |
| 2001-12-26 05:50 | 0.23 | 54 | 2001-12-26 11:15 | 780 | 2001-12-26 19:05 | 0.55 | 08 |
| 2001-12-30 22:25 | 0.75 | -90 | 2001-12-31 16:20 | 108 | 2001-12-31 23:20 | 1.04 | 08 |
| 2002-01-10 10:35 | 0.79 | -90 | 2002-01-11 05:30 | 92 | 2002-01-12 12:50 | 2.09 | 08 |
| 2002-02-20 06:55 | 0.04 | 72 | 2002-02-20 07:55 | 14 | 2002-02-20 12:10 | 0.22 | 08 |
| 2002-04-21 01:45 | 0.90 | 37 | 2002-04-21 23:20 | 2520 | 2002-04-22 14:30 | 1.53 | 08 |
| 2002-07-07 13:15 | 0.28 | 100 | 2002-07-07 19:55 | 23 | 2002-07-08 10:45 | 0.90 | 08 |

Table 2—Continued

| Onset time | Rise (days) | Longitude (degrees) | Peak time | 10 MeV peak flux (pfu) | Fitting end | Fitting duration (days) | GOES |
|---|---|---|---|---|---|---|---|
| 2002-08-22 03:05 | 0.27 | 62 | 2002-08-22 09:40 | 36 | 2002-08-23 01:25 | 0.93 | 08 |
| 2002-11-09 15:45 | 0.58 | 29 | 2002-11-10 05:40 | 404 | 2002-11-10 10:10 | 0.77 | 08 |
| 2003-05-31 03:15 | 0.15 | 65 | 2003-05-31 06:45 | 27 | 2003-05-31 16:20 | 0.55 | 08 |
| 2003-12-02 13:10 | 0.22 | 95 | 2003-12-02 18:20 | 24 | 2003-12-03 00:15 | 0.46 | 10 |
| 2004-04-11 06:25 | 0.51 | 47 | 2004-04-11 18:45 | 36 | 2004-04-12 06:50 | 1.02 | 11 |
| 2004-09-19 18:20 | 0.28 | 58 | 2004-09-20 01:00 | 57 | 2004-09-20 07:50 | 0.56 | 11 |
| 2004-11-01 06:15 | 0.08 | 120 | 2004-11-01 08:05 | 63 | 2004-11-01 16:05 | 0.41 | 11 |
| 2005-01-17 10:50 | 0.26 | 5 | 2005-01-17 17:10 | 4730 | 2005-01-18 04:55 | 0.75 | 10 |
| 2005-01-20 06:50 | 0.06 | 61 | 2005-01-20 08:10 | 1860 | 2005-01-20 19:55 | 0.55 | 11 |
| 2005-06-16 20:55 | 0.25 | 87 | 2005-06-17 03:00 | 44 | 2005-06-17 17:10 | 0.84 | 11 |
| 2005-07-14 07:25 | 0.85 | 80 | 2005-07-15 03:45 | 134 | 2005-07-16 04:55 | 1.90 | 11 |
| 2005-08-22 19:35 | 0.63 | 60 | 2005-08-23 10:45 | 337 | 2005-08-23 19:30 | 1.00 | 11 |
| 2006-12-06 18:00 | 1.03 | -79 | 2006-12-07 18:40 | 1980 | 2006-12-08 13:50 | 1.83 | 11 |
| 2006-12-13 02:55 | 0.27 | 23 | 2006-12-13 09:25 | 698 | 2006-12-13 22:05 | 0.80 | 11 |
| 2011-03-07 22:50 | 0.38 | 59 | 2011-03-08 08:00 | 50 | 2011-03-09 13:20 | 1.60 | 13 |
| 2011-06-07 07:30 | 0.45 | 64 | 2011-06-07 18:20 | 73 | 2011-06-08 09:40 | 1.09 | 13 |
| 2011-08-09 08:15 | 0.04 | 83 | 2011-08-09 09:15 | 29 | 2011-08-09 20:55 | 0.53 | 15 |
| 2011-11-26 08:25 | 0.70 | 49 | 2011-11-27 01:20 | 68 | 2011-11-27 20:05 | 1.49 | 15 |
| 2012-01-27 19:00 | 0.28 | 71 | 2012-01-28 01:50 | 796 | 2012-01-29 02:35 | 1.32 | 13 |
| 2012-03-13 18:00 | 0.11 | 62 | 2012-03-13 20:45 | 469 | 2012-03-14 07:20 | 0.56 | 13 |
| 2012-05-17 02:00 | 0.11 | 89 | 2012-05-17 04:35 | 311 | 2012-05-17 14:55 | 0.54 | 15 |
| 2012-07-07 00:40 | 0.30 | 50 | 2012-07-07 07:45 | 25 | 2012-07-08 04:55 | 1.18 | 13 |
| 2012-09-28 01:15 | 0.15 | 41 | 2012-09-28 04:45 | 28 | 2012-09-28 14:45 | 0.56 | 13 |
| 2013-04-11 08:50 | 0.33 | -12 | 2013-04-11 16:45 | 114 | 2013-04-12 07:55 | 0.96 | 13 |
| 2013-05-22 14:05 | 0.65 | 70 | 2013-05-23 05:45 | 1655 | 2013-05-23 20:25 | 1.26 | 15 |
| 2013-12-28 19:10 | 0.17 | -7 | 2013-12-28 23:15 | 29 | 2013-12-29 10:50 | 0.65 | 13 |
| 2014-01-06 08:15 | 0.32 | 99 | 2014-01-06 16:00 | 42 | 2014-01-07 06:10 | 0.91 | 13 |
| 2014-01-07 15:40 | 1.50 | 11 | 2014-01-09 03:40 | 1026 | 2014-01-09 15:05 | 1.98 | 13 |
| 2015-06-18 05:00 | 0.41 | 90 | 2015-06-18 14:45 | 17 | 2015-06-19 15:50 | 1.45 | 13 |
| 2015-10-29 03:30 | 0.27 | 99 | 2015-10-29 10:00 | 24 | 2015-10-29 21:50 | 0.76 | 13 |

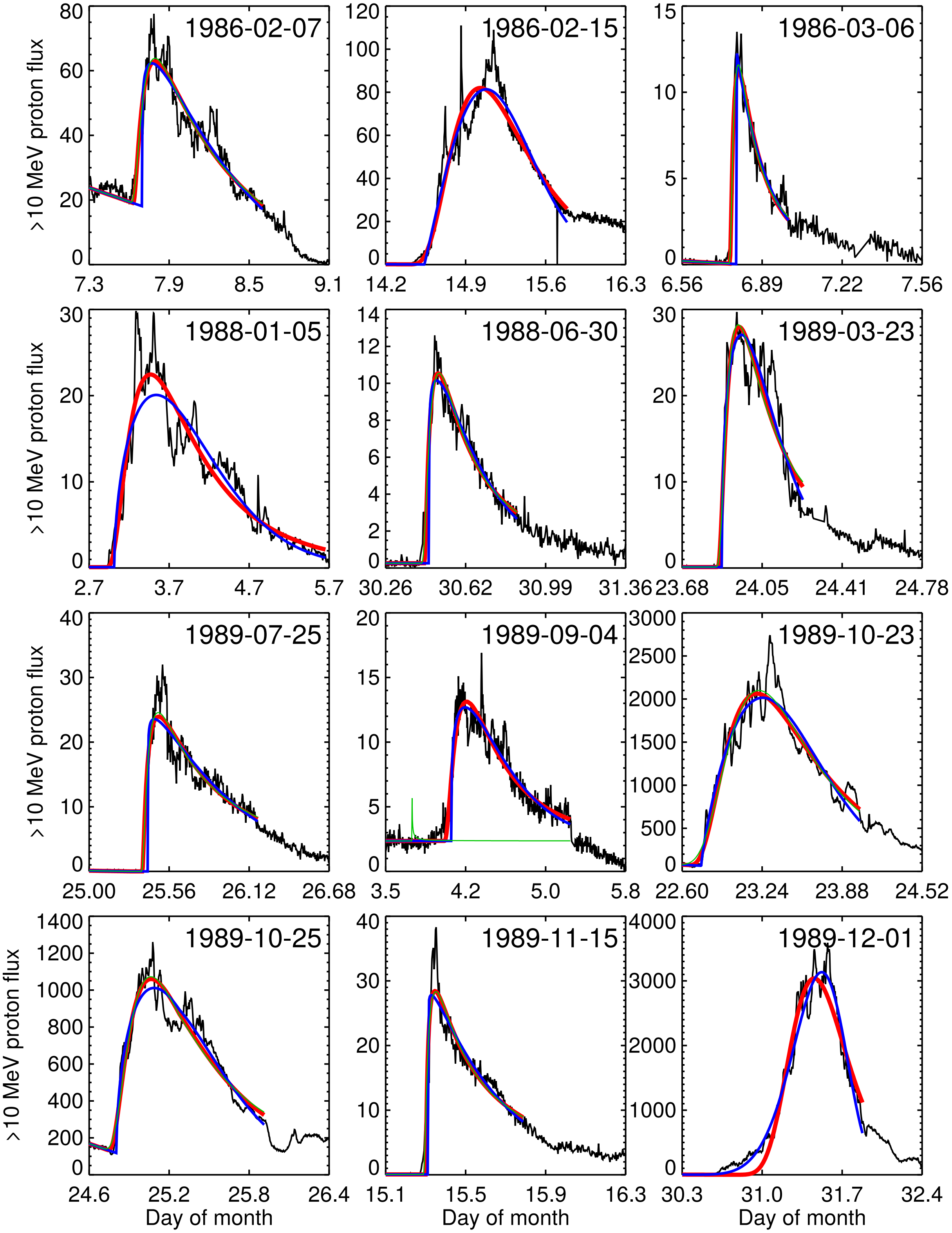
1986-02-07
1986-02-15
1986-03-06
1988-01-05
1988-06-30
1989-03-23
1989-07-25
1989-09-04
1989-10-23
1989-10-25
1989-11-15
1989-12-01
>10 MeV proton flux
Day of month

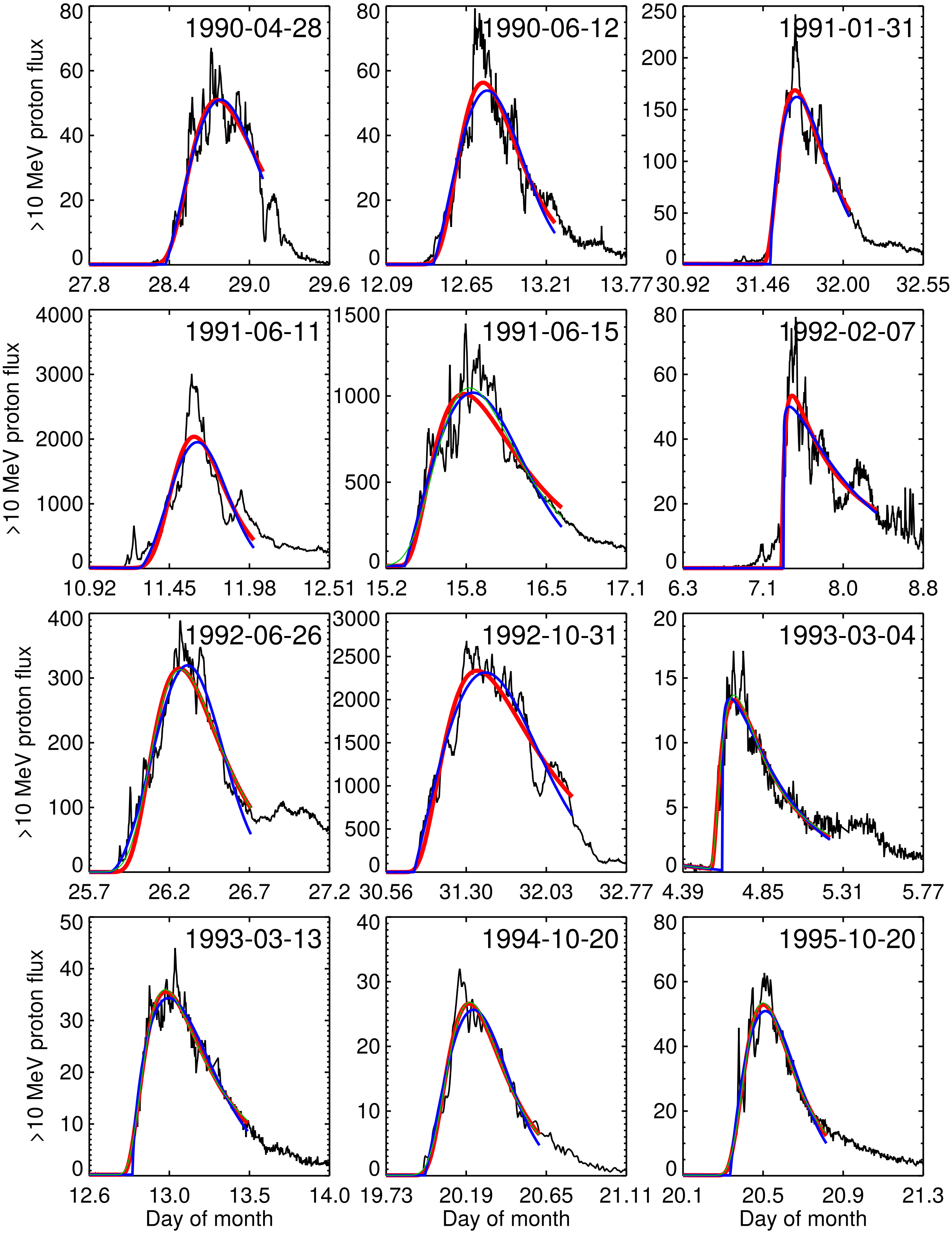
1990-04-28
1990-06-12
1991-01-31
1991-06-11
1991-06-15
1992-02-07
1992-06-26
1992-10-31
1993-03-04
1993-03-13
1994-10-20
1995-10-20
>10 MeV proton flux
Day of month

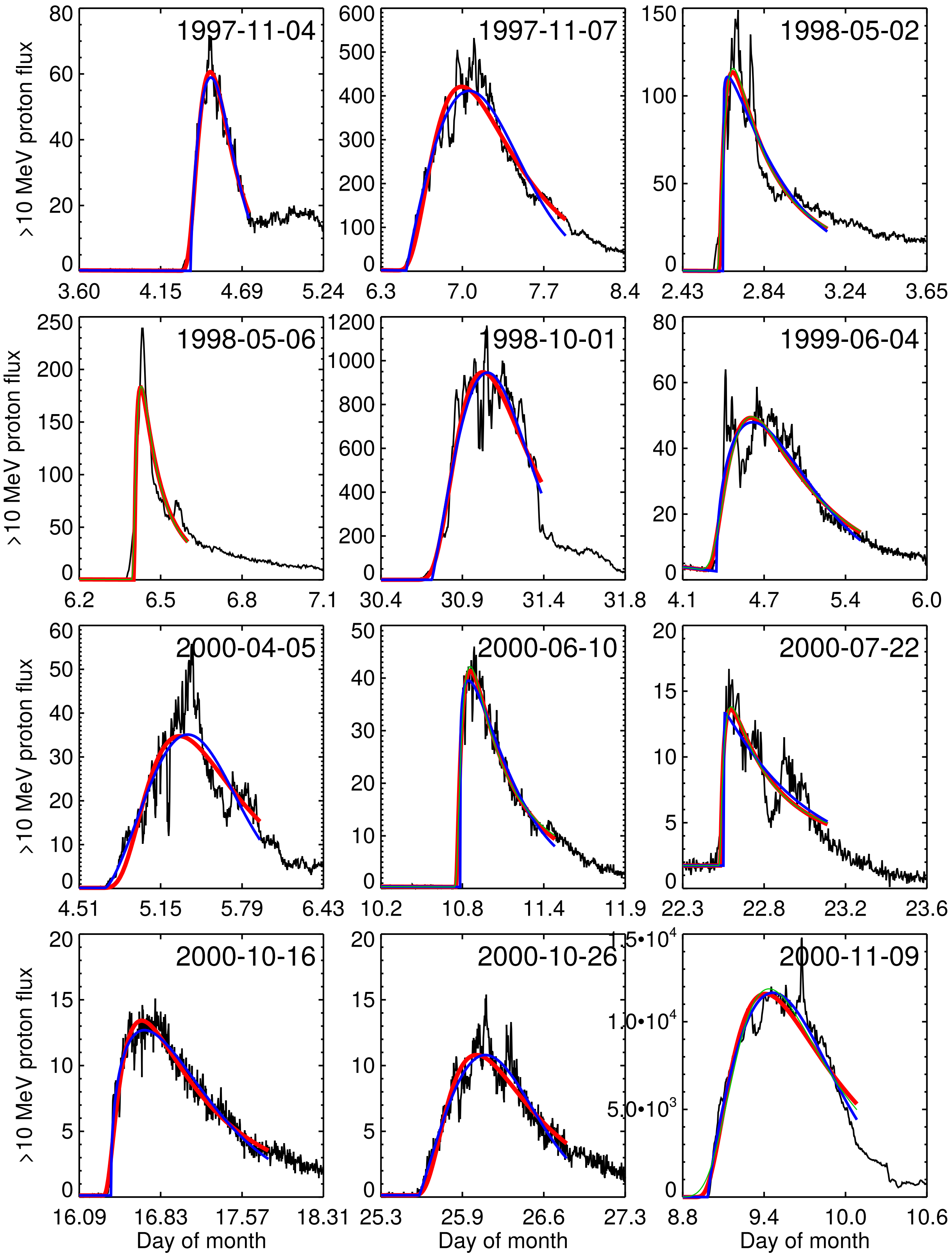
1997-11-04
1997-11-07
1998-05-02
1998-05-06
1998-10-01
1999-06-04
2000-04-05
2000-06-10
2000-07-22
2000-10-16
2000-10-26
2000-11-09
>10 MeV proton flux
Day of month

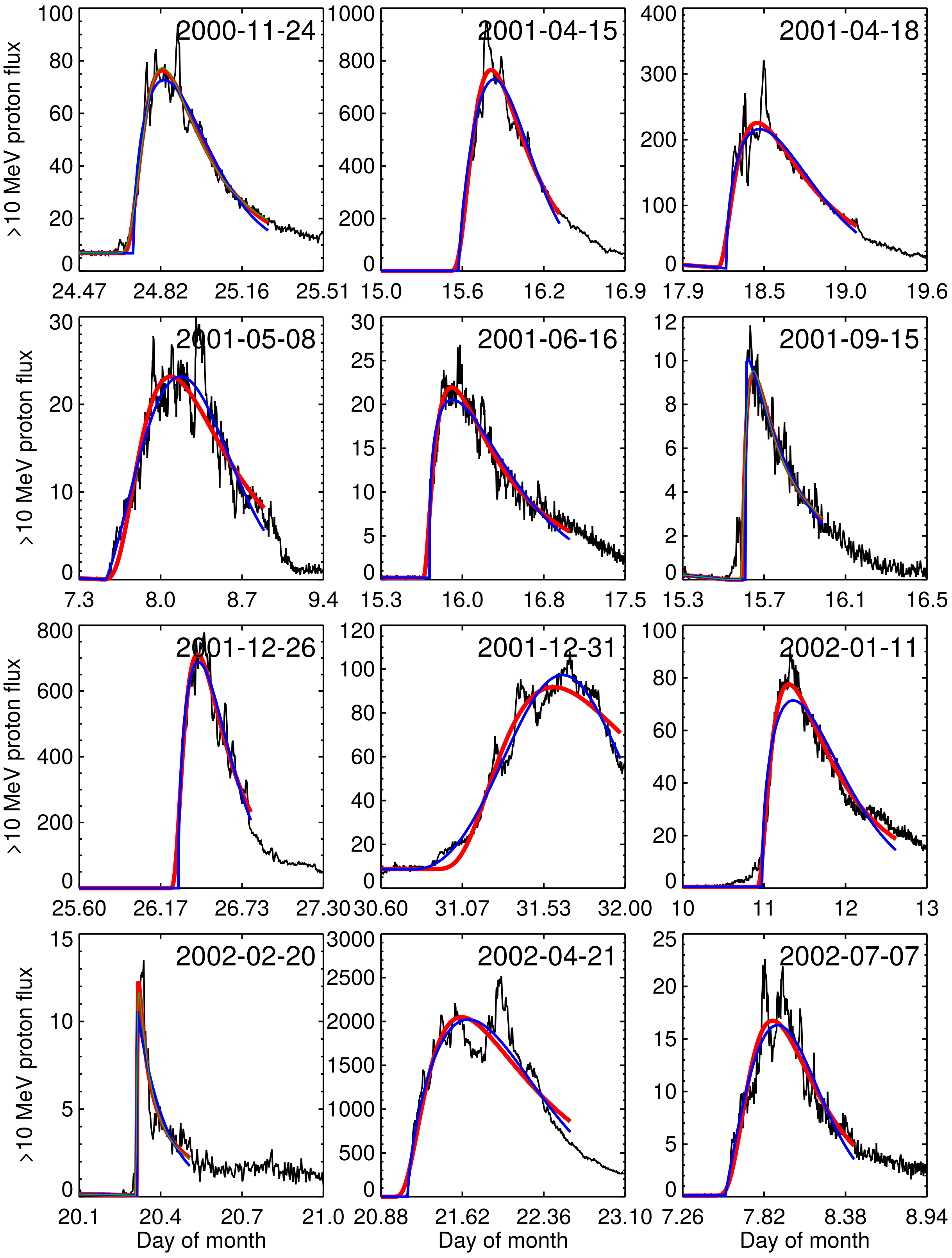
2000-11-24
2001-04-15
2001-04-18
2001-05-08
2001-06-16
2001-09-15
2001-12-26
2001-12-31
2002-01-11
2002-02-20
2002-04-21
2002-07-07
>10 MeV proton flux
Day of month

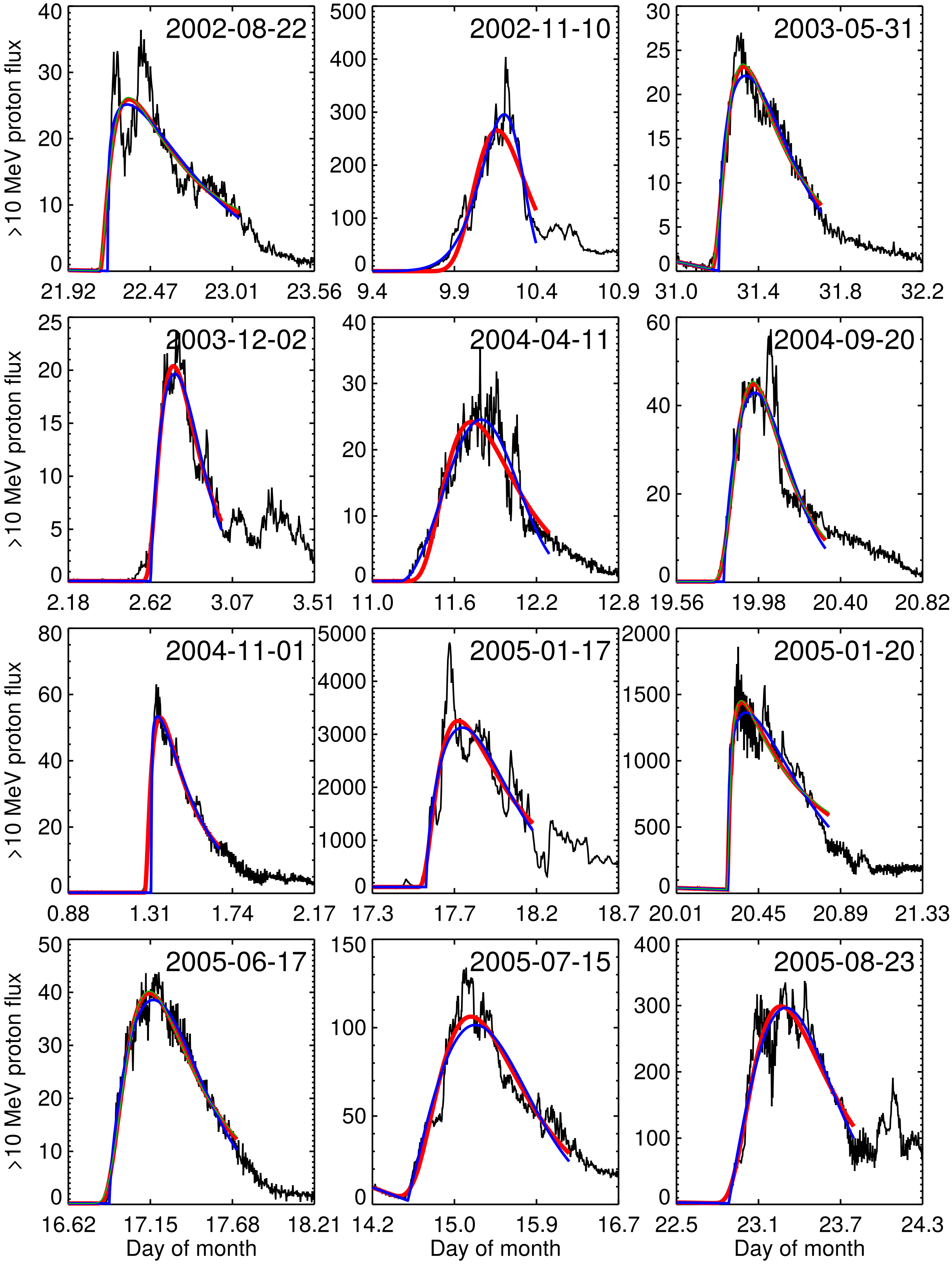
2002-08-22
2002-11-10
2003-05-31
2003-12-02
2004-04-11
2004-09-20
2004-11-01
2005-01-17
2005-01-20
2005-06-17
2005-07-15
2005-08-23
>10 MeV proton flux
Day of month
21.92 22.47 23.01 23.56
9.4 9.9 10.4 10.9
31.0 31.4 31.8 32.2
2.18 2.62 3.07 3.51
11.0 11.6 12.2 12.8
19.56 19.98 20.40 20.82
0.88 1.31 1.74 2.17
17.3 17.7 18.2 18.7
20.01 20.45 20.89 21.33
16.62 17.15 17.68 18.21
14.2 15.0 15.9 16.7
22.5 23.1 23.7 24.3

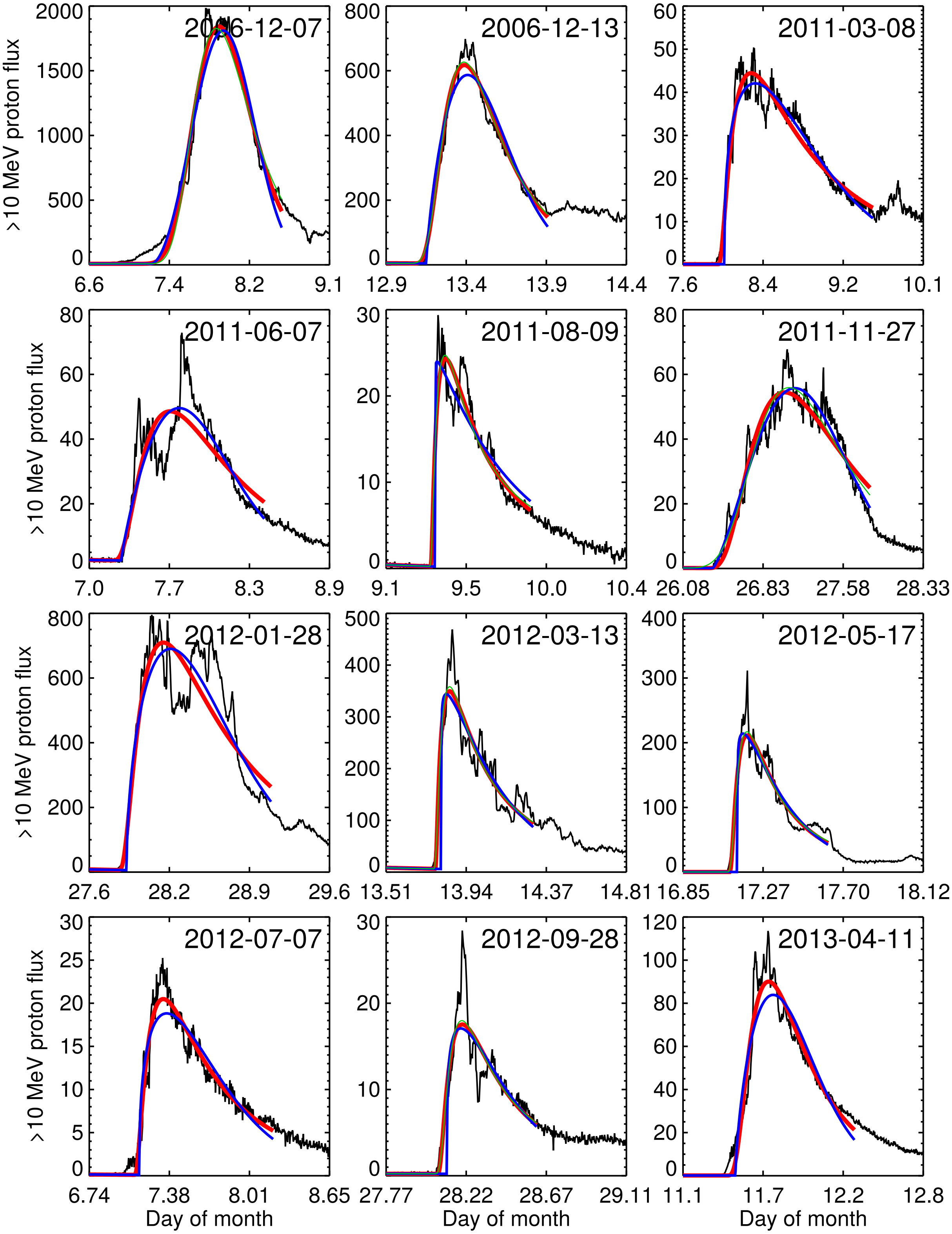
2006-12-07
2006-12-13
2011-03-08
2011-06-07
2011-08-09
2011-11-27
2012-01-28
2012-03-13
2012-05-17
2012-07-07
2012-09-28
2013-04-11
>10 MeV proton flux
Day of month
0 500 1000 1500 2000
0 200 400 600 800
0 10 20 30 40 50 60
0 20 40 60 80
0 10 20 30
0 100 200 300 400 500
0 100 200 300 400
0 5 10 15 20 25 30
0 20 40 60 80 100 120
6.6 7.4 8.2 9.1
12.9 13.4 13.9 14.4
7.6 8.4 9.2 10.1
7.0 7.7 8.3 8.9
9.1 9.5 10.0 10.4
26.08 26.83 27.58 28.33
27.6 28.2 28.9 29.6
13.51 13.94 14.37 14.81
16.85 17.27 17.70 18.12
6.74 7.38 8.01 8.65
27.77 28.22 28.67 29.11
11.1 11.7 12.2 12.8

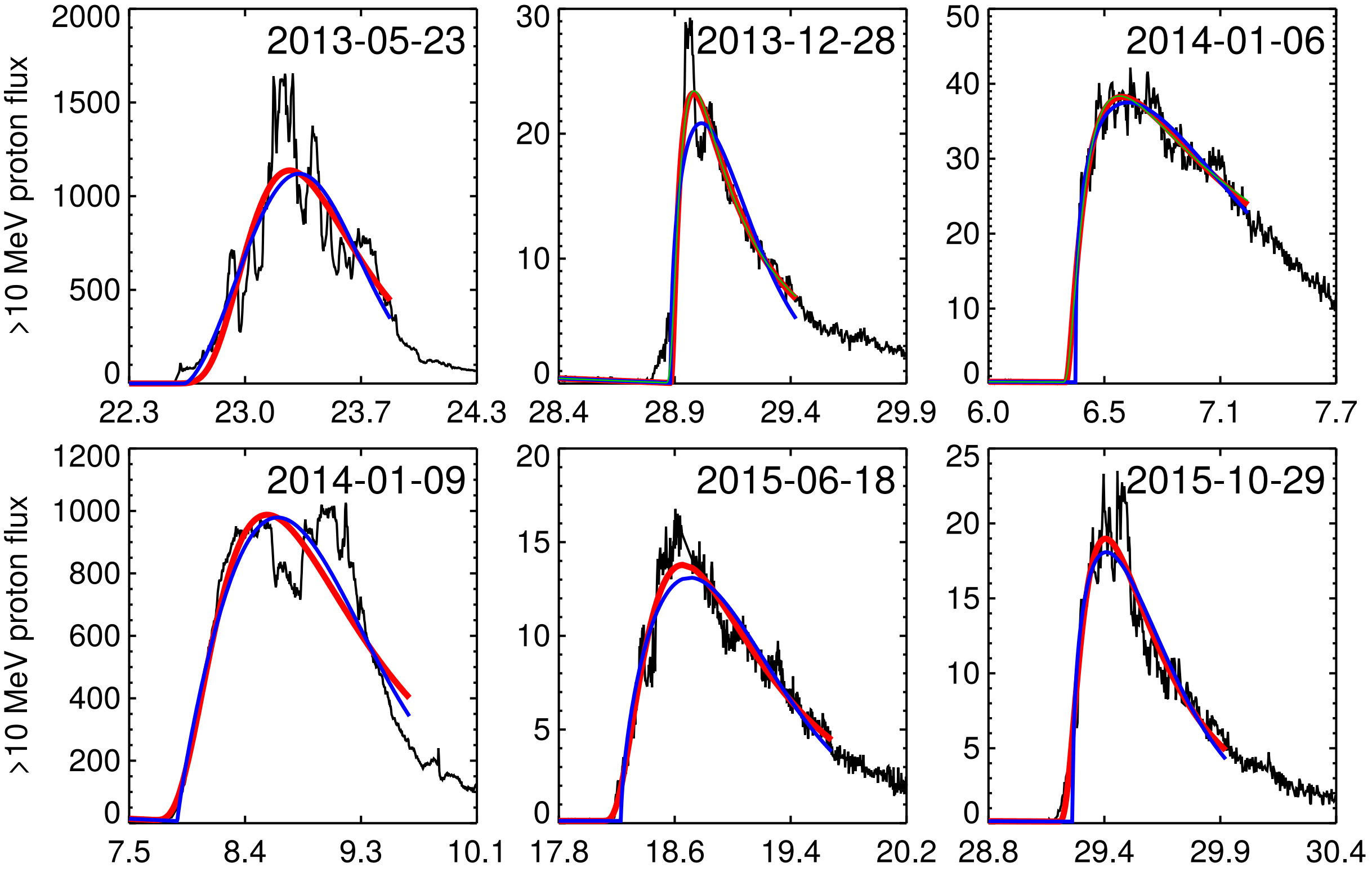

2013-05-23
2013-12-28
2014-01-06
2014-01-09
2015-06-18
2015-10-29
>10 MeV proton flux
>10 MeV proton flux
2000
1500
1000
500
0
22.3
23.0
23.7
24.3
30
20
10
0
28.4
28.9
29.4
29.9
50
40
30
20
10
0
6.0
6.5
7.1
7.7
1200
1000
800
600
400
200
0
7.5
8.4
9.3
10.1
20
15
10
5
0
17.8
18.6
19.4
20.2
25
20
15
10
5
0
28.8
29.4
29.9
30.4

## Appendix C: Additional figures

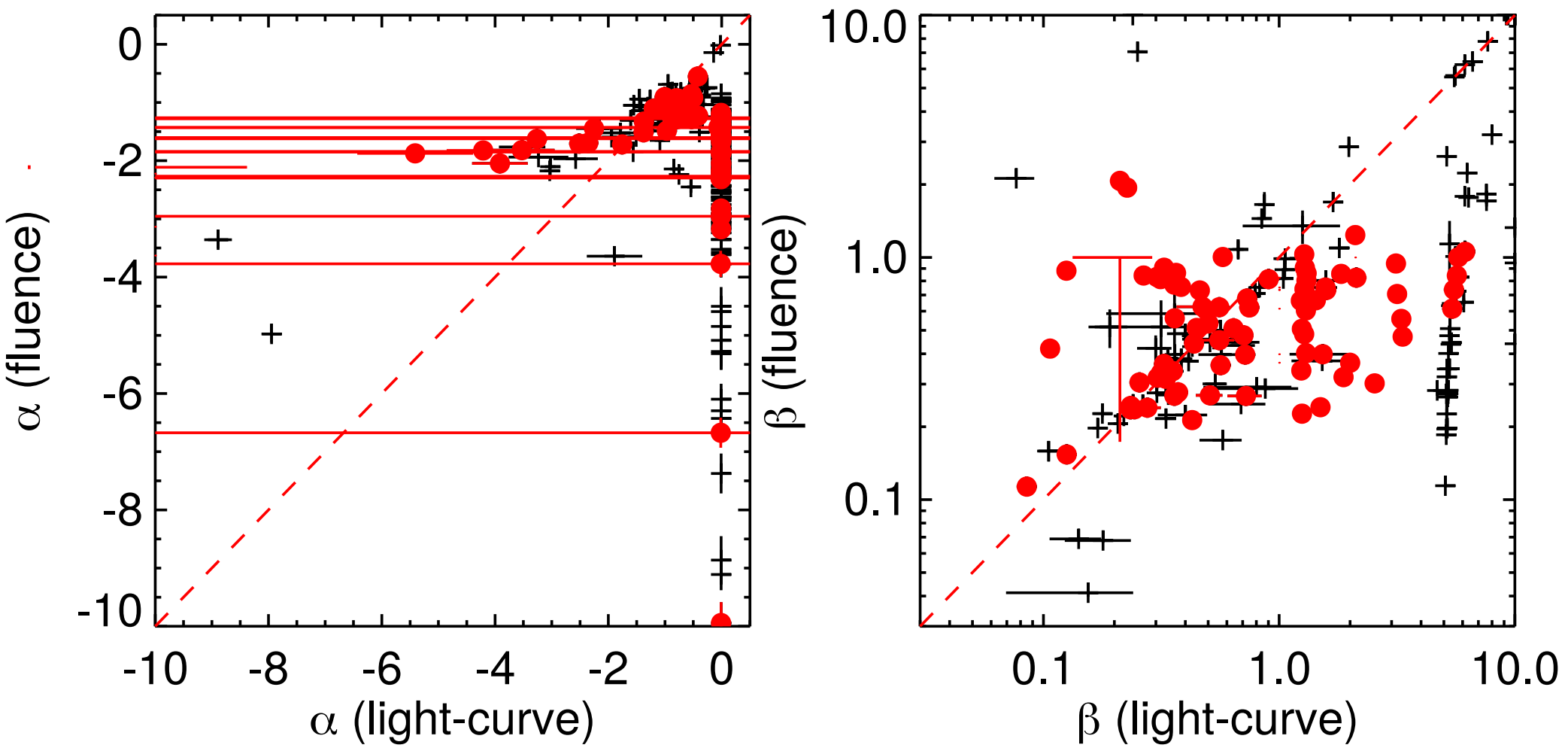

Fig. 39.— A comparison of the fitted shape parameters $\alpha$ (negative by definition for inverse Weibull; left panel) and $\beta$ (right panel) for inverse Weibull fits (with floating $t_0$) to the light-curve (horizontal axis) and cumulative fluence (vertical axis) for common events. Filled red circles show results for the "good" events listed in App. B, while plus symbols represent all other events.

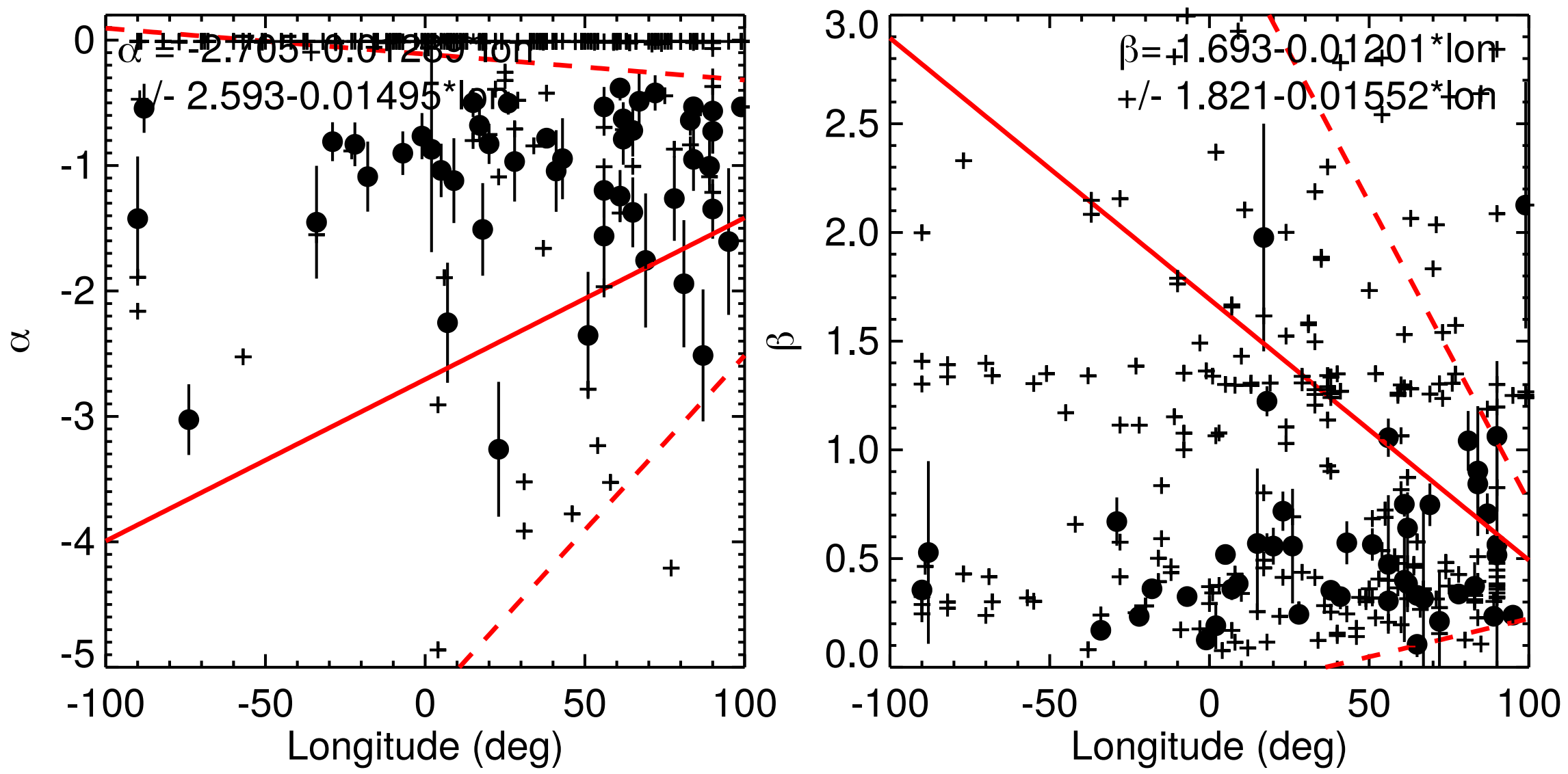

Fig. 40.— The longitudinal dependence of $\alpha$ and $\beta$ from inverse Weibull fits to 55 SEP light curves. The format matches the description of Fig. 20.

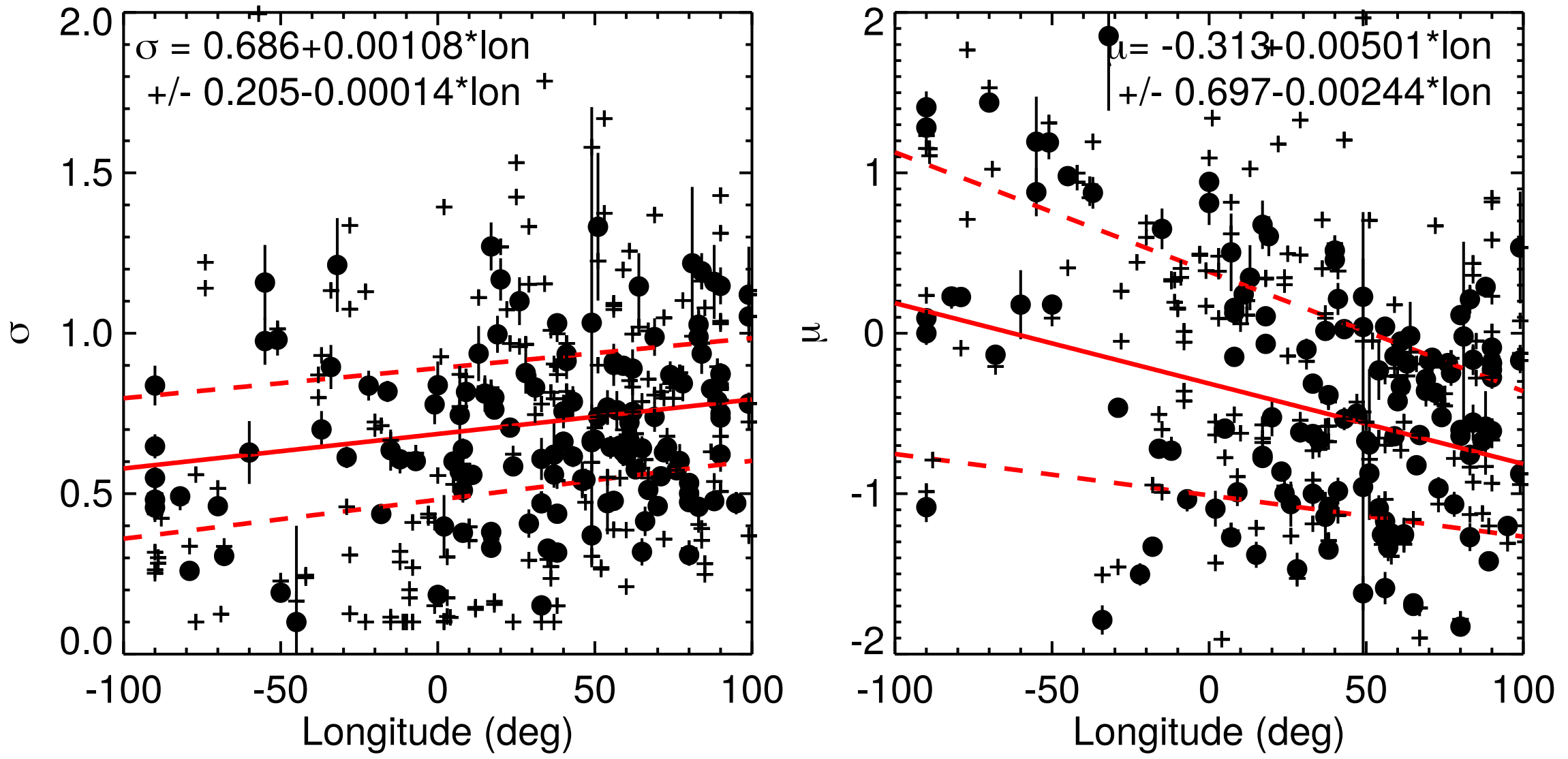

Fig. 41.— The longitudinal dependence of $\sigma$ and $\mu$ from lognormal fits to 139 SEP light curves, chosen to be unique events with good fits to the light curves (filled circles; error bars are plotted but are mostly small). In these fits $t_0$ is fixed at $t_{on}$.

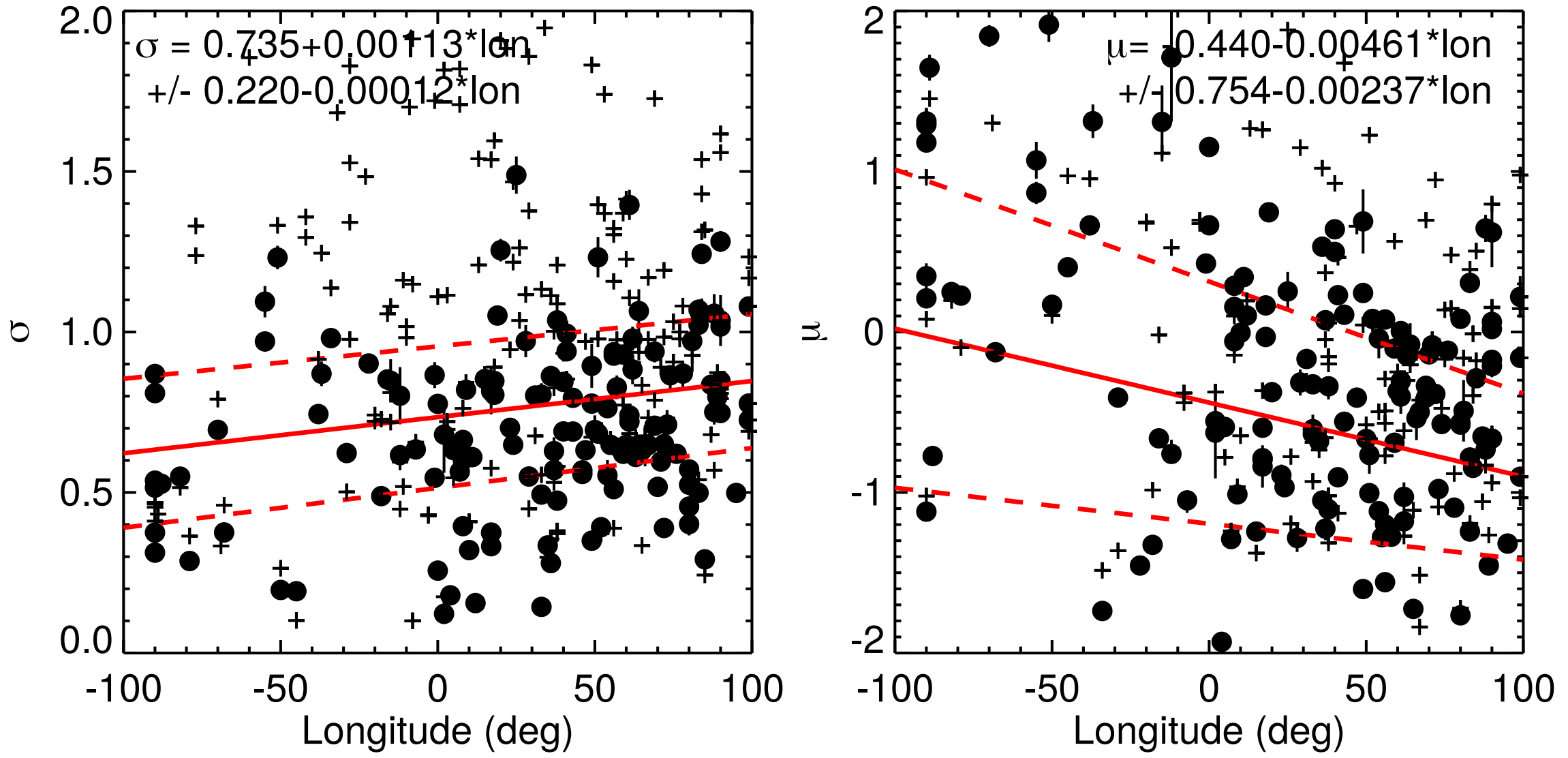

Fig. 42.— The longitudinal dependence of $\sigma$ and $\mu$ from lognormal fits to 141 SEP cumulative fluences, chosen to be unique events with good fits to the light curves (filled circles). In these fits $t_0$ is fixed at $t_{on}$. Small crosses represent fits to the remaining datasets. Other features are described in the caption to Fig. 41.

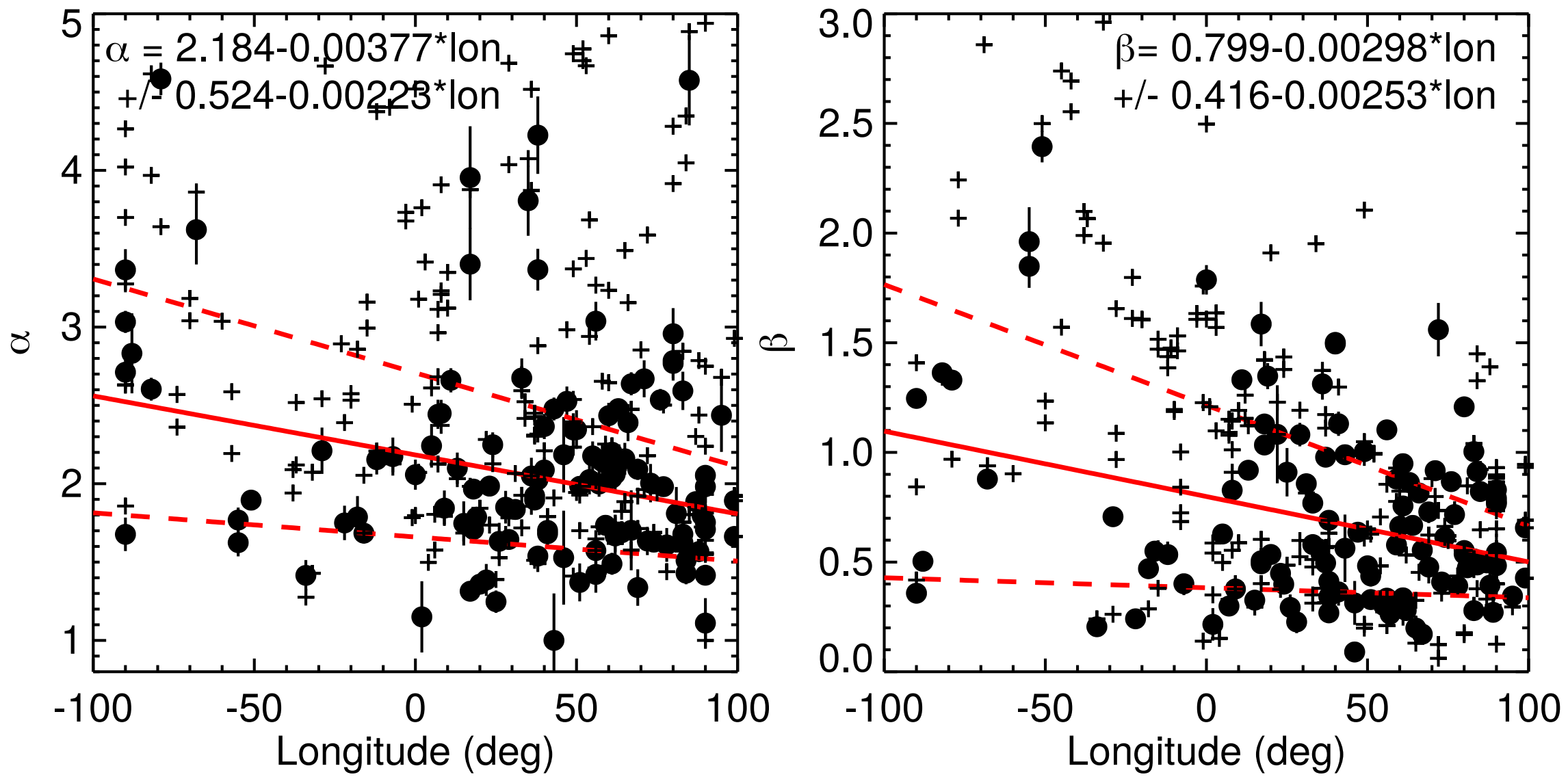

Fig. 43.— The longitudinal dependence of $\alpha$ and $\beta$ from Weibull fits to 118 SEP light curves. The format matches the description of Fig. 41.

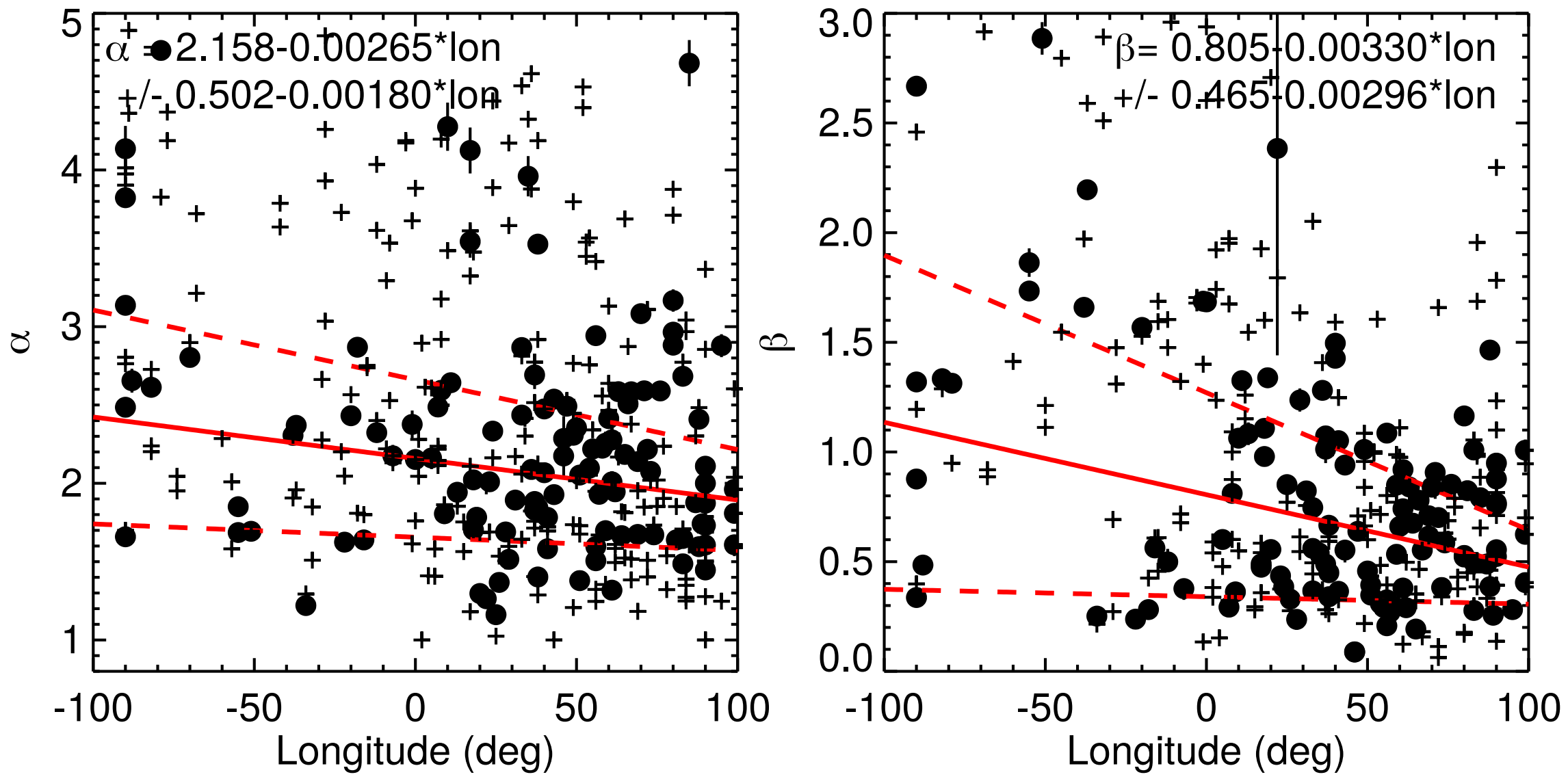

Fig. 44.— The longitudinal dependence of $\alpha$ and $\beta$ from Weibull fits to 116 SEP fluences. The format matches the description of Fig. 41.

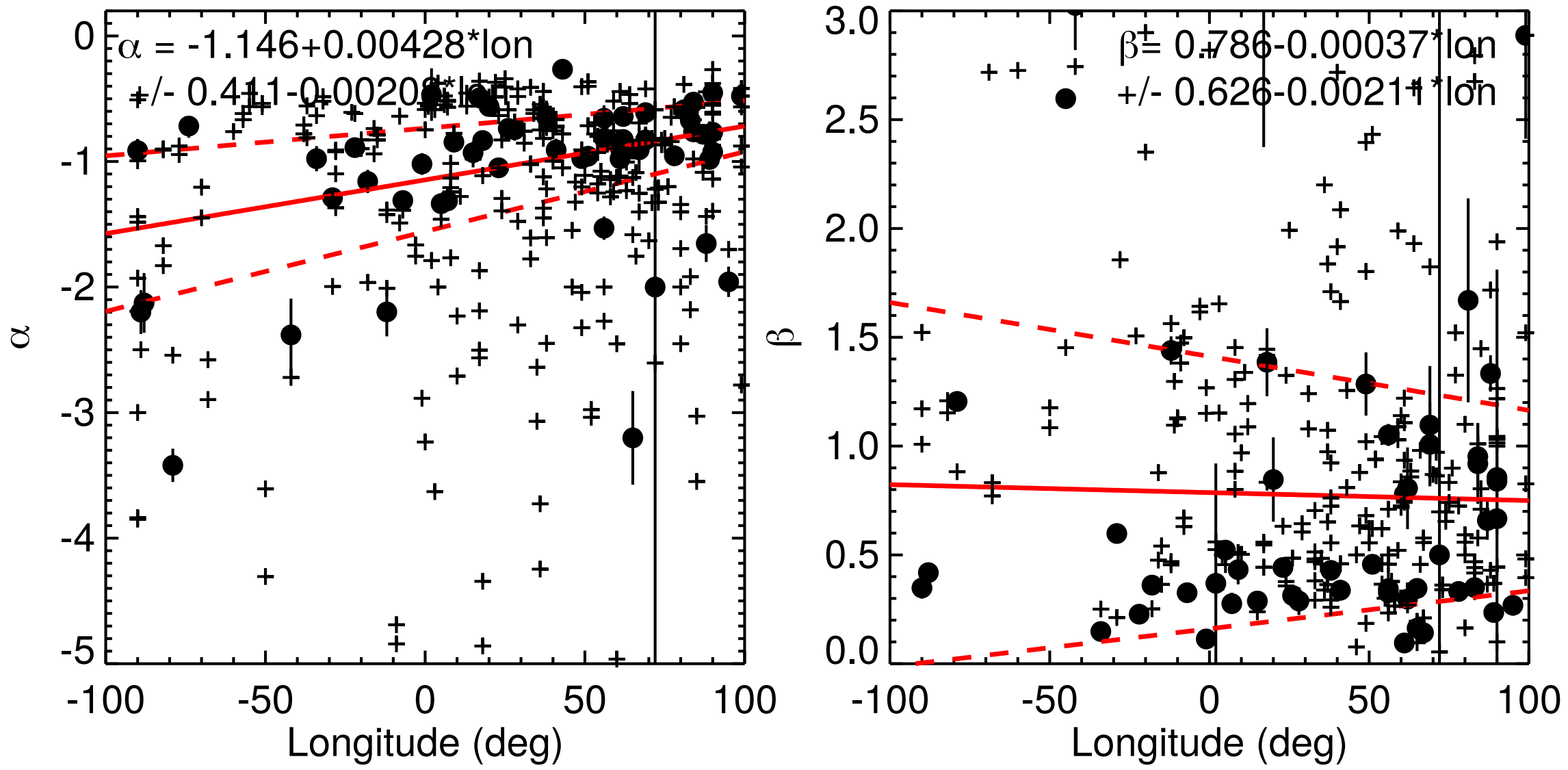

Fig. 45.— The longitudinal dependence of $\alpha$ and $\beta$ from inverse Weibull fits to 55 SEP light curves. The format matches the description of Fig. 41.

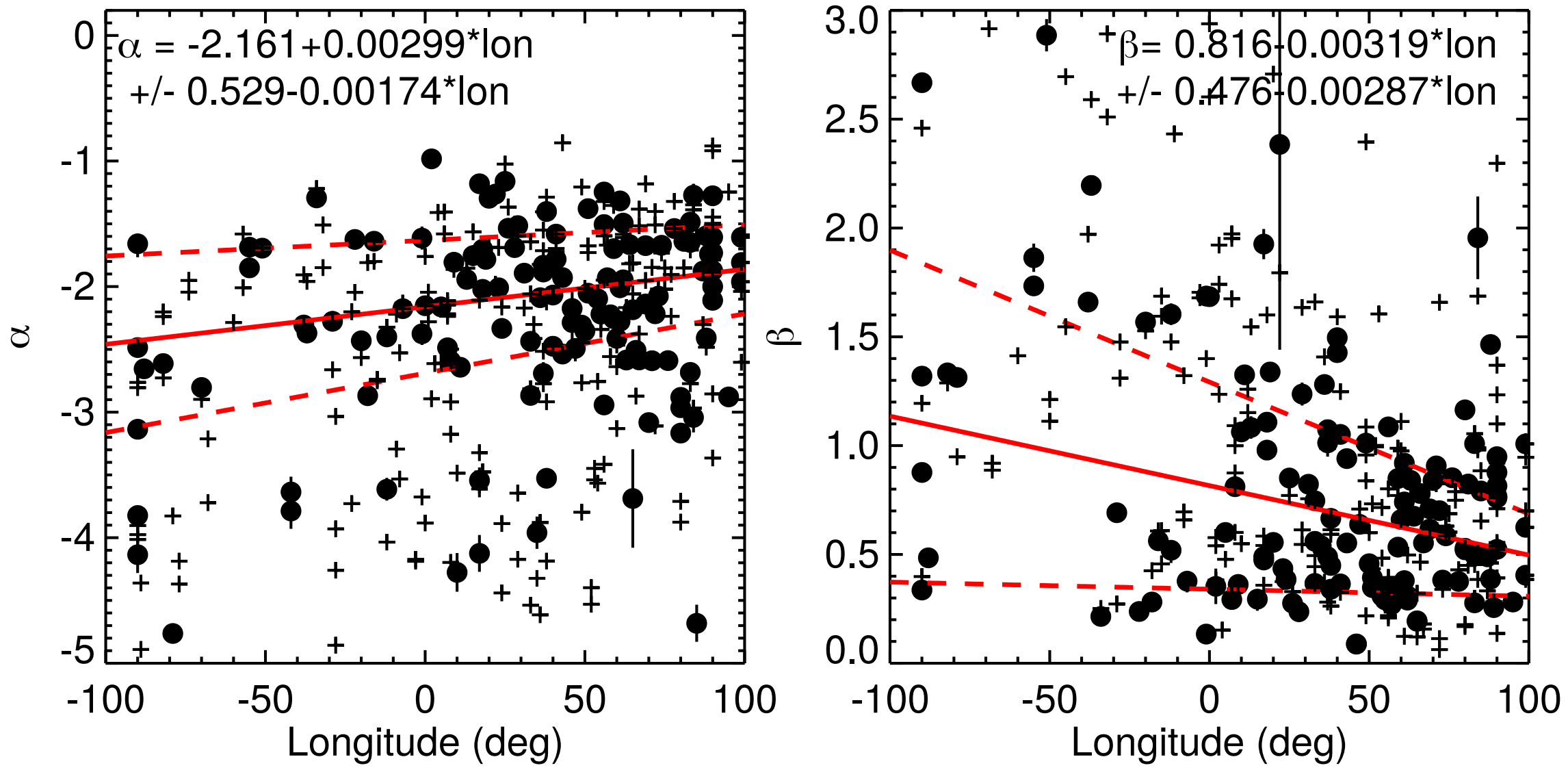

Fig. 46.— The longitudinal dependence of $\alpha$ and $\beta$ from Weibull fits to 126 SEP fluences. The format matches the description of Fig. 41.

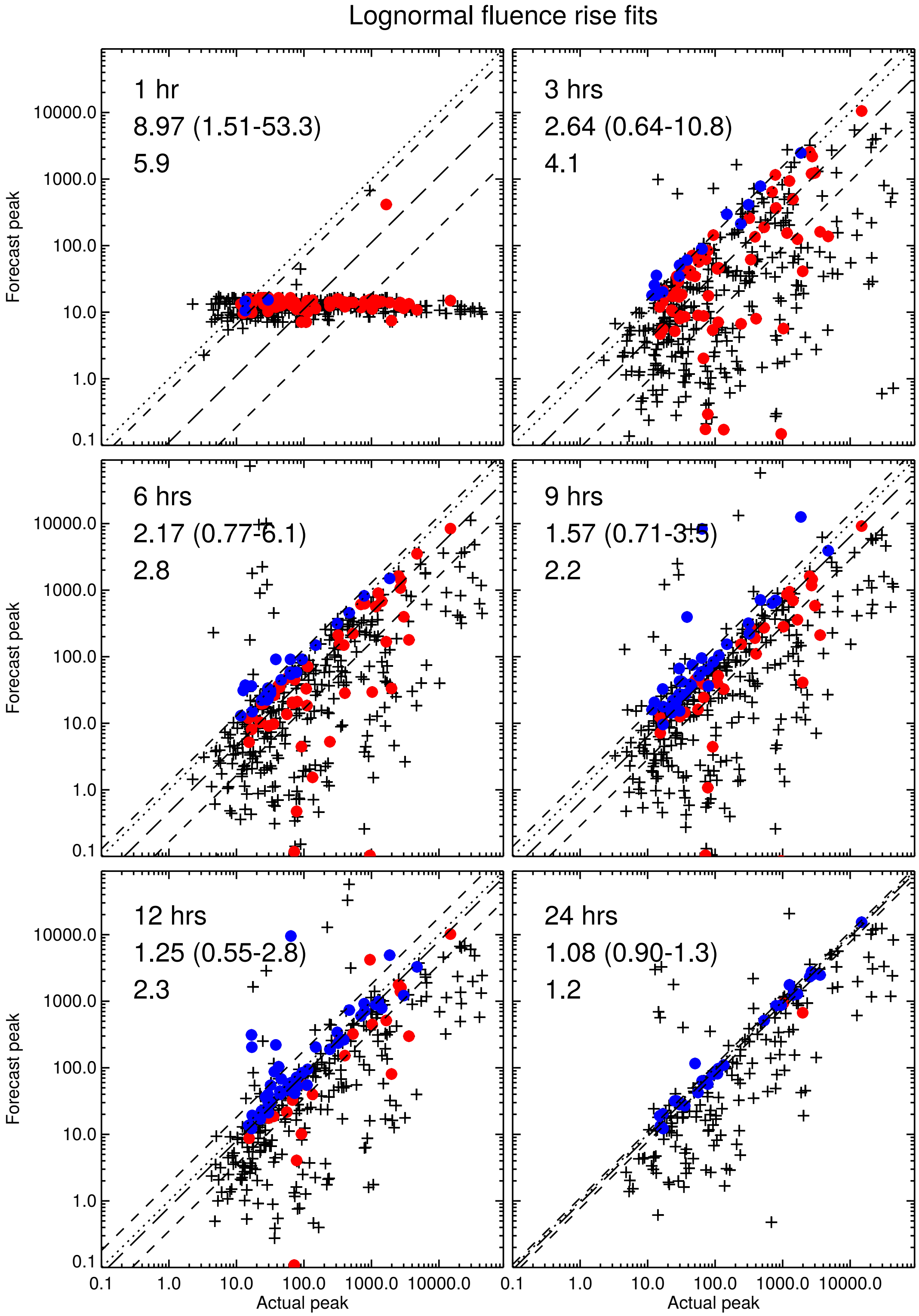

Fig. 47.— Comparison of the forecast maximum flux versus the actual peak flux for lognormal fits to cumulative fluence with fixed $\sigma$ and $\mu$ and floating $t_0$.

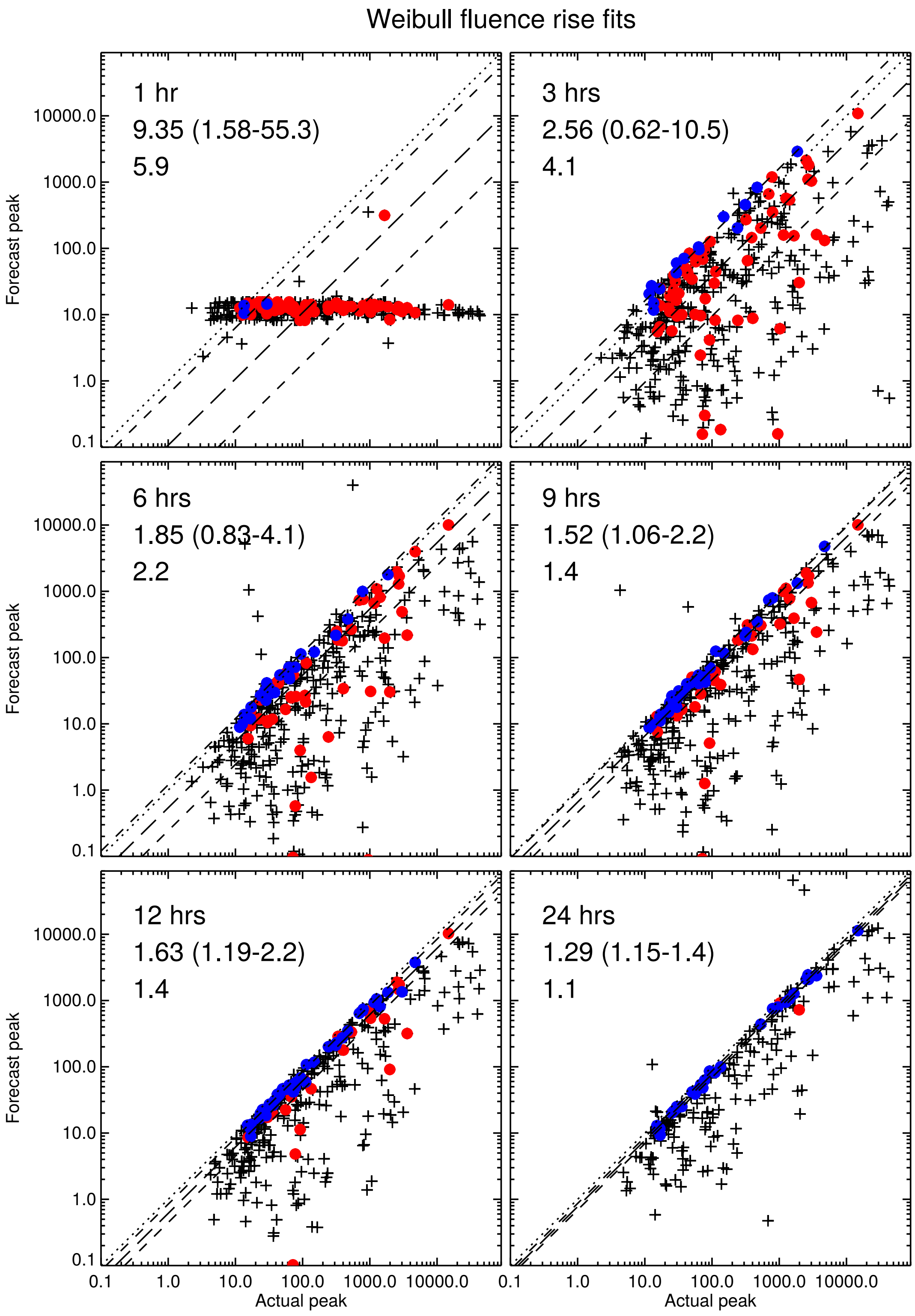

Fig. 48.— Comparison of the forecast maximum flux versus the actual peak flux for Weibull fits to cumulative fluence with fixed $\alpha$ and $\beta$ and floating $t_0$.

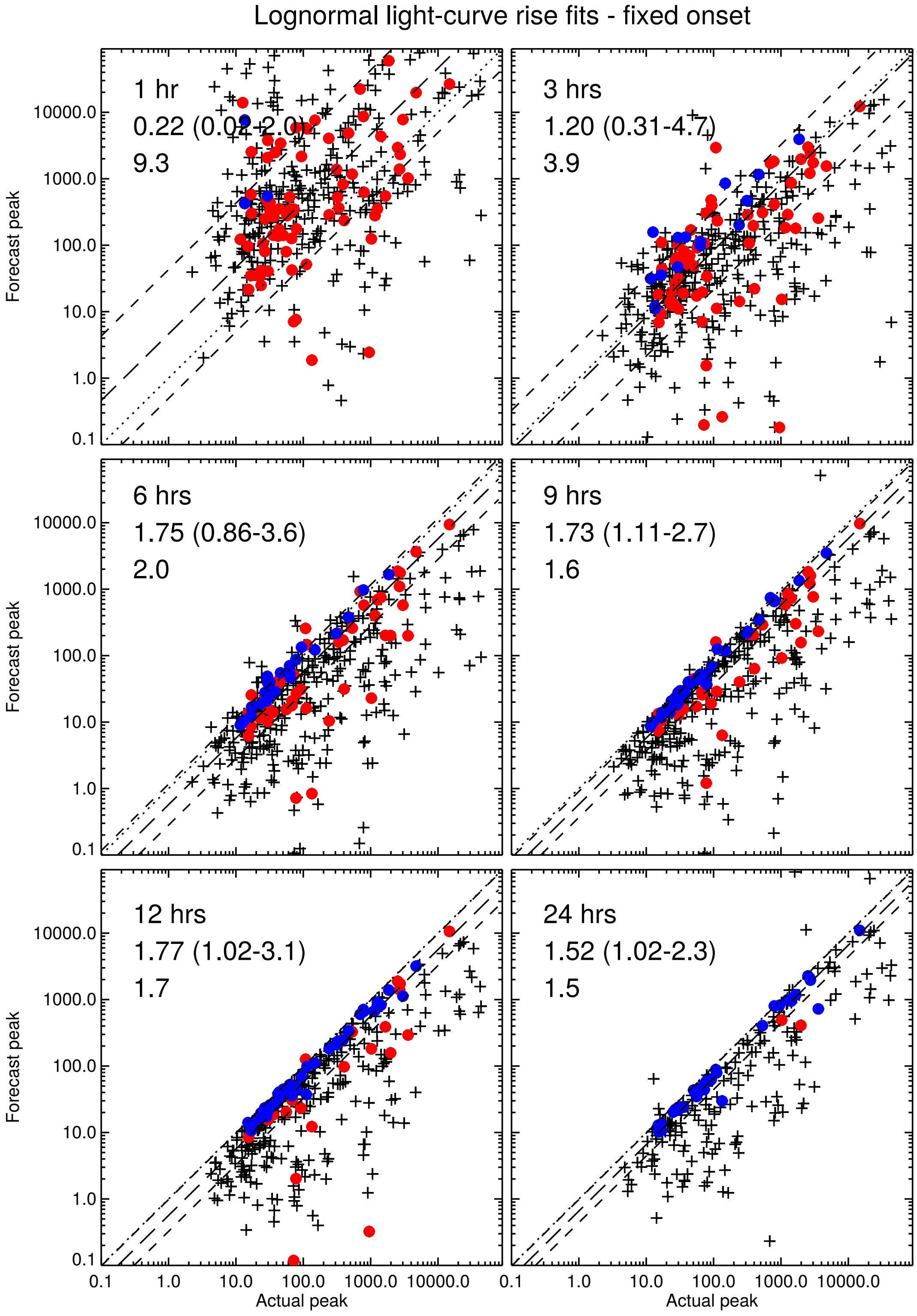

Fig. 49.— Comparison of the forecast maximum flux versus the actual peak flux for lognormal fits to light curves with fixed $\sigma$, $\mu$ and $t_0$.

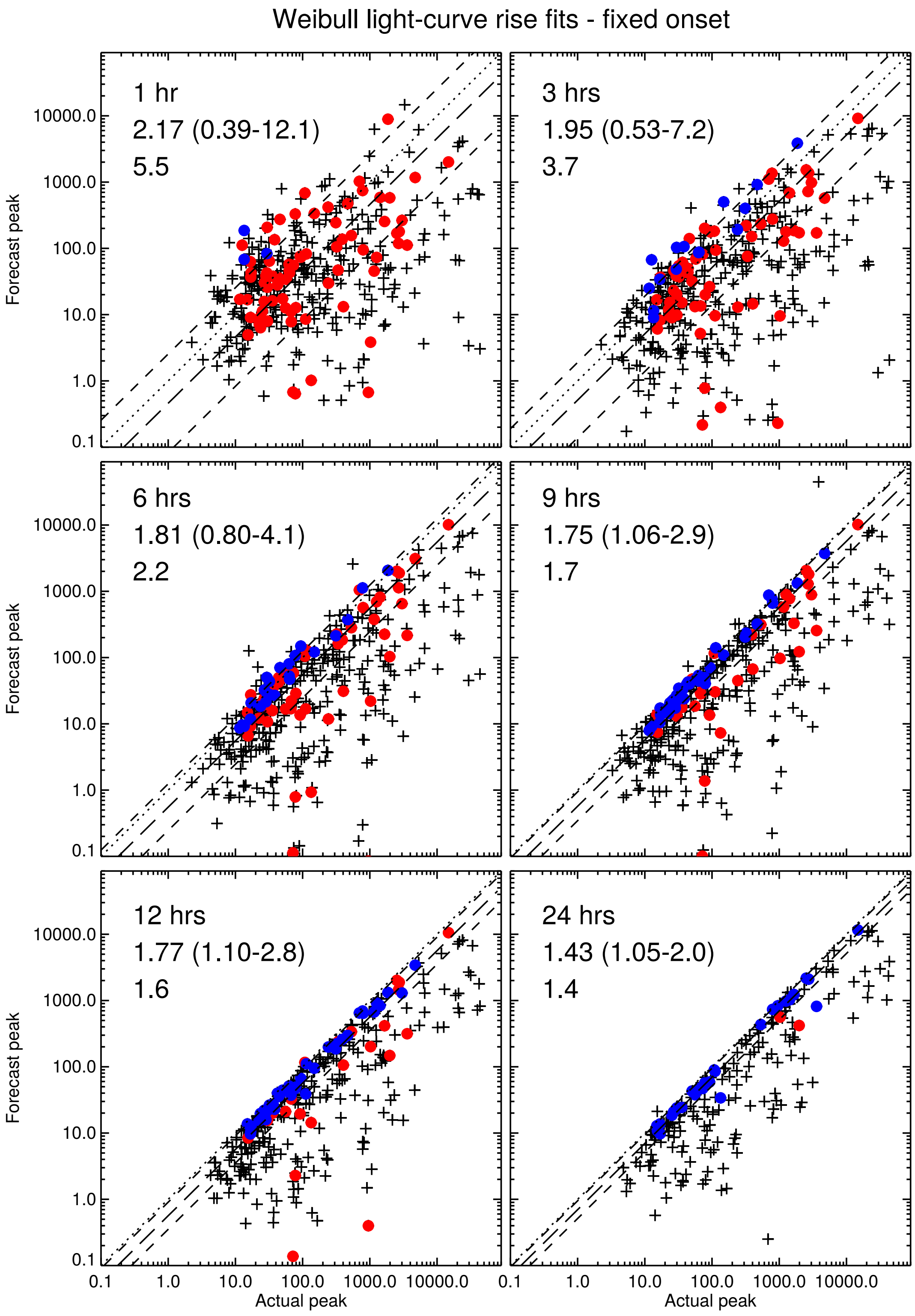

Fig. 50.— Comparison of the forecast maximum flux versus the actual peak flux for Weibull fits to light curves with fixed $\alpha$, $\beta$ and $t_0$.

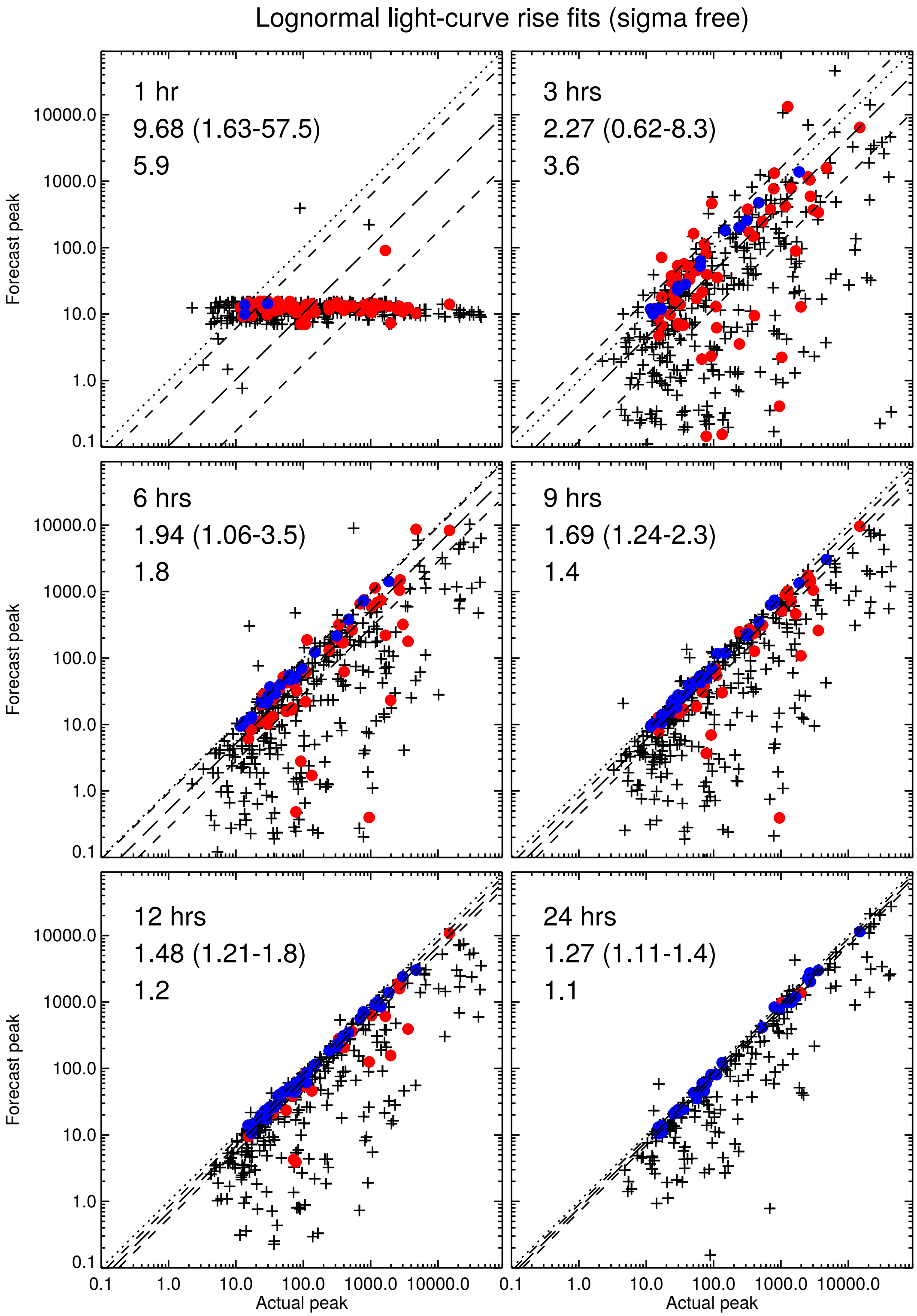

Fig. 51.— Comparison of the forecast maximum flux (vertical axis) versus the actual peak flux (horizontal axis) at 6 different times for lognormal fits to light curves with fixed $\mu$ and floating $\sigma$ and $t_0$.

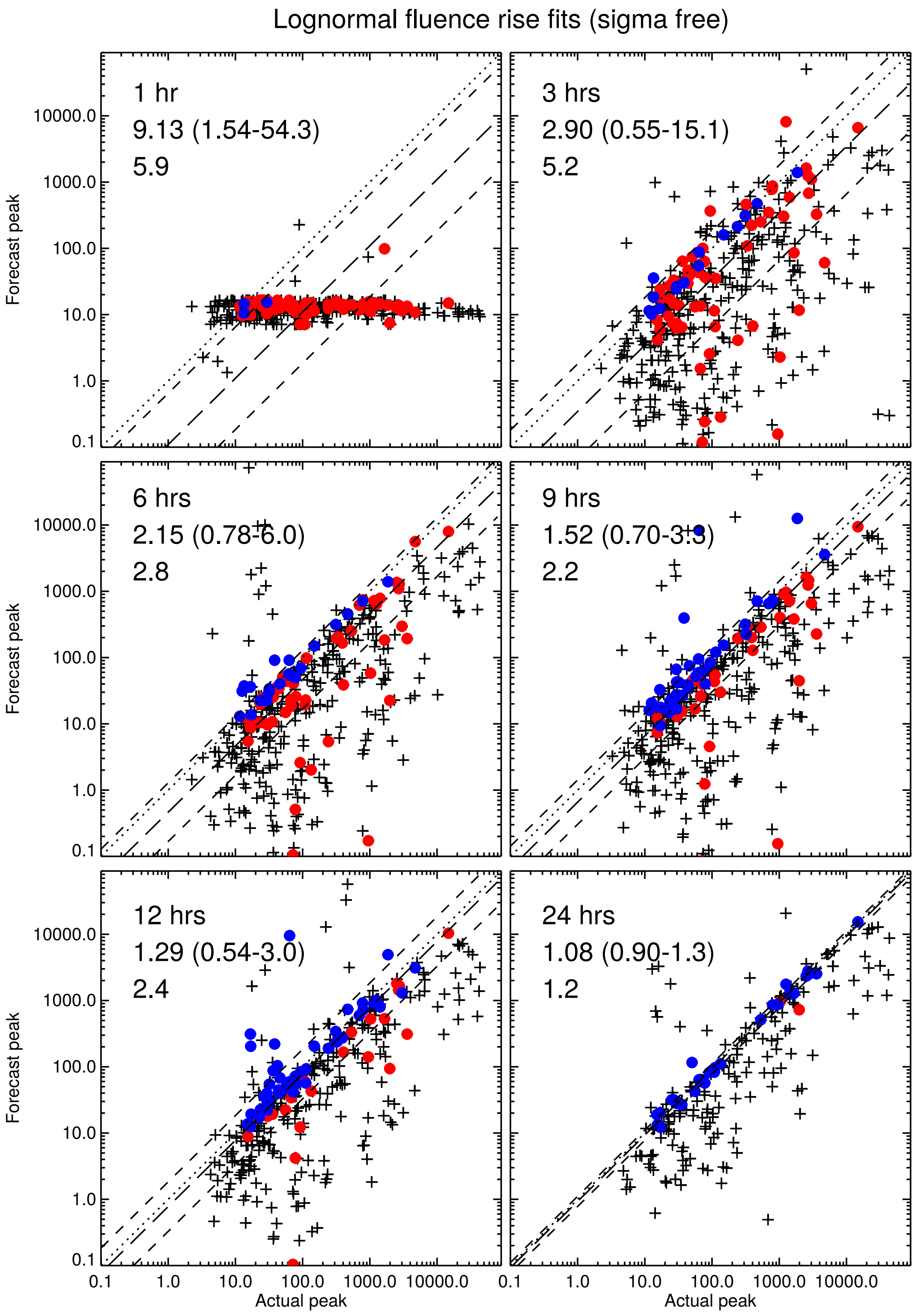

Fig. 52.— Comparison of the forecast maximum flux versus the actual peak flux for lognormal fits to cumulative fluence with fixed $\mu$ and floating $\sigma$ and $t_0$.

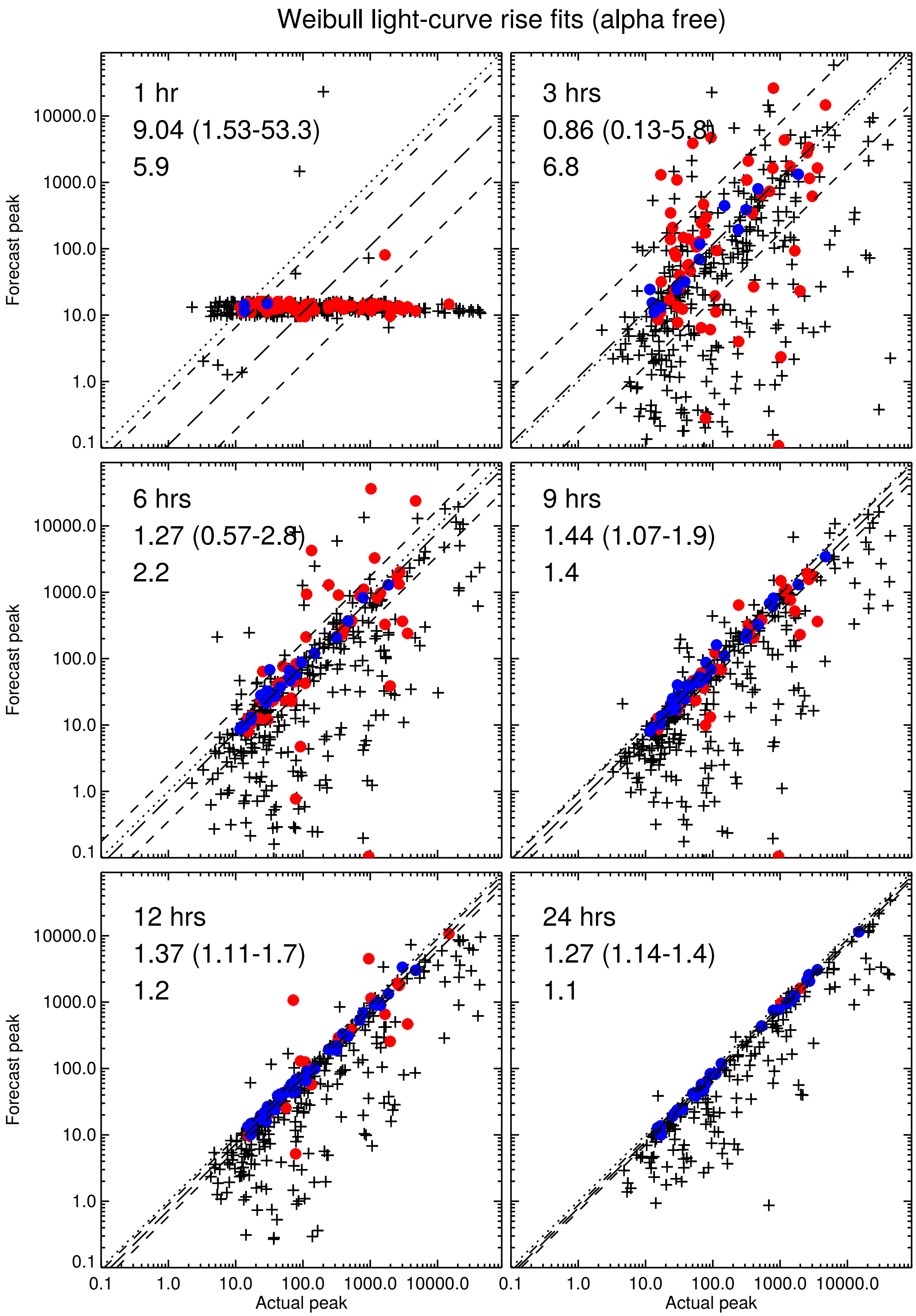

Fig. 53.— Comparison of the forecast maximum flux versus the actual peak flux for Weibull fits to light curves with fixed $\beta$ and floating $\alpha$ and $t_0$.

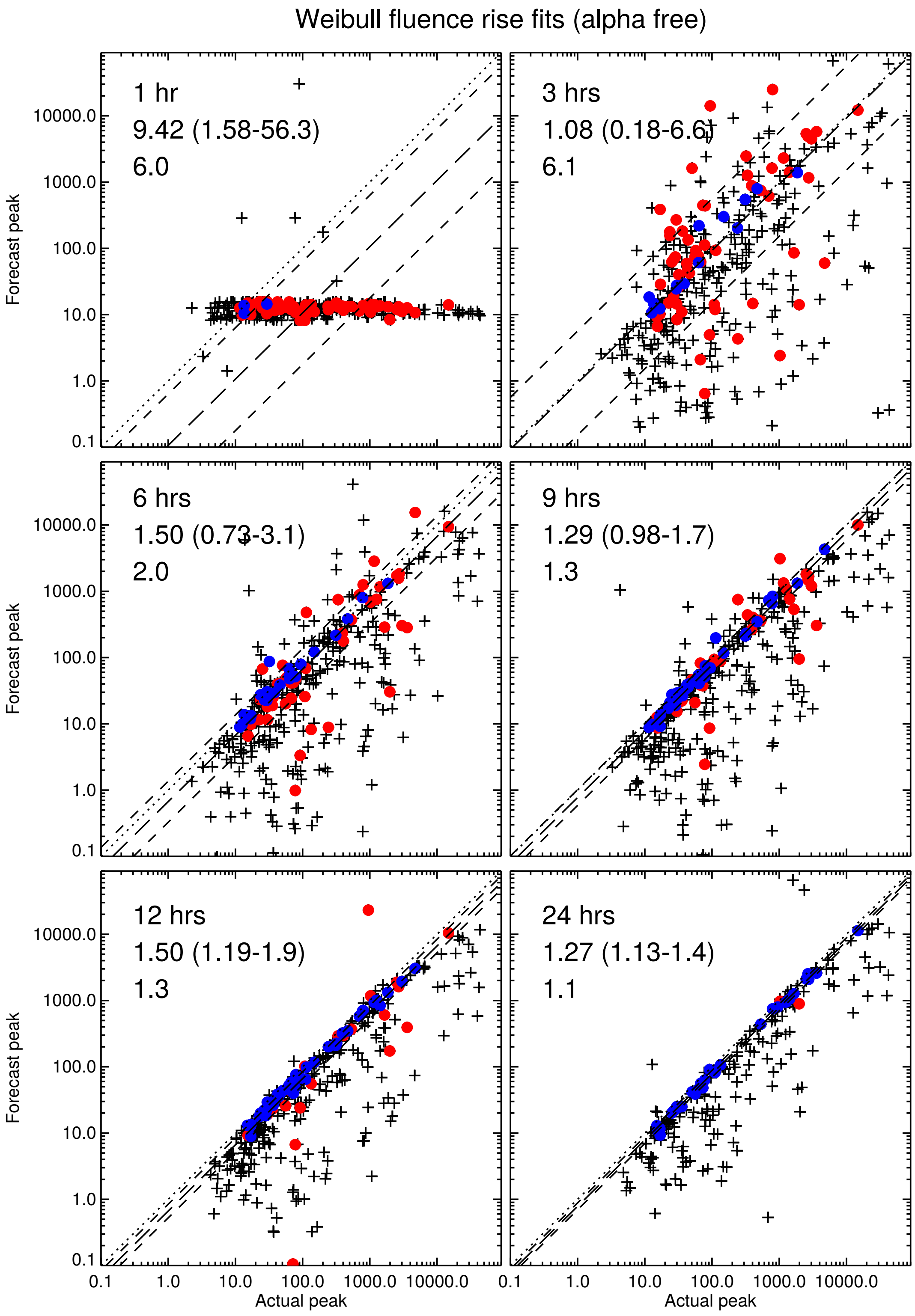

Fig. 54.— Comparison of the forecast maximum flux versus the actual peak flux for Weibull fits to cumulative fluence with fixed $\beta$ and floating $\alpha$ and $t_0$.

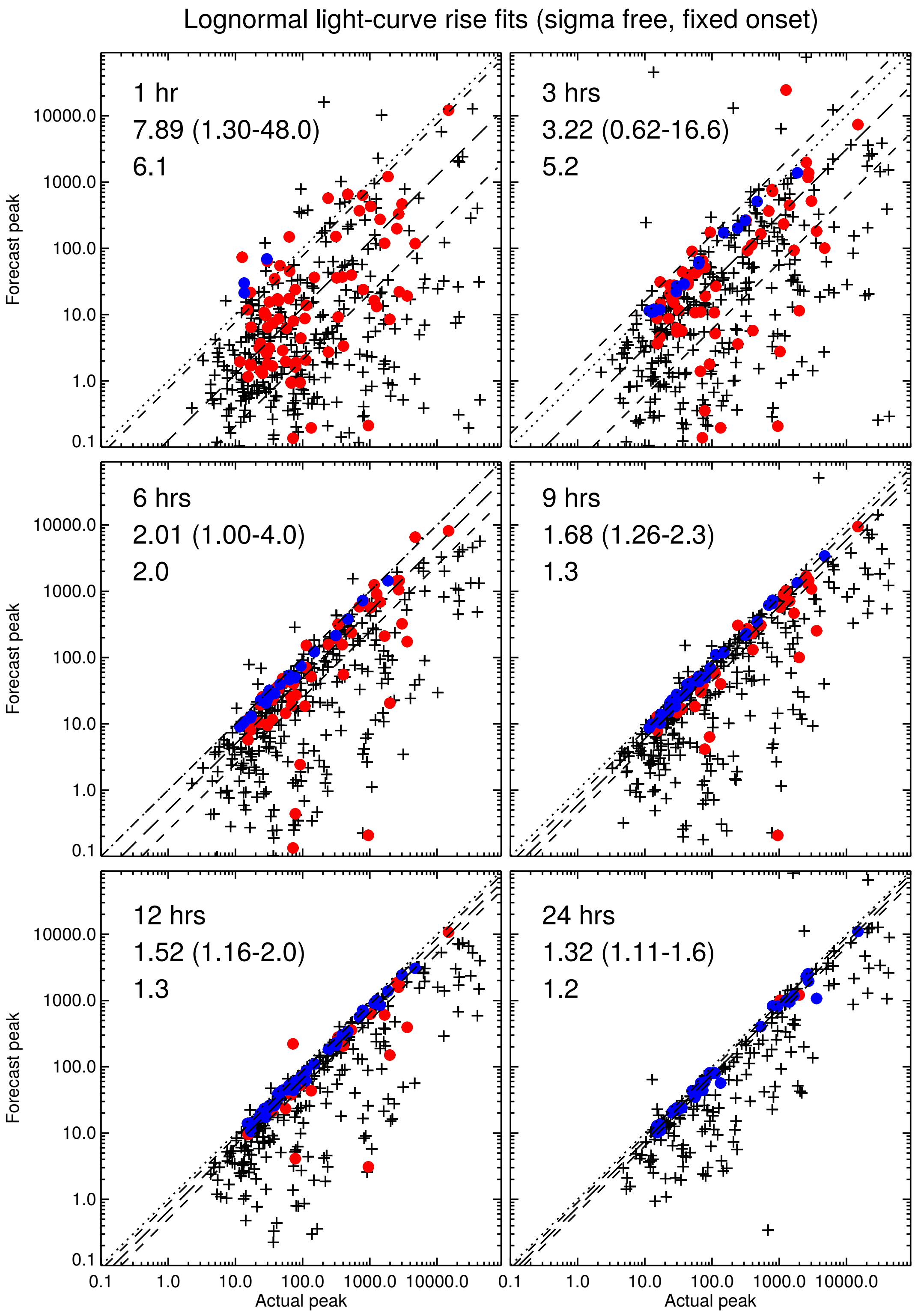

Fig. 55.— Comparison of the forecast maximum flux versus the actual peak flux for lognormal fits to light curves with fixed $\mu$ and $t_0$ and floating $\sigma$.

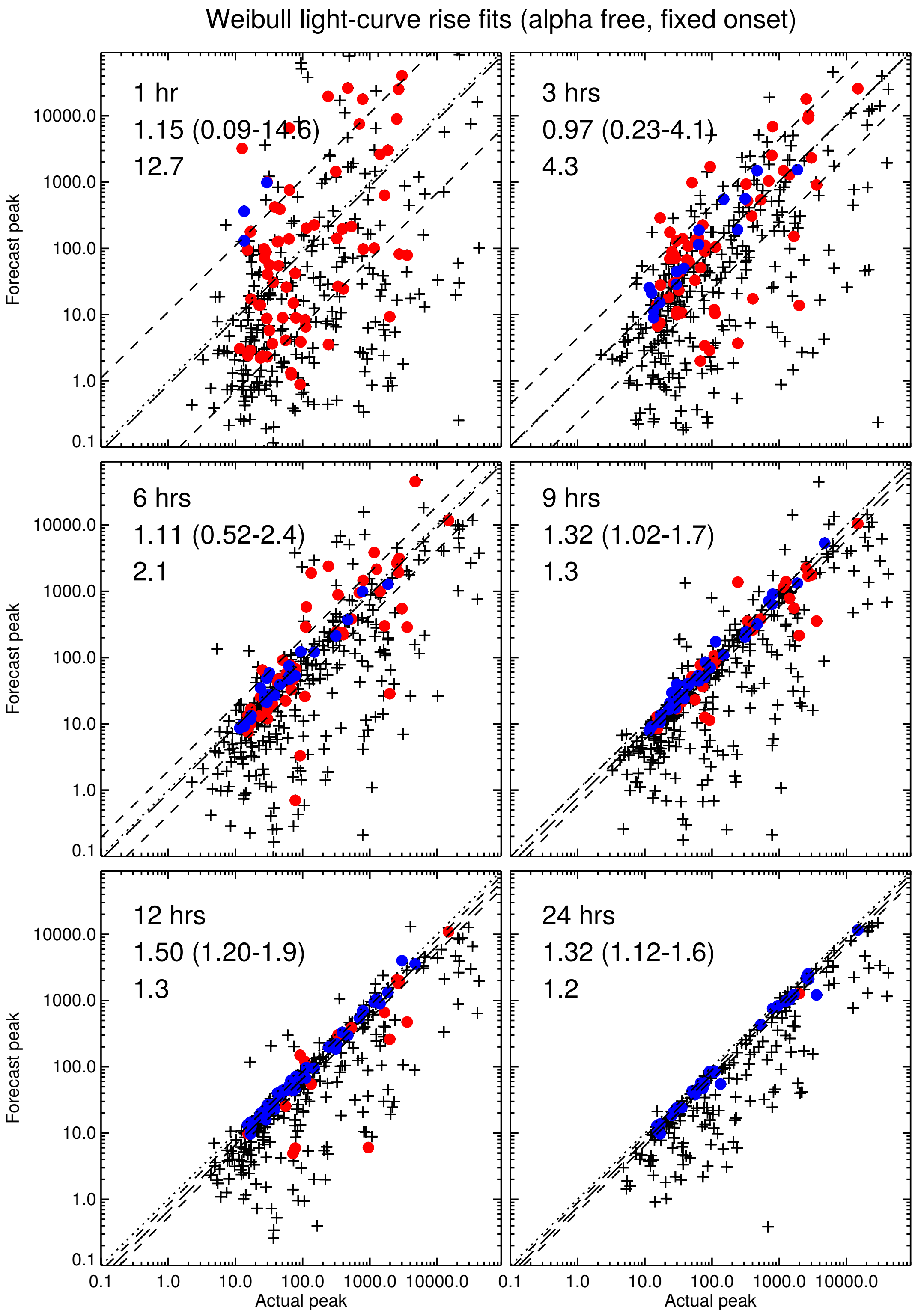

Fig. 56.— Comparison of the forecast maximum flux versus the actual peak flux for Weibull fits to light curves with fixed $\beta$ and $t_0$ and floating $\alpha$.

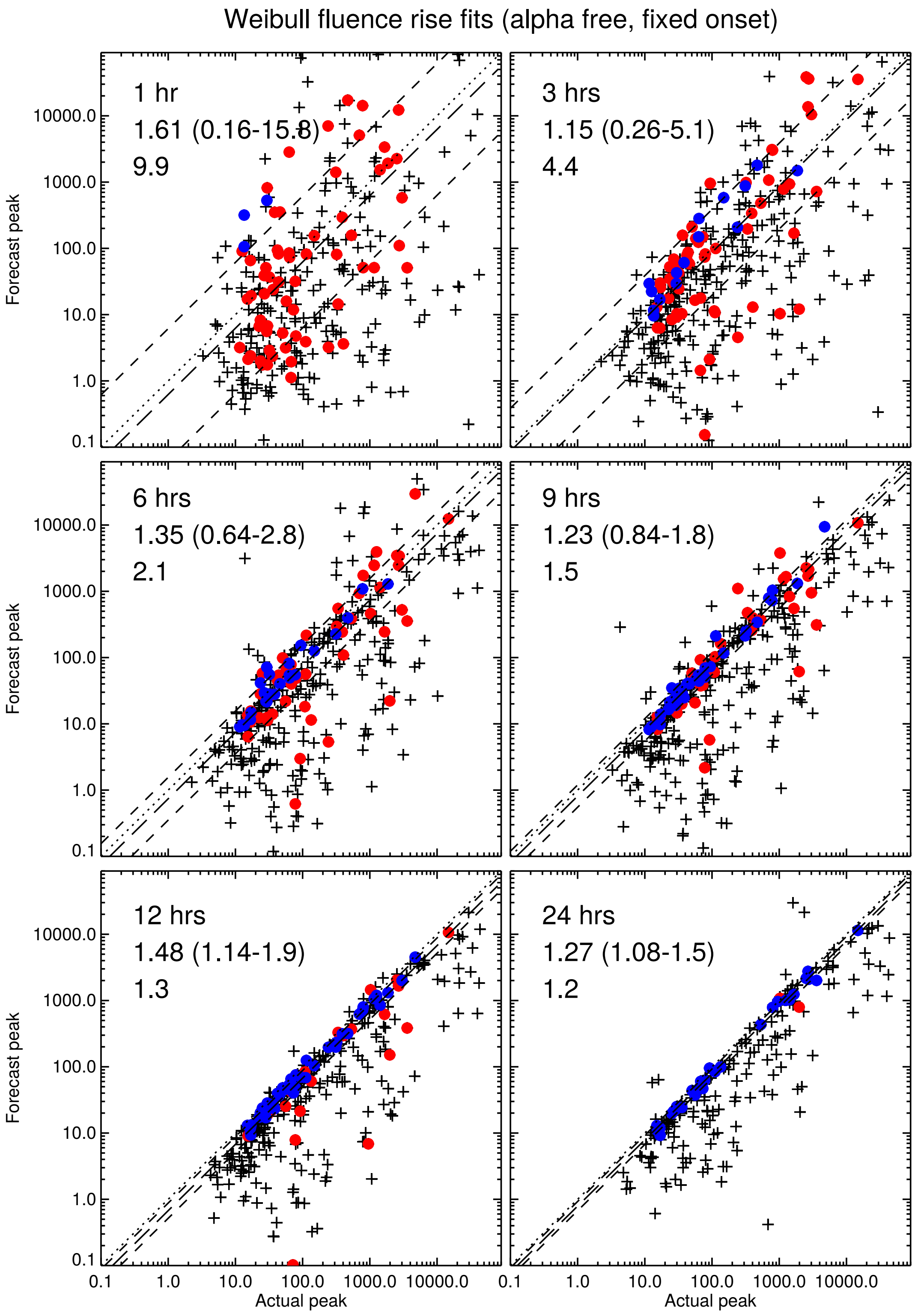

Fig. 57.— Comparison of the forecast maximum flux versus the actual peak flux for Weibull fits to cumulative fluence with fixed $\beta$ and $t_0$ and floating $\alpha$.

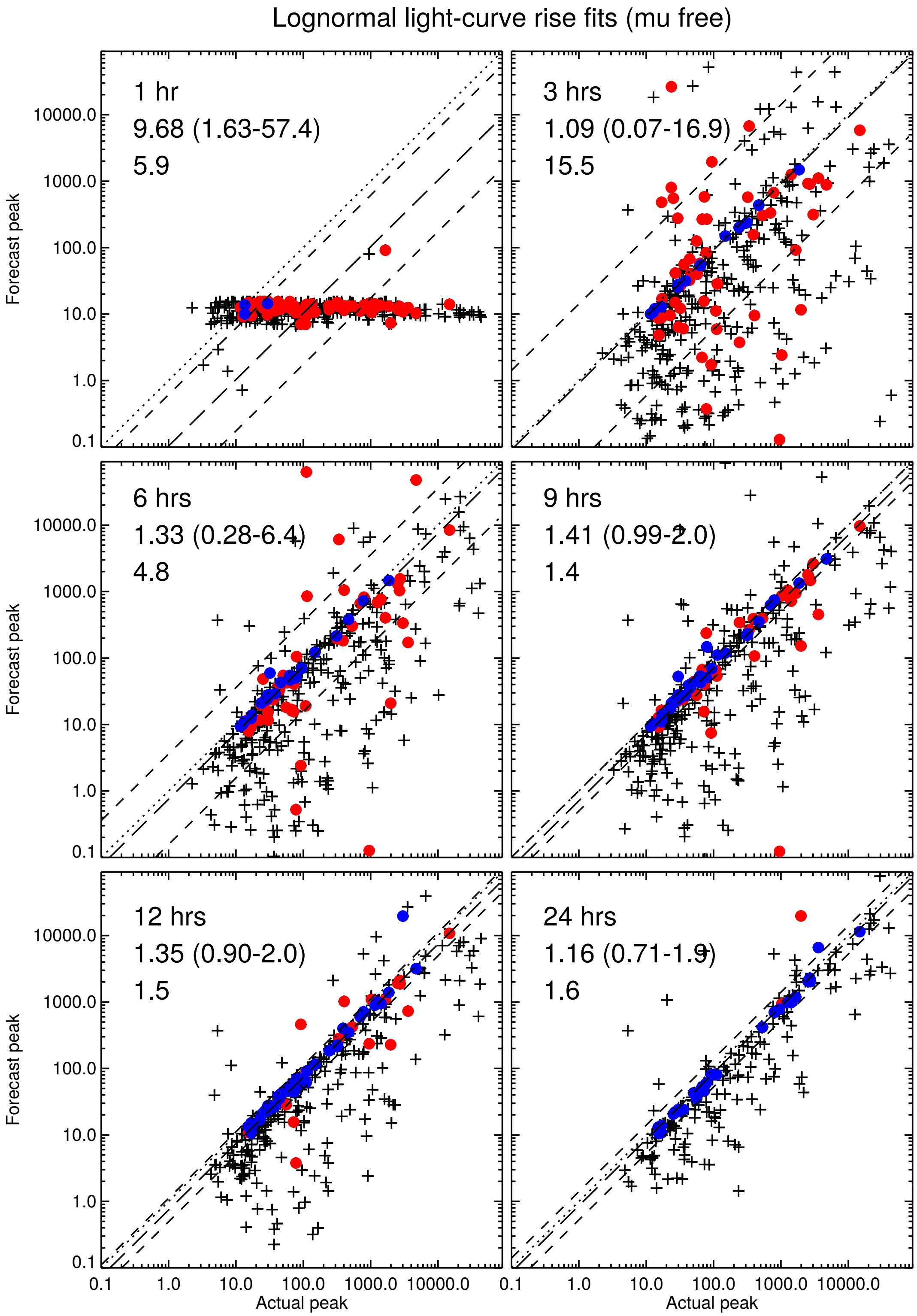

Fig. 58.— Comparison of the forecast maximum flux (vertical axis) versus the actual peak flux (horizontal axis) at 6 different times for lognormal fits to light curves with fixed $\sigma$ and floating $\mu$ and $t_0$.

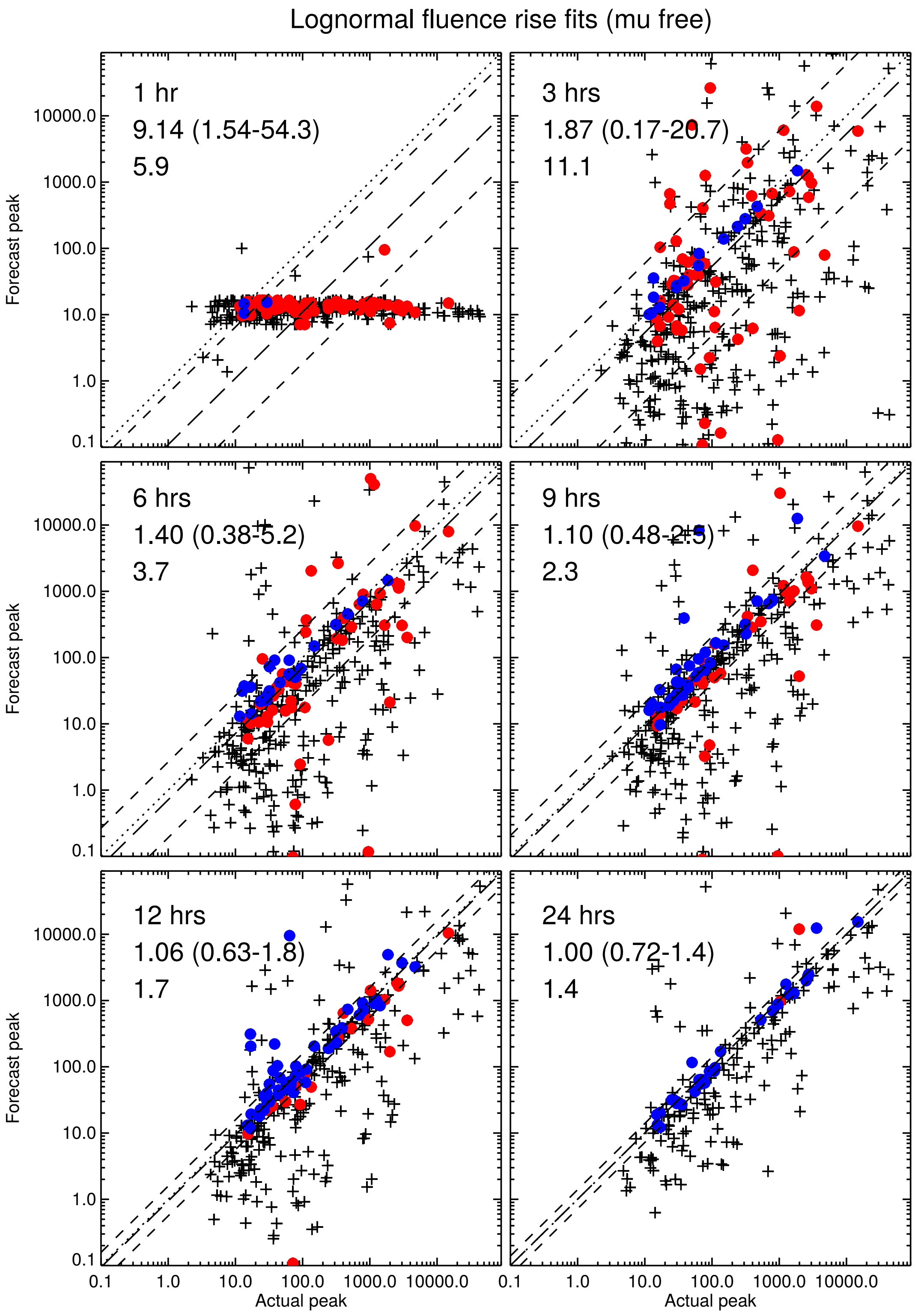

Fig. 59.— Comparison of the forecast maximum flux versus the actual peak flux for lognormal fits to cumulative fluence with fixed $\sigma$ and floating $\mu$ and $t_0$.

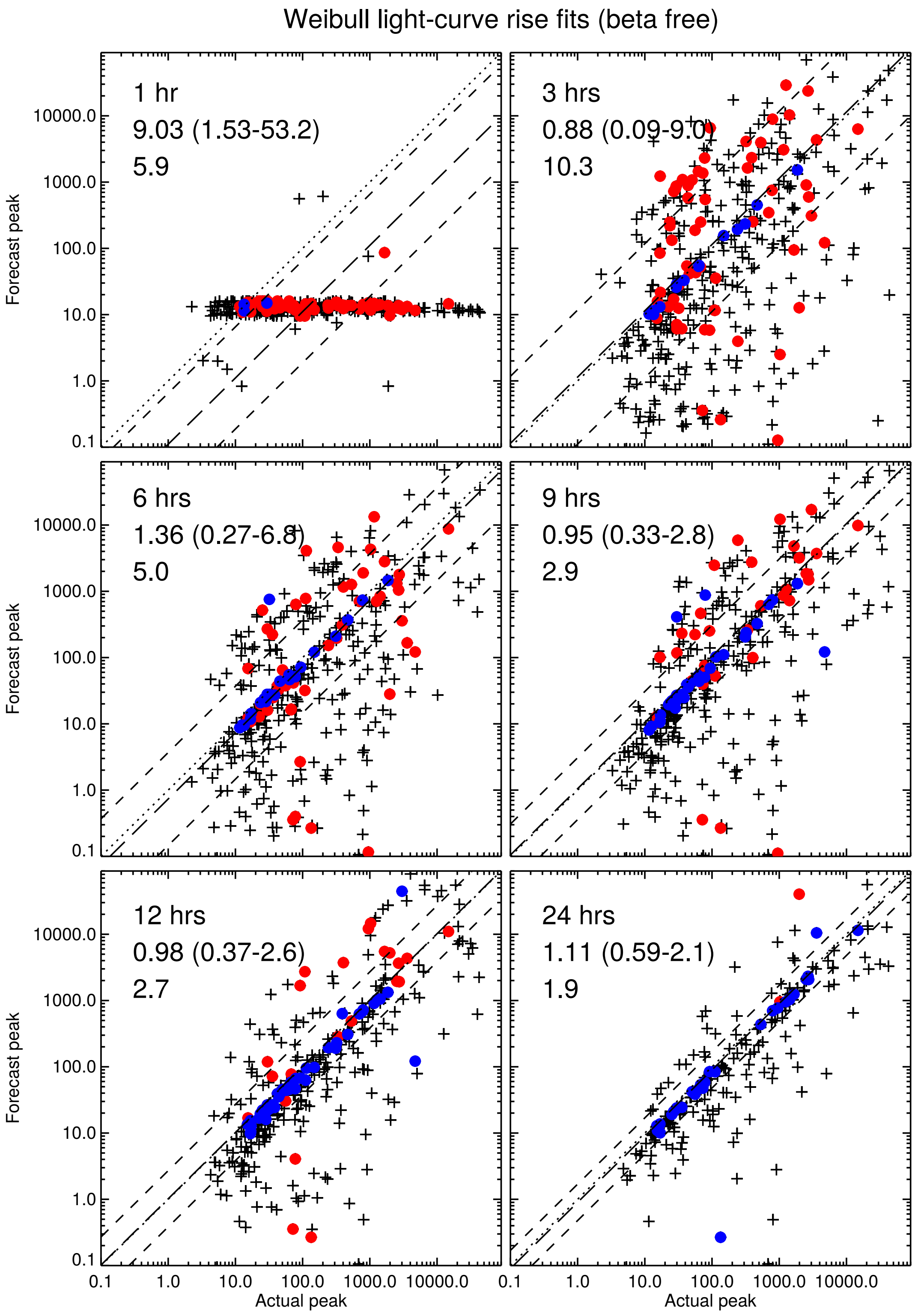

Fig. 60.— Comparison of the forecast maximum flux versus the actual peak flux for Weibull fits to light curves with fixed $\alpha$ and floating $\beta$ and $t_0$.

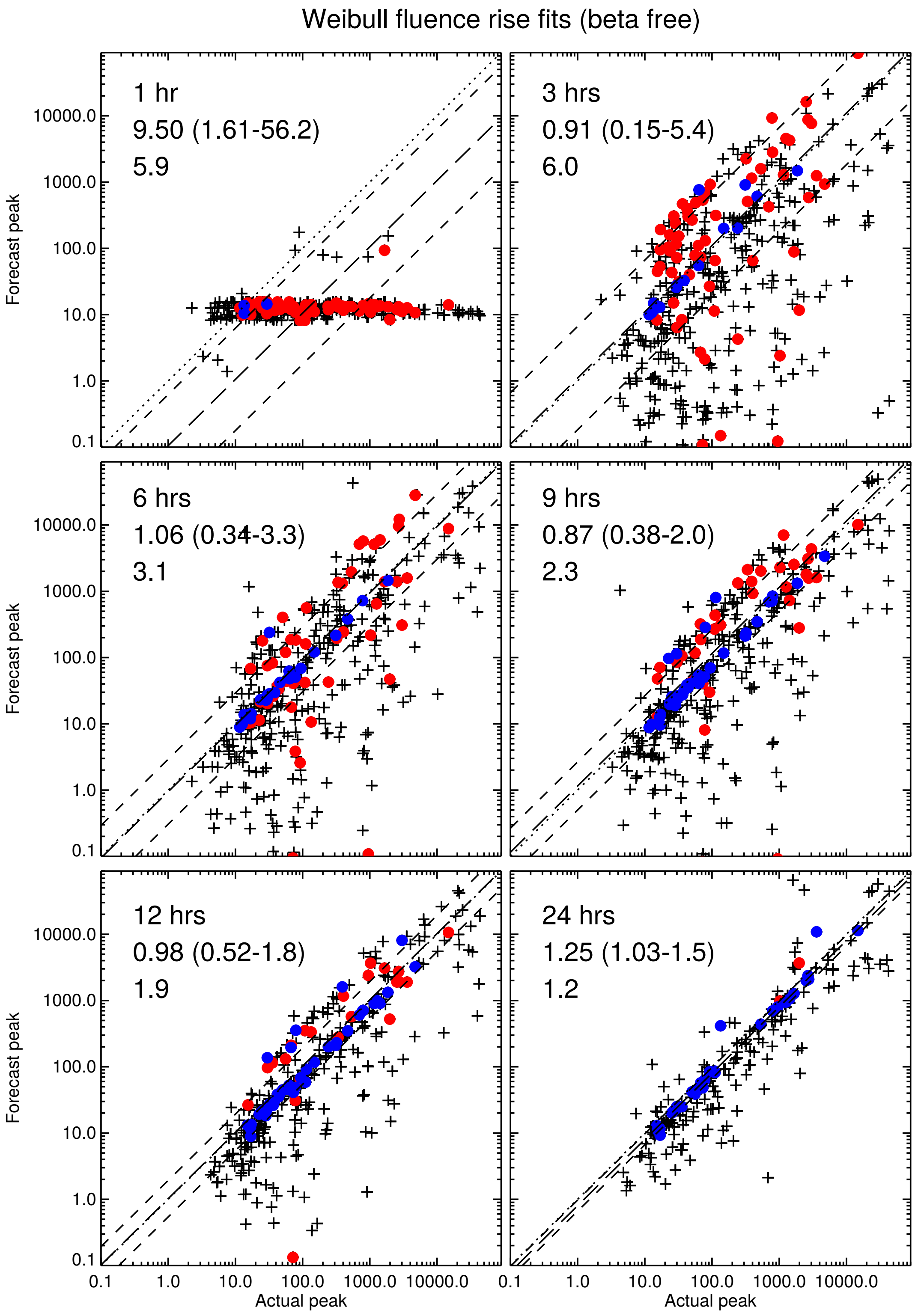

Fig. 61.— Comparison of the forecast maximum flux versus the actual peak flux for Weibull fits to cumulative fluence with fixed $\alpha$ and floating $\beta$ and $t_0$.

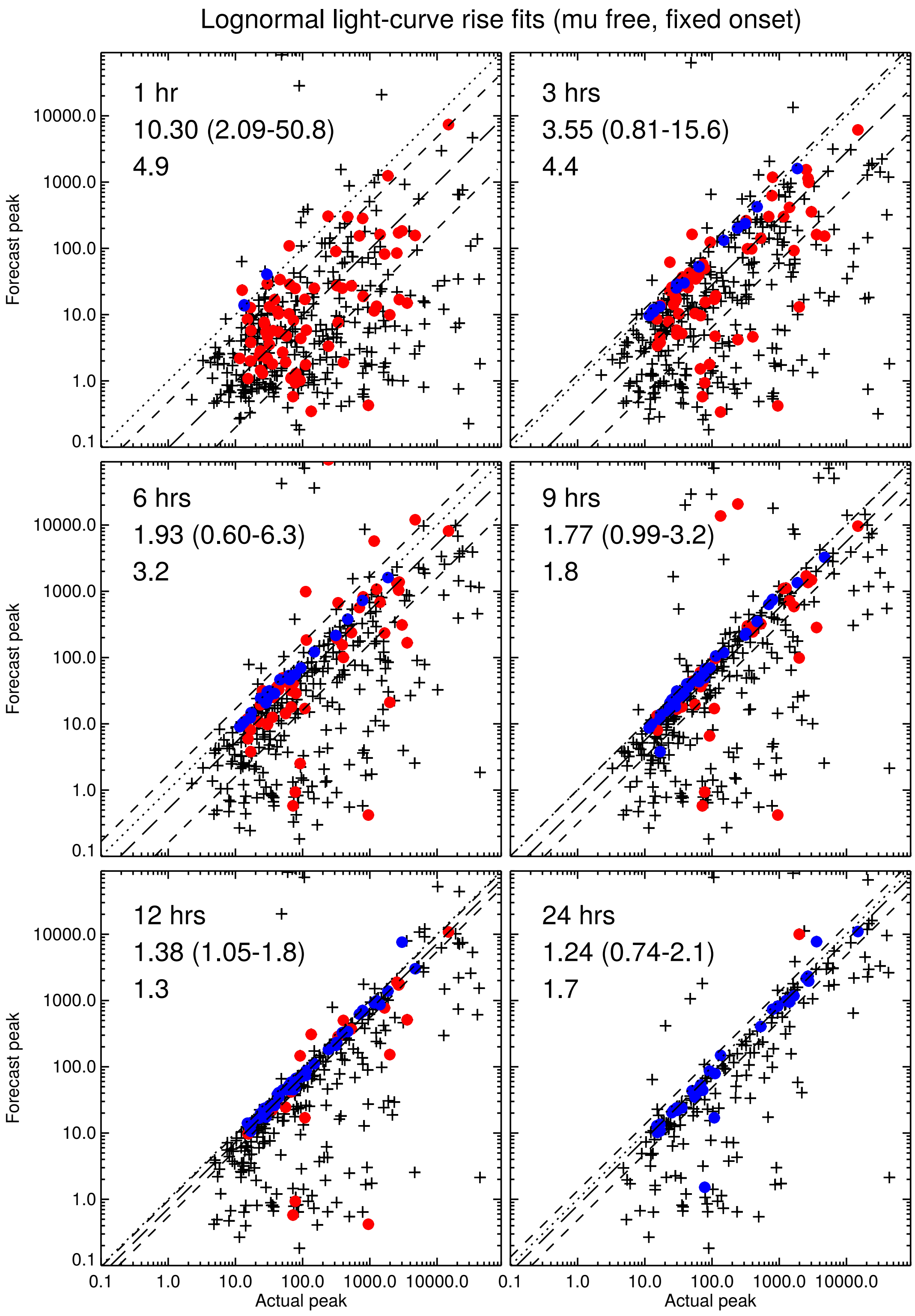

Fig. 62.— Comparison of the forecast maximum flux versus the actual peak flux for lognormal fits to light curves with fixed $\sigma$ and $t_0$ and floating $\mu$.

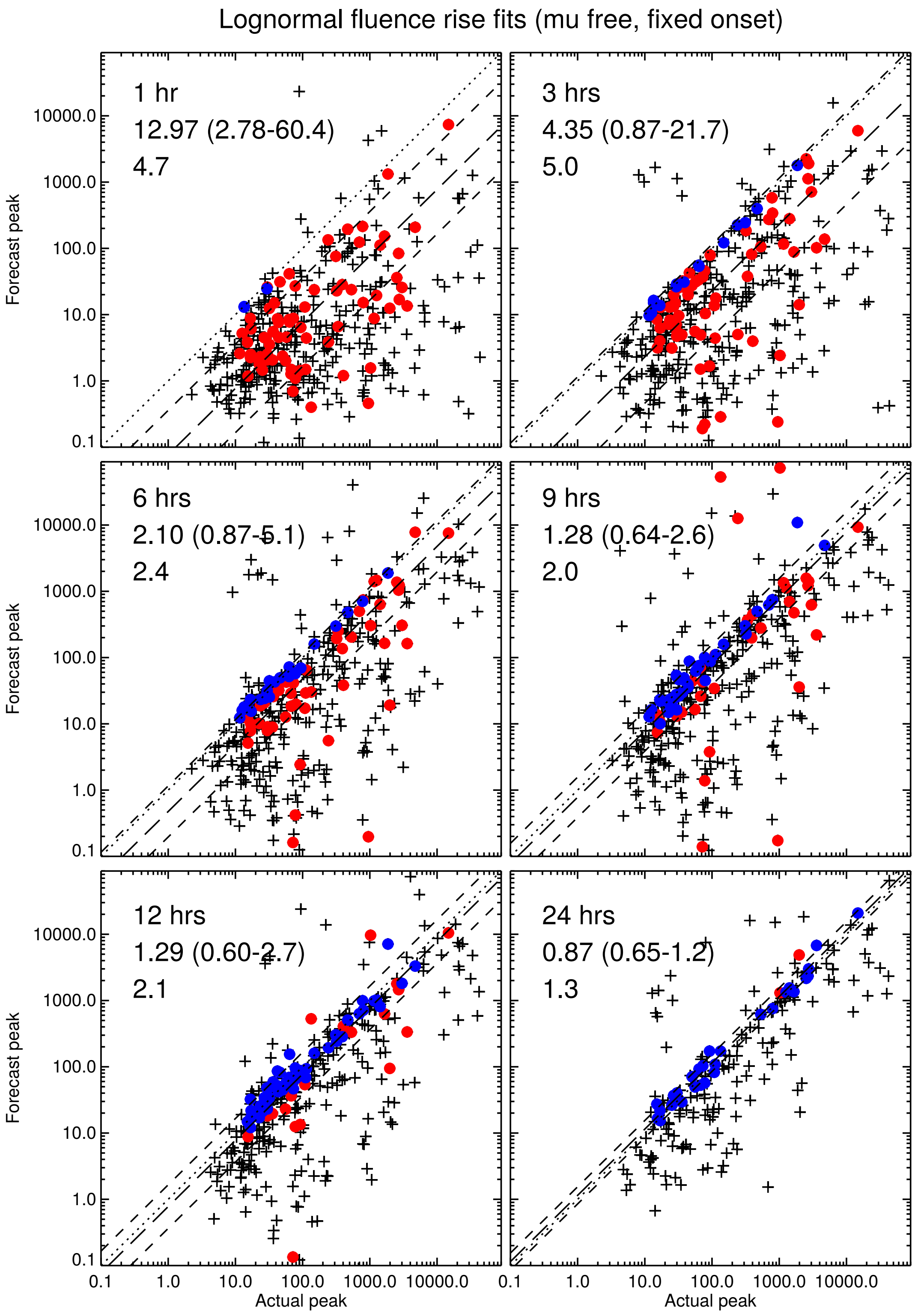

Fig. 63.— Comparison of the forecast maximum flux versus the actual peak flux for lognormal fits to cumulative fluence with fixed $\sigma$ and $t_0$ and floating $\mu$.

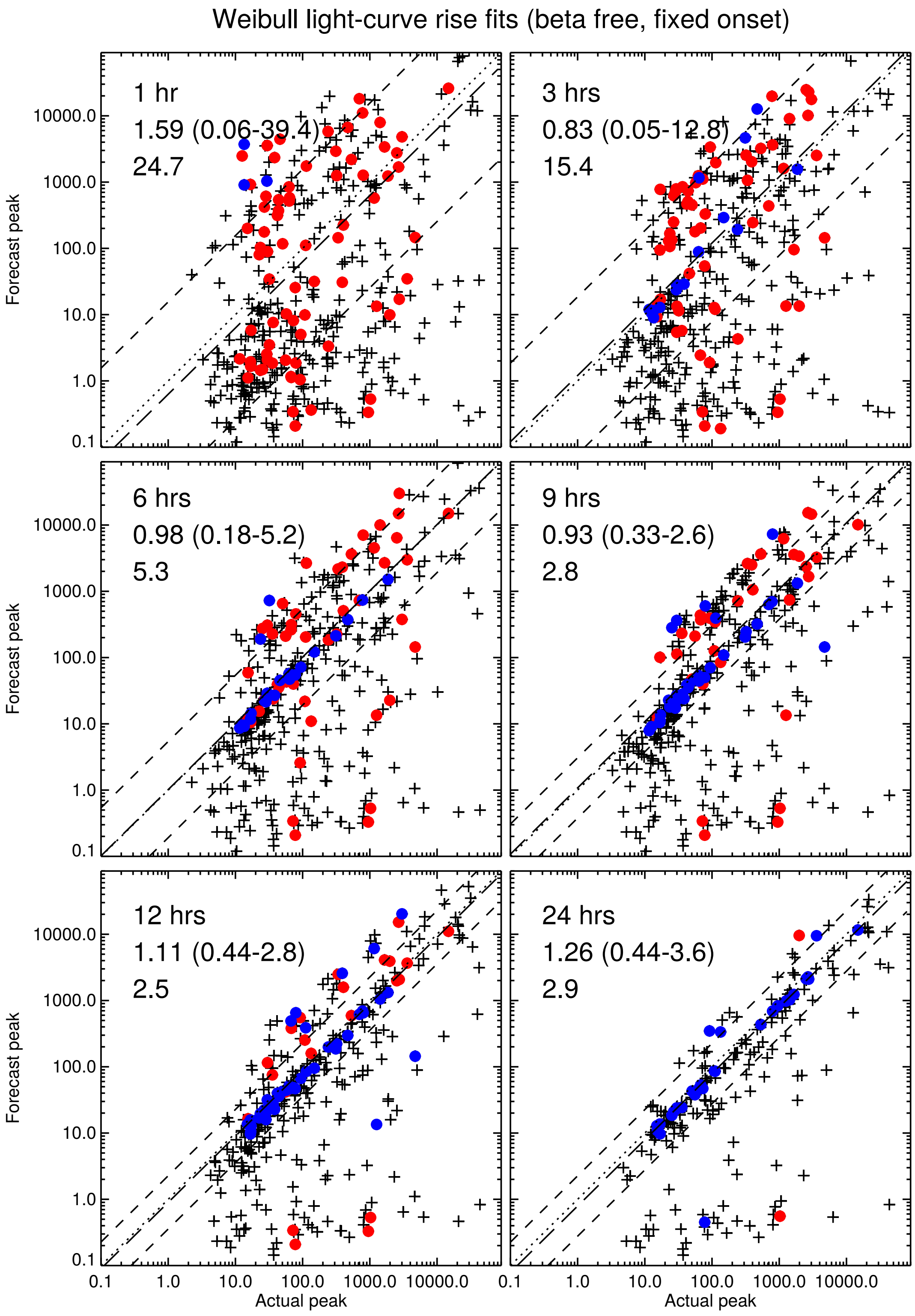

Fig. 64.— Comparison of the forecast maximum flux versus the actual peak flux for Weibull fits to light curves with fixed $\alpha$ and $t_0$ and floating $\beta$.

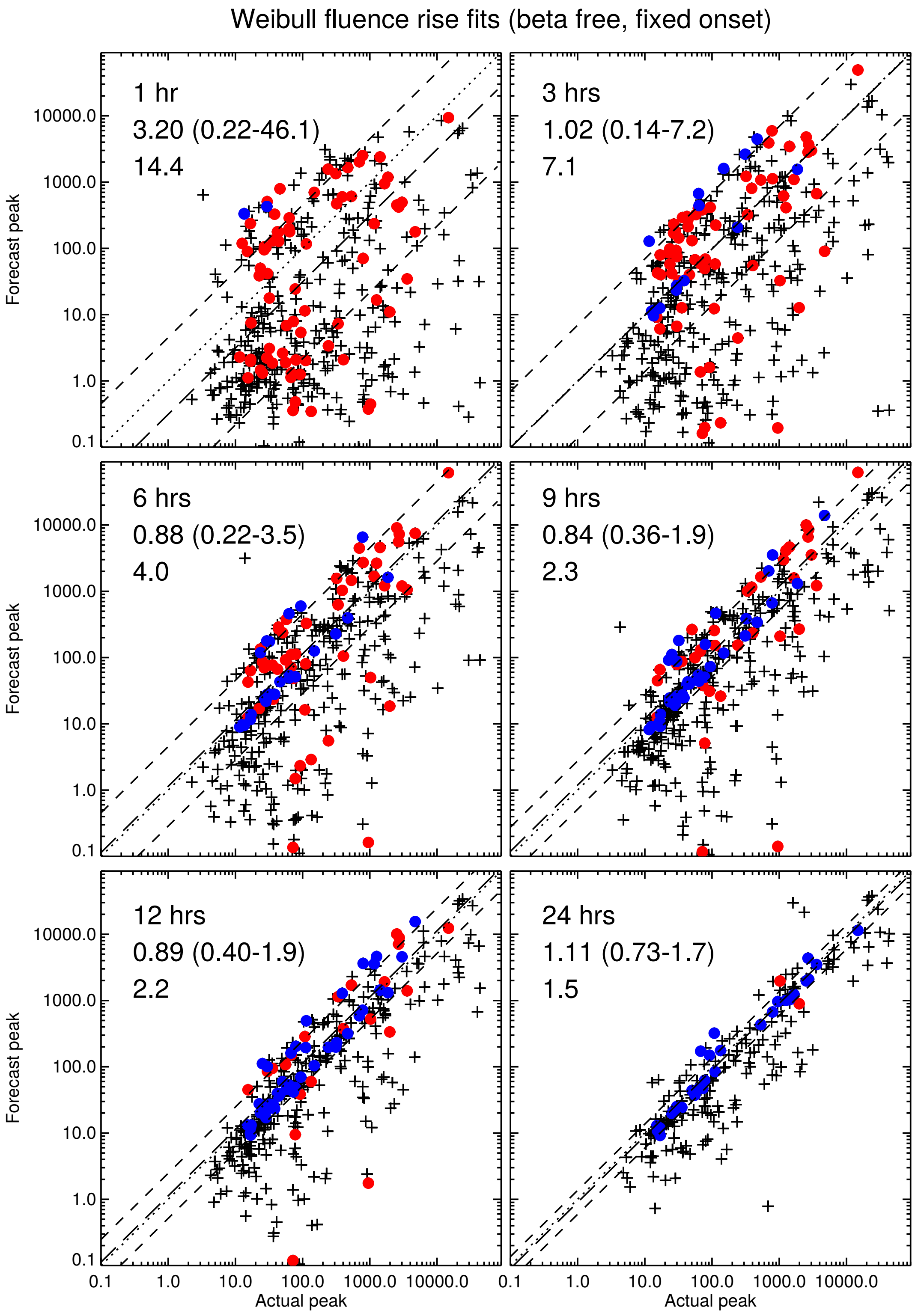

Fig. 65.— Comparison of the forecast maximum flux versus the actual peak flux for Weibull fits to cumulative fluence with fixed $\alpha$ and $t_0$ and floating $\beta$.

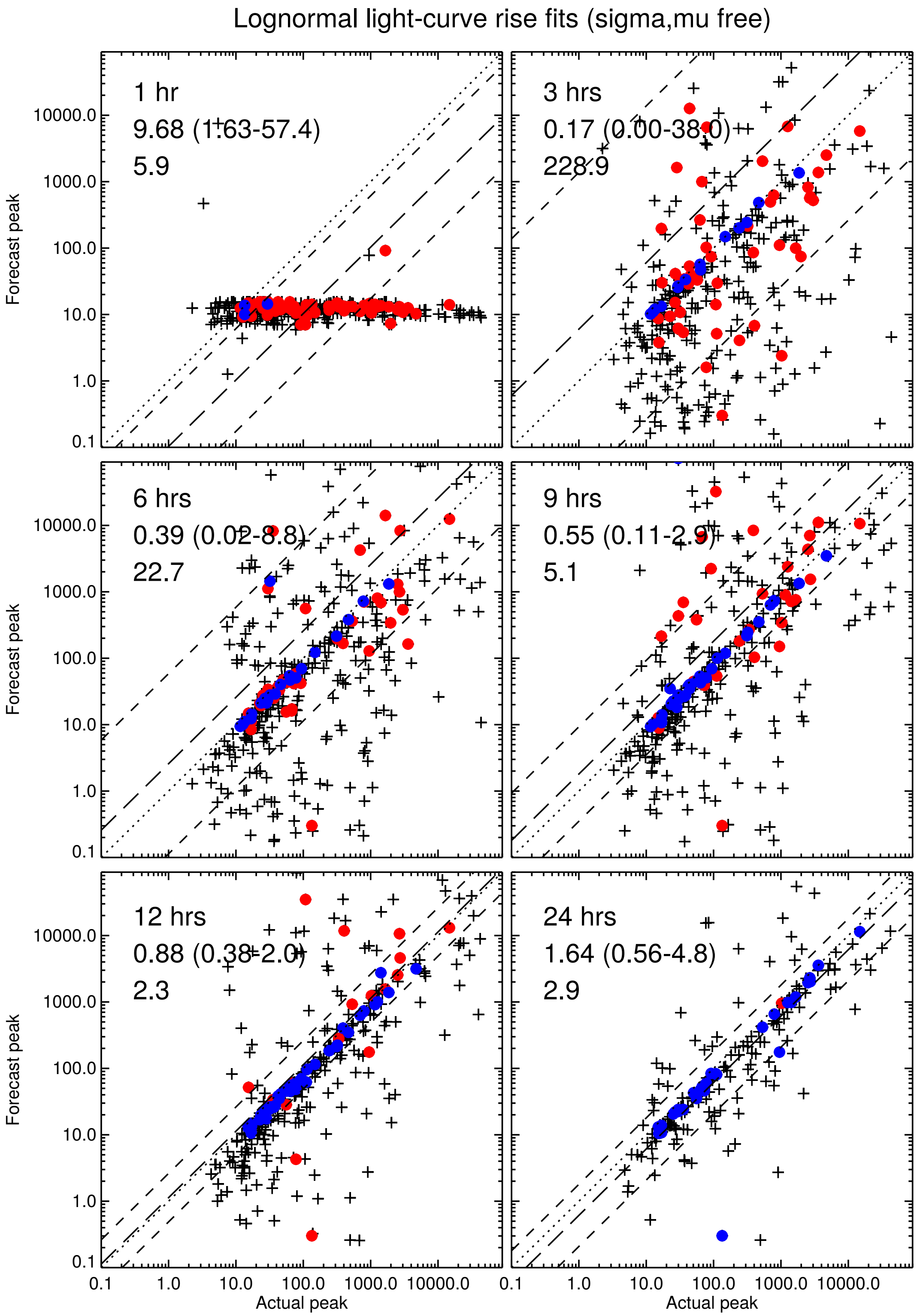

Fig. 66.— Comparison of the forecast maximum flux (vertical axis) versus the actual peak flux (horizontal axis) at 6 different times for lognormal fits to light curves with floating $\sigma$, $\mu$ and $t_0$.

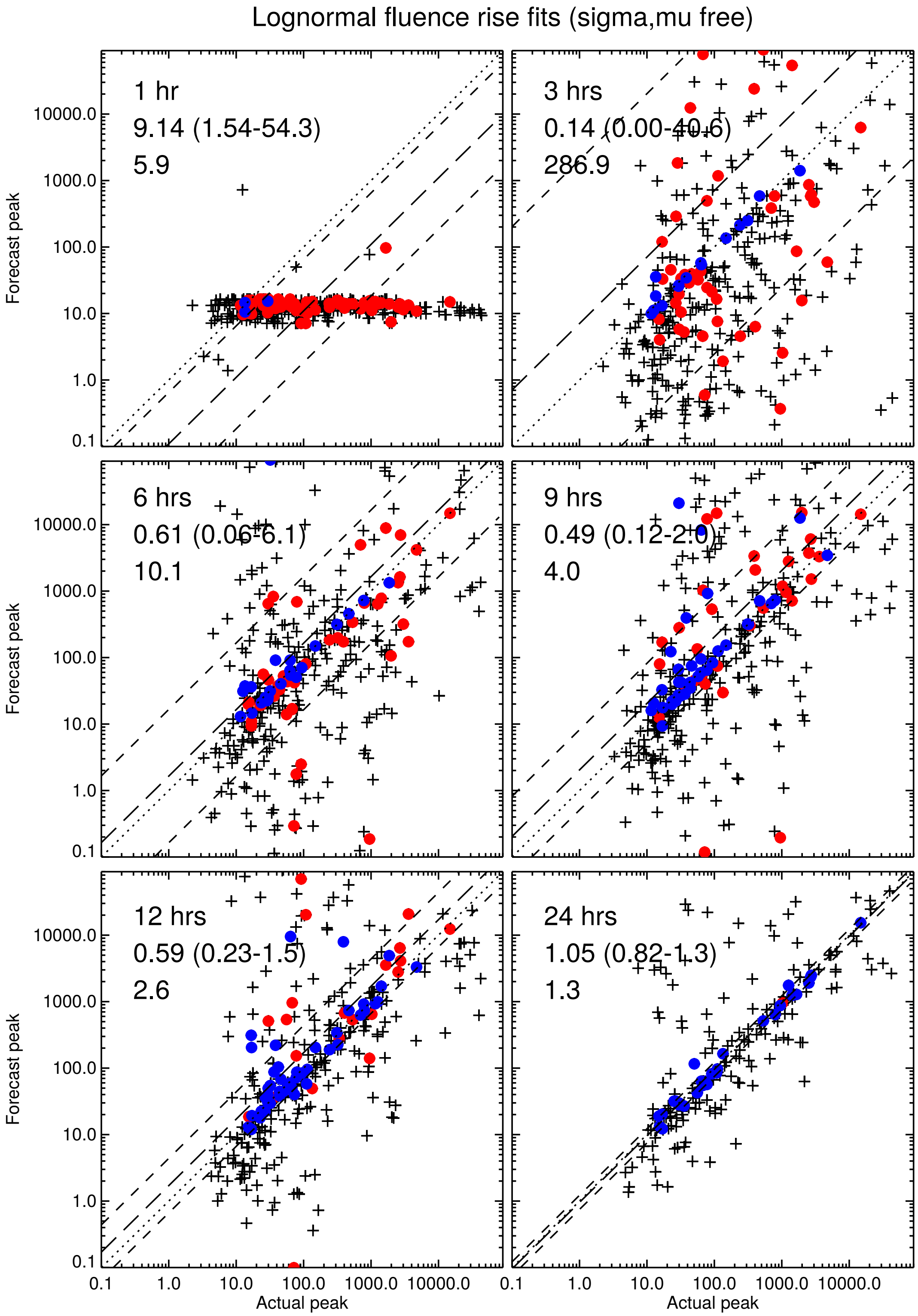

Fig. 67.— Comparison of the forecast maximum flux versus the actual peak flux for lognormal fits to cumulative fluence with floating $\sigma$, $\mu$ and $t_0$.

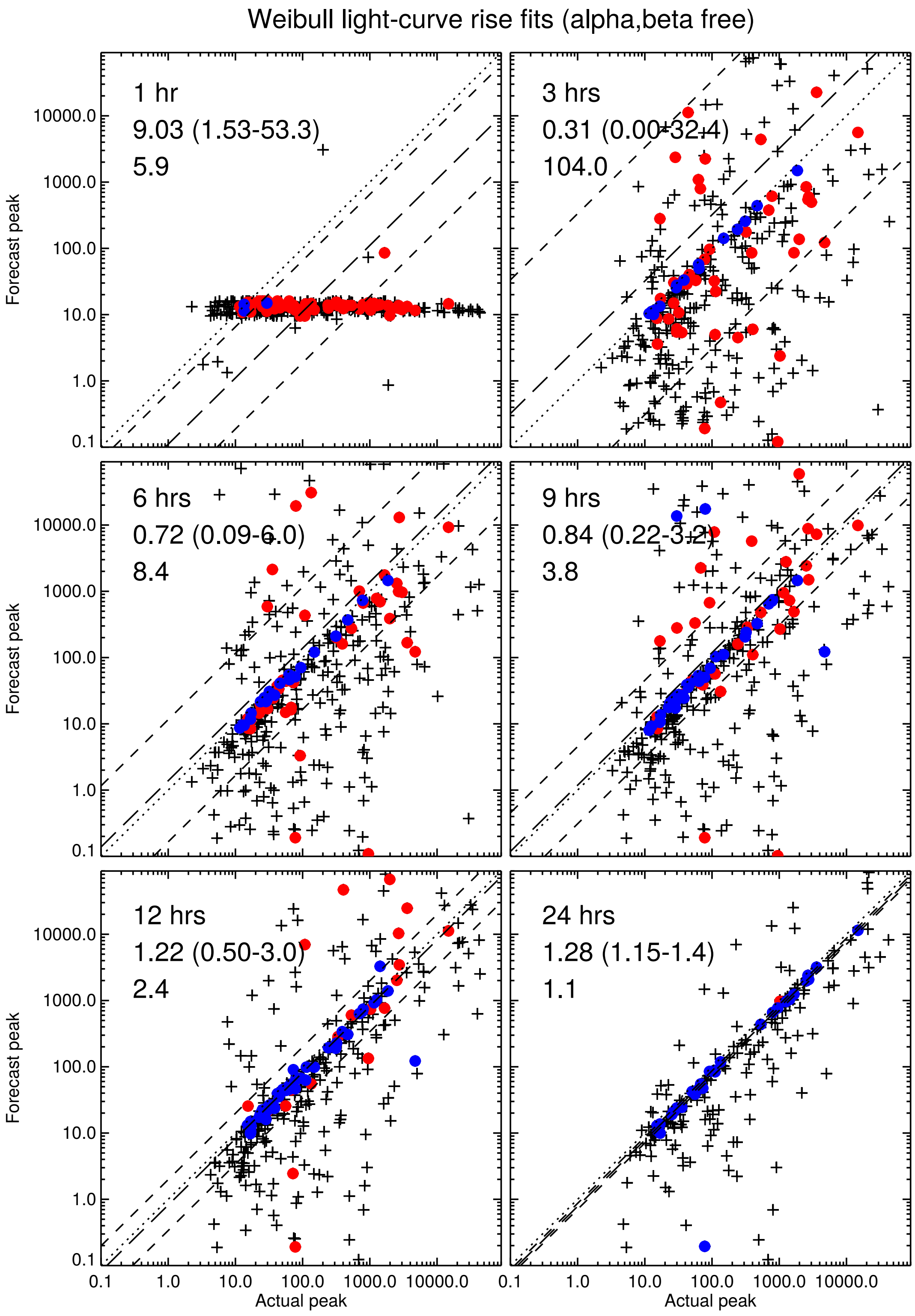

Fig. 68.— Comparison of the forecast maximum flux versus the actual peak flux for Weibull fits to light curves with floating $\alpha$, $\beta$ and $t_0$.

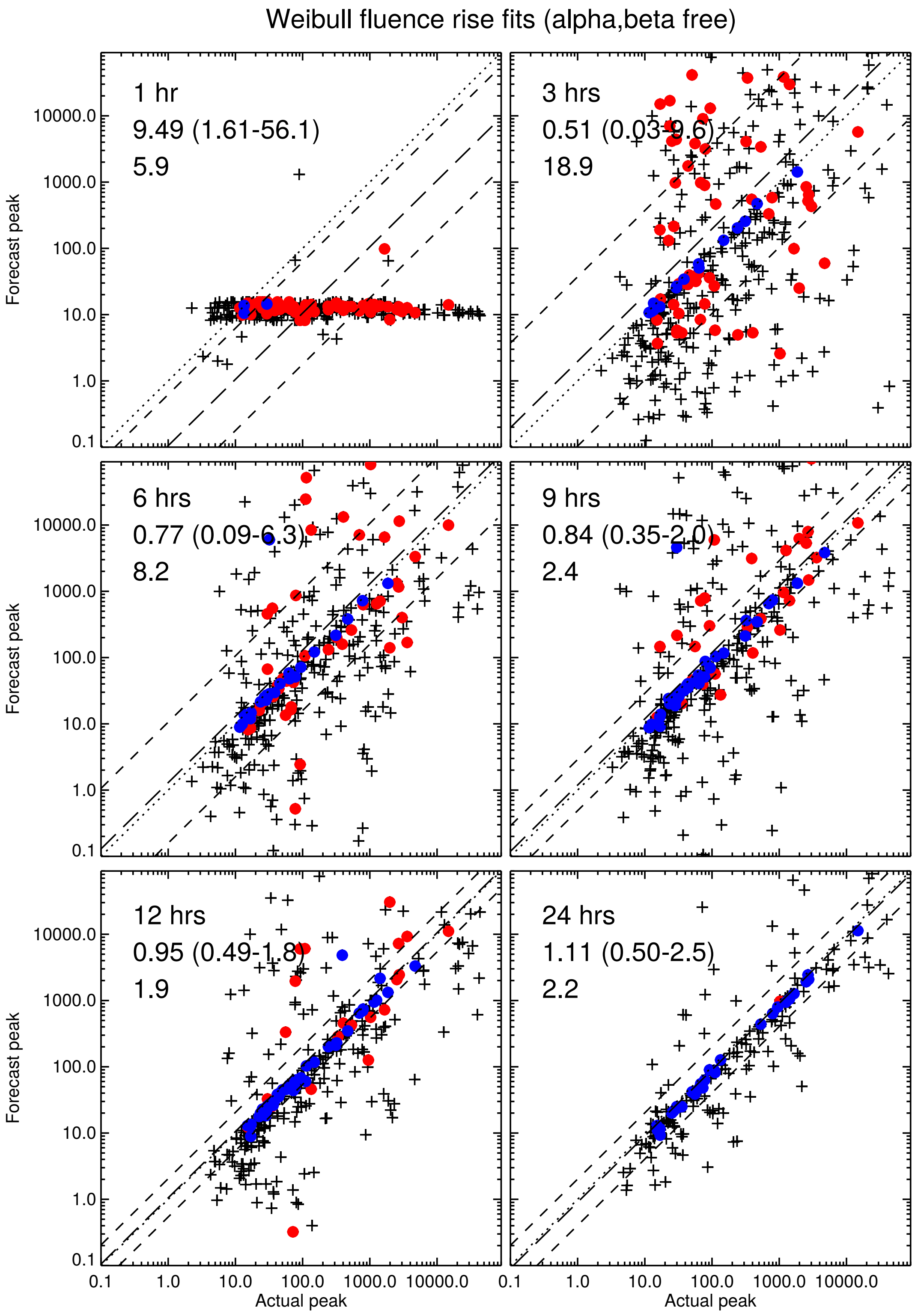

Fig. 69.— Comparison of the forecast maximum flux versus the actual peak flux for Weibull fits to cumulative fluence with floating $\alpha$, $\beta$ and $t_0$.

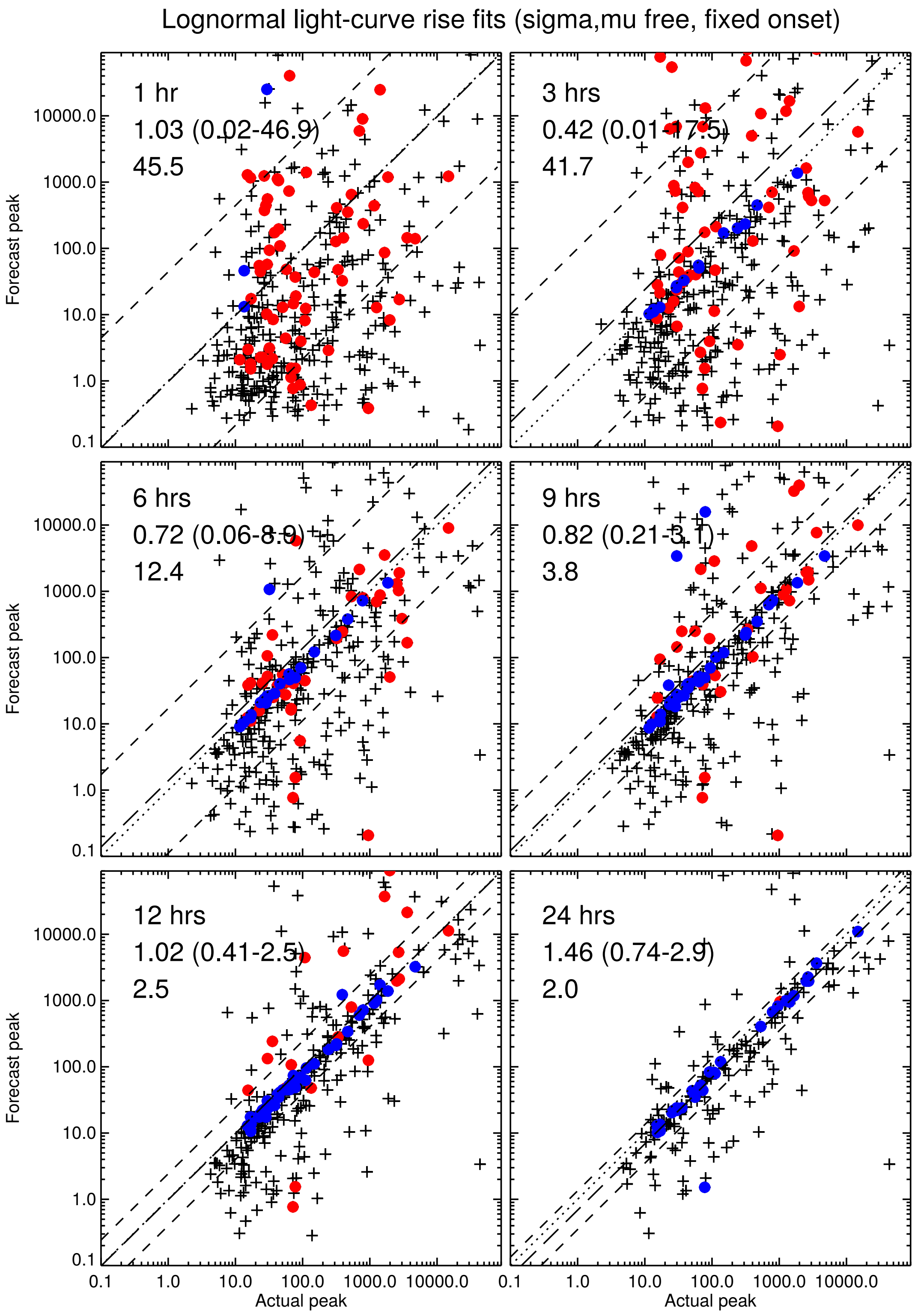

Fig. 70.— Comparison of the forecast maximum flux versus the actual peak flux for lognormal fits to light curves with fixed $t_0$ and floating $\sigma$ and $\mu$.

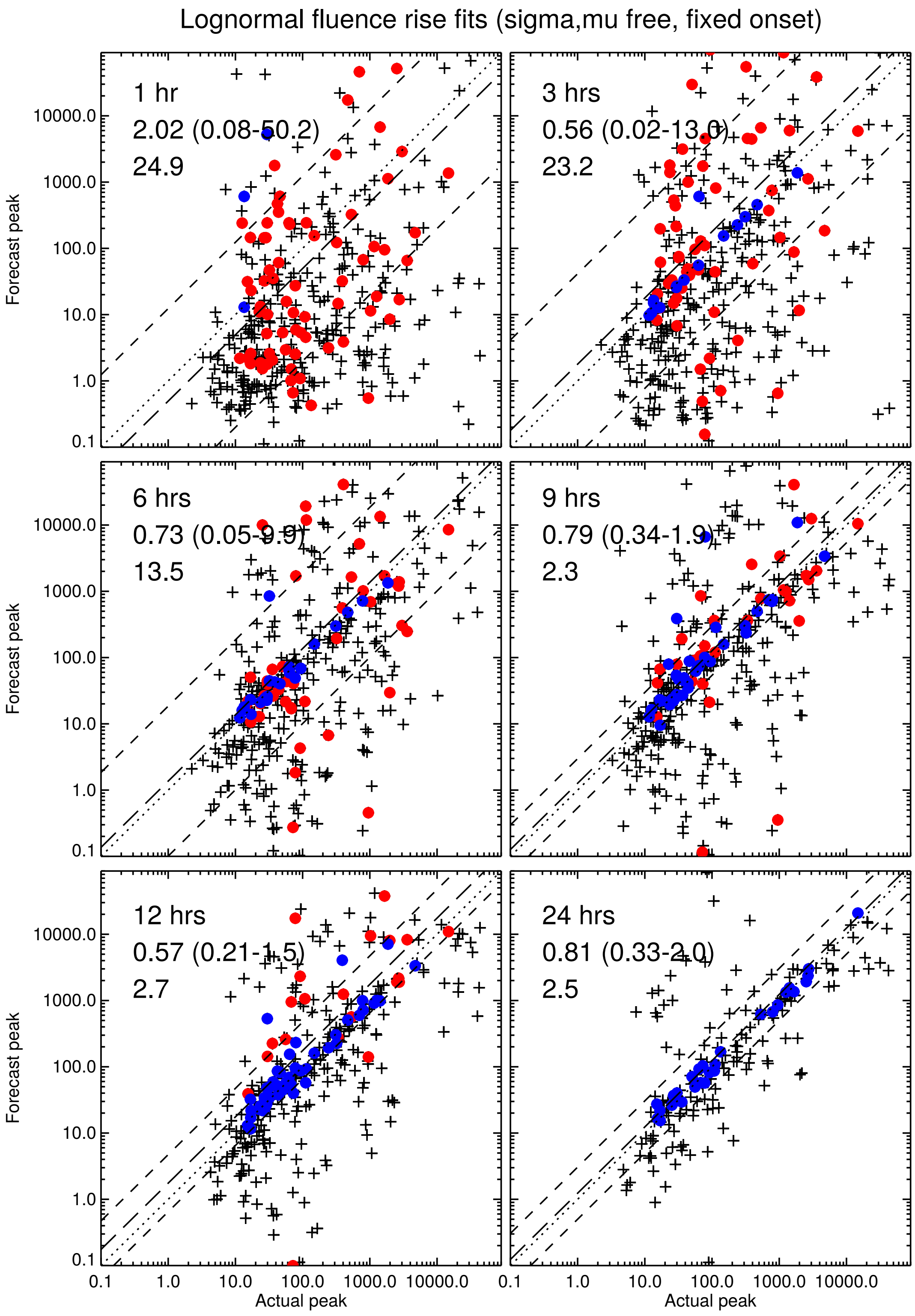

Fig. 71.— Comparison of the forecast maximum flux versus the actual peak flux for lognormal fits to cumulative fluence with fixed $t_0$ and floating $\sigma$ and $\mu$.

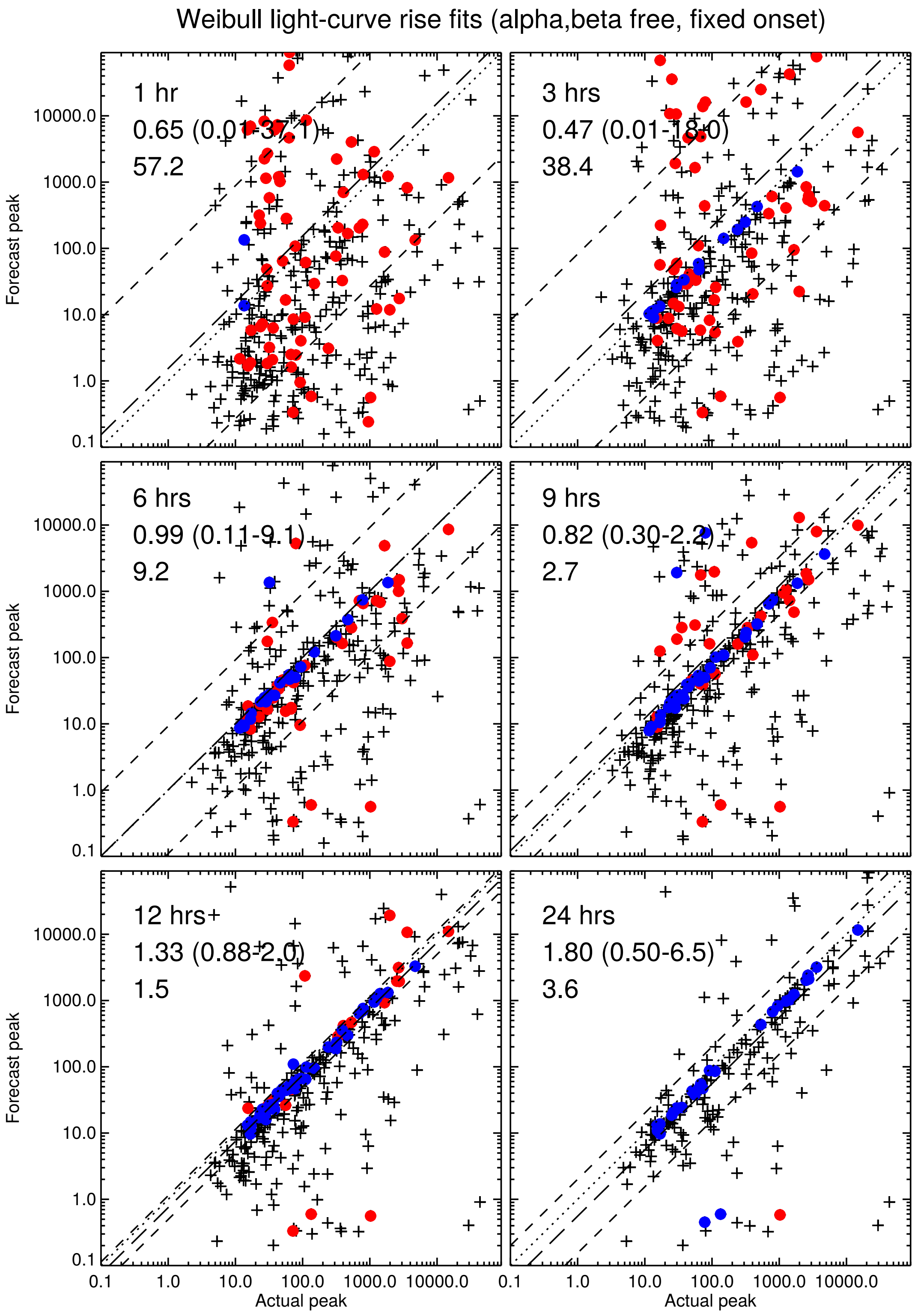

Fig. 72.— Comparison of the forecast maximum flux versus the actual peak flux for Weibull fits to light curves with fixed $t_0$ and floating $\alpha$ and $\beta$.

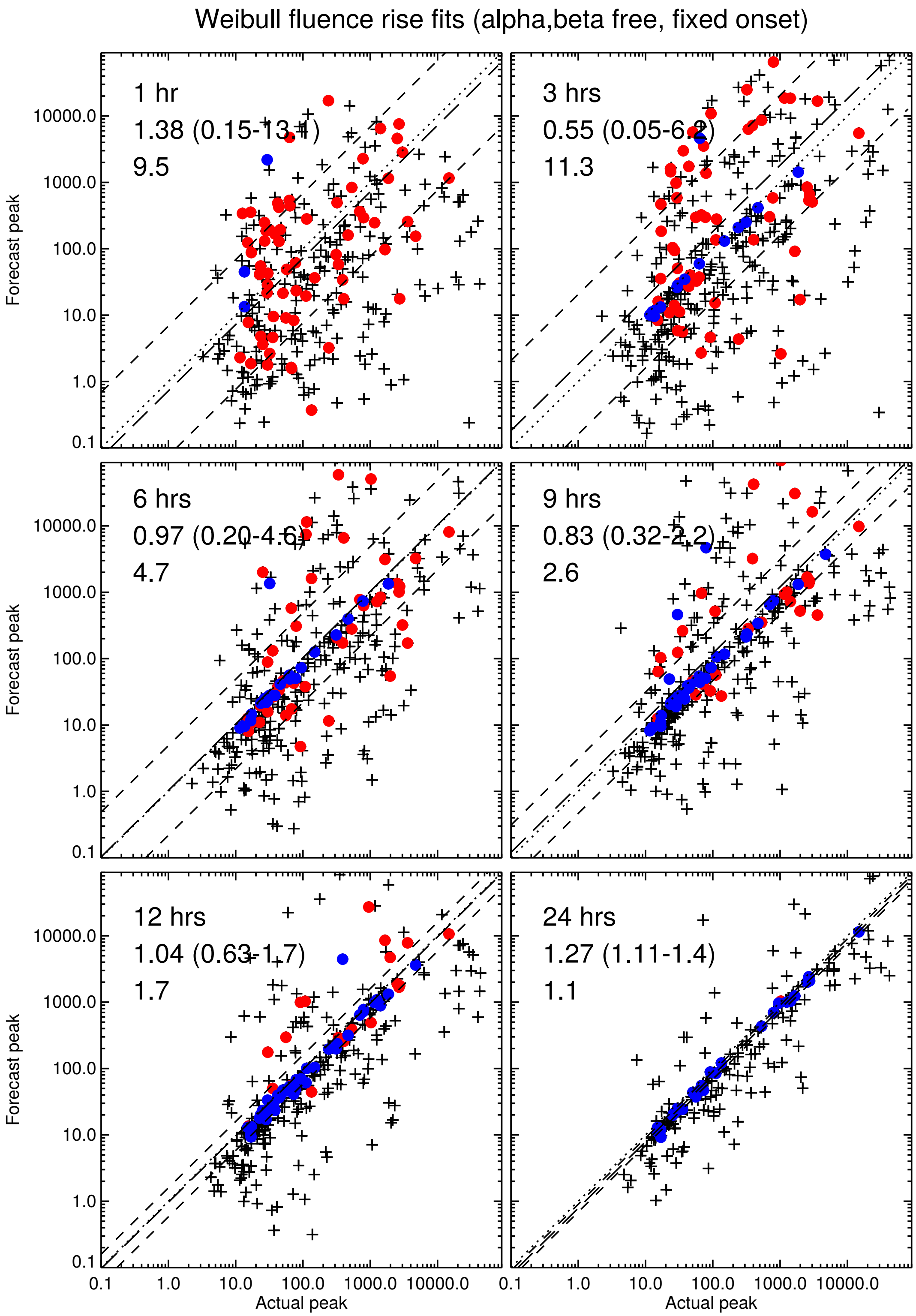

Fig. 73.— Comparison of the forecast maximum flux versus the actual peak flux for Weibull fits to cumulative fluence with fixed $t_0$ and floating $\alpha$ and $\beta$.

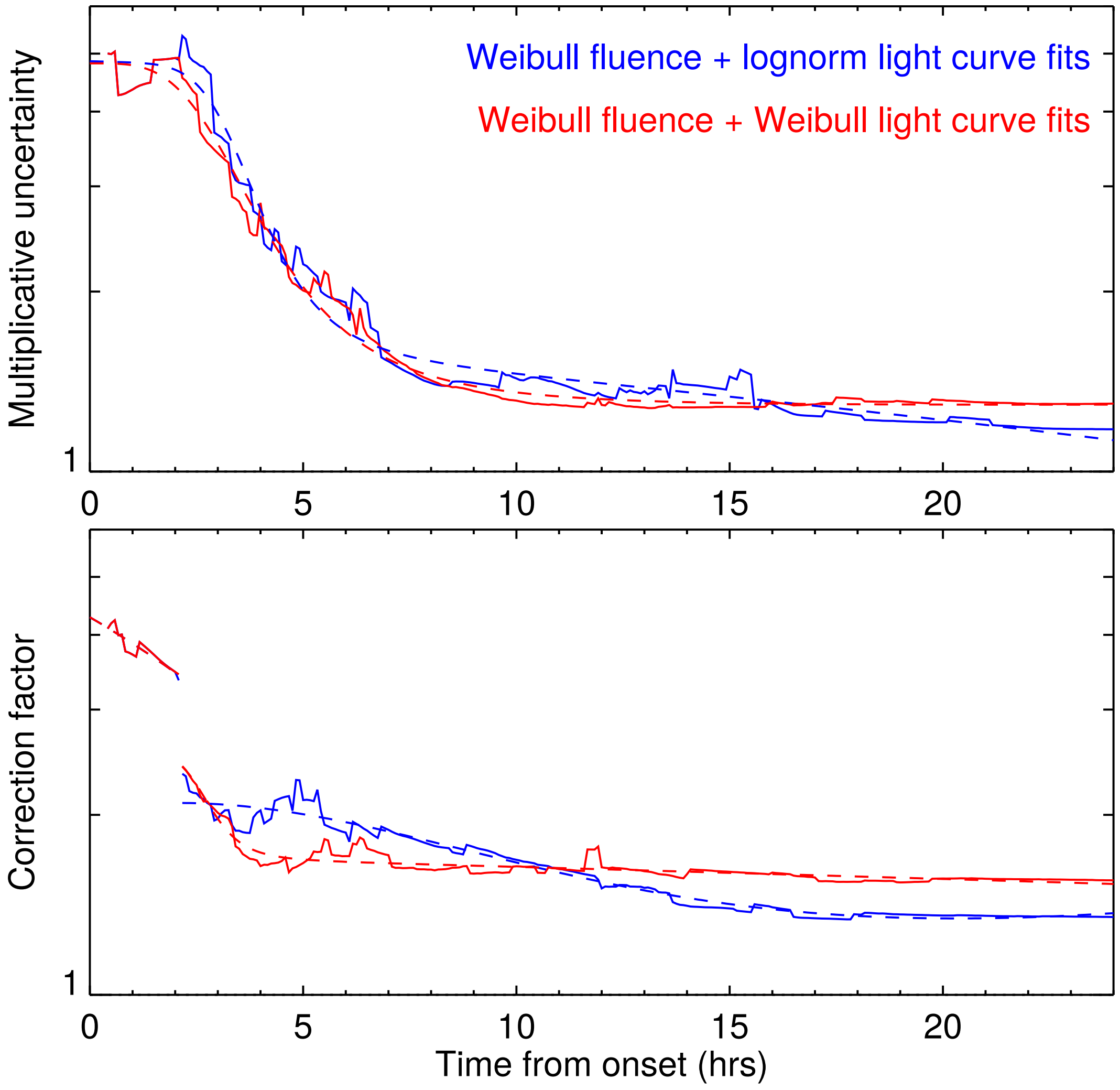

Fig. 74.— The multiplicative uncertainty in the correction factor (upper panel) and the corresponding correction factor (lower panel) versus time for the lognormal (blue) and Weibull (red) versions of the test light-curve forecasting process. Dashed lines present the analytic fits to the curves from equations (5)-(8).